\documentclass[iop]{emulateapj}

\usepackage{natbib}
\usepackage{amsmath, amssymb}
\usepackage{times}
\usepackage{apjfonts, graphicx, graphics}
\usepackage[breaklinks,colorlinks,urlcolor=blue,citecolor=blue]{hyperref}
\usepackage[all]{hypcap}
\usepackage{aas_macros}
\usepackage{subfigure}
\graphicspath{{figs_scalrel/}}

\shorttitle{X-ray scaling relations of early-type galaxies}
\shortauthors{Iu.V. Babyk et al.}

\begin{document}

% TITLE PAGE
\title{X-ray scaling relations of early-type galaxies}
\author{Iu.~V. Babyk$^{1,2,\ast}$}
\author{B.~R. McNamara$^{1,3}$}
\author{P.~E.~J. Nulsen$^{4,5}$}
\author{M.~T. Hogan$^{1,3}$}
\author{A.~N. Vantyghem$^{1}$}
\author{H.~R. Russell$^{6}$}
\author{F.~A. Pulido$^{1}$}
\author{A.~C. Edge$^{7}$}

\affil{
    $^{1}$ Department of Physics and Astronomy, University of Waterloo, 200 University Ave W, Waterloo, ON N2L 3G1, Canada\\
$^{2}$ Main Astronomical Observatory of NAS of Ukraine, 27 Academica Zabolotnogo St, 03143, Kyiv, Ukraine\\
$^{3}$ Perimeter Institute for Theoretical Physics, 31 Caroline St N, Waterloo, ON, N2L 2Y5, Canada \\
$^{4}$ Harvard-Smithsonian Center for Astrophysics, 60 Garden Str, Cambridge, MA 02138, USA \\
$^{5}$ ICRAR, University of Western Australia, 35 Stirling Hwy, Crawley, WA 6009, Australia \\
$^{6}$ Institute of Astronomy, Madingley Road, Cambridge, CB3 0HA, UK \\
$^{7}$ Department of Physics, University of Durham, South Road, Durham DH1 3LE, United Kingdom \\
    \\
}

%
% THEN INTO THE MAIN PAPER
%

\begin{abstract}
\hspace{0.5cm} X-ray luminosity, temperature, gas mass, total mass, and their scaling relations are derived for 94 early-type galaxies using archival $Chandra$ X-ray Observatory observations. Consistent with earlier studies, the scaling relations, $L_X \propto T^{4.5\pm0.2}$, $M \propto T^{2.4\pm0.2}$, and $L_X \propto M^{2.8\pm0.3}$,  are significantly steeper than expected from  self similarity.  This steepening indicates that their atmospheres are heated above the level expected from gravitational infall alone.  Energetic feedback from nuclear black holes and supernova explosions are likely heating agents. The tight $L_X - T$ correlation for low-luminosities systems (i.e., below 10$^{40}$ erg/s) are at variance with hydrodynamical simulations which generally predict higher temperatures for low luminosity galaxies. We also investigate the relationship between total mass and pressure, $Y_X = M_g \times T$, finding $M \propto Y_{X}^{0.45\pm0.04}$. We explore the gas mass to total mass fraction in early-type galaxies and find a range of $0.1-1.0\%$. We find no correlation between the gas-to-total mass fraction with temperature or total mass. Higher stellar velocity dispersions and higher metallicities are found in hotter, brighter, and more massive atmospheres. X-ray core radii derived from $\beta$-model fitting are used to characterize the degree of core and cuspiness of hot atmospheres.
\end{abstract}

\keywords{
    galaxies: elliptical and lenticular, X-rays: galaxies 
}

\altaffiltext{*}{
    \href{mailto:ibabyk@uwaterloo.ca, babikyura@gmail.com}{ibabyk@uwaterloo.ca, babikyura@gmail.com}
}

\section{Introduction}\label{sec:1}
In the $\Lambda$CDM cosmogony, small dark matter density fluctuations grew through the influence of gravity to create today's  massive dark matter haloes. Assuming structure developed primarily by gravitational forces and that cooling was negligible, the gas temperature, luminosity, and halo mass should scale as self-similar power laws (\citealt{Kaiser:86}). Atmospheric temperatures in the bremsstrahlung regime ($kT \gtrsim$ 2 keV) should scale with mass as $T \propto M^{2/3}$.  Likewise, X-ray luminosity should scale with temperature as $L_X \propto T^2$. Numerical simulations of hot atmospheres that respond to gravity alone have confirmed the self-similarity of the X-ray scaling relations (\citealt{Evrard:96, Bryan:98, Thomas:01, Voit:02, Voitr:05}). However, observation has revealed significant deviations from self-similarity. The clearest departures are the slopes which steepen toward lower masses.

%Besides the well-established Fundamental Plane, Faber-Jackson, Kormendy and color-magnitude scaling relations, the X-ray galaxy scaling relations can be intensively used as a probe of galaxy structure and evolution as well \citep{Khos:04}. Recent X-ray scaling relation researches have been performed mostly for the relations between optical and X-ray luminosity which links the stellar mass with their X-ray luminosity, and for the $L_X-T$ as well. They have been widely used to explore the origin and evolution of the interstellar medium (ISM) of early-type galaxies \citep[e.g., see ][for more details]{Fabbiano:89, White:91, David:91}. ETGs (elliptical and lenticular, E \& S0) host atmospheres of hot diffuse gas \citep{Pellegrini:94, Sarazin:97}. They are mostly old systems and consist of old, red giant stars. However, they also demonstrate a little star formation activity \citep{Su:13}. Moreover, recent observational results showed the detection of cold molecular gas and dust in ETGs (\citealt{Huang:09}, for more details see review by \citealt{Sarzi:08}). According to the self-similar model, the physical properties of different size objects should scale similarly, i.e. since galaxy clusters, galaxy groups and galaxies contain hot gas, the halo properties should be similar \citep{Navarro:97, OSullivan_sample:03}. 

Galaxy clusters scale as $L_X \propto T^{2.7-3.0}$ and $T \propto M^{1.5-1.7}$ \citep{Vikhlinin:06, Vikhlinin:09}. Groups and early-type galaxies (ETGs) depart more significantly from self-similarity, $L_X \propto T^{3-5}$ and $T \propto M^{2-3}$ \citep[][and references therein]{Boroson:10, Kim:13}.  These departures indicate that processes beyond gravity alone, such as radiative cooling, supernova heating, and feedback by active galactic nuclei (AGN) are significant. Their relative contributions are unclear. However, these processes affect low mass systems the most due to the lower gravitational binding energy of their atmospheres (\citealt{Giodini:13}). 

Numerical simulations that incorporate radio-AGN feedback are able to reproduce the observed X-ray scaling relations of clusters and ETGs quiet well (\citealt{Borgani:05, Borgani:06, Sijacki:06, Puchwein:08, Borgani:09, Booth:10, Schaye:10}), at least on large scales. Our understanding of radio-AGN feedback has advanced rapidly over the past decade (see e.g. \citealt{McNamara:07, McNamara:12}) primarily through X-ray spectral-imaging studies of hot atmospheres (\citealt{Birzan:04, Rafferty:06, Shin:16, Hlavacek:15}).  X-ray cavities inflated by radio jets embedded in hot atmospheres provide an accurate measure of feedback energetics.  Combined with studies of sound waves and weak shock fronts (\citealt{Nulsen:05, Borgani:05, Borgani:06, Borgani:09, Randall:15, Fabian:17, Forman:17}),  this work has shown that radio AGN release enough energy to offset radiative cooling while affecting the X-ray scaling relations in clusters \citep{Main:17}.   
%By extending studies of X-ray scaling relations from clusters to early type galaxies we are able to
%further probe the long-term impact of AGN feedback on hot atmospheres and galaxy evolution.

Studies of the scaling relations have often targeted galaxy clusters and groups \citep{OSullivan:01, David:06, Boroson:10, Babyk:14a, Babyk_book:15, Su:15, Goulding:16}.  Less massive ETGs are more difficult to study and thus have not received the same level of attention.  Nevertheless, the X-ray scaling relations of ETG are sensitive to the  origin and evolution of the interstellar medium of ETGs and their host galaxies \citep[see e.g.][for more details]{Bender:89, Fabbiano:89, Mathews:90, White:91, David:91, Mathews:03, Khos:04}. However, the X-ray analysis of ETGs is complicated by X-ray emission from low-mass X-ray binaries (LMXBs) and other stellar sources that contribute to the total X-ray emission. Their contributions must be estimated reliably to obtain meaningful measurements of the atmospheric properties \citep{Revnivtsev:07a, Revnivtsev:08a, Boroson:10, Kim:13, Kim:15}. The immediate aim of this paper is to do just this in a longer-term effort to evaluate the degree to which AGN feedback and supernova explosions affect galaxy evolution over several decades in halo mass.

Recent studies of ETG scaling relations have concerned small samples exploring $L_X-T$, $L_X-M$, and $L_X-L_B$. Here, we perform uniform analysis 94 ETGs within 5$r_e$\footnote{Here, $r_e$ is the half-light radius.} taken from the $Chandra$ archive.
%The 5$r_e$ was chosen to be consistent with previous measurements (e.g., \citet{Mathews:06, Kim:13, Forbes:16, Alabi:16} performed their results within 5$r_e$). \citet{Alabi:16} argued that dark matter within 5$r_e$ is expected to dominant over the enclosed mass. Thanks to the high resolution of Chandra instruments, we are able to perform careful as possible spectral analysis, to find the contribution of each component in the total X-ray radiation (i.e., hot gas, CVs, ABs, unresolved LMXBs, etc.), and, that is more significant, to subtract an X-ray emission produced by number of components. 
We investigate these scaling relations in addition to $M-M_g$ and $L_X-\sigma_c$, and the first time for ETGs, $M-M_g \times T$. We further study the structural and dynamical properties of ETGs.

The paper is organized as follows. The review of our sample and data analysis is presented in Section~\ref{sec:2}. Section~\ref{sec_spec_anal} describes the spectral analysis. The surface brightness, density, and mass calculations are described in Section~\ref{sec_mass}. Section~\ref{sec_rel} describes the results of the luminosity-temperature, mass-temperature, luminosity-mass, and mass$-Y_X$ scaling relations. The results are discussed in Section~\ref{sec_disc} and our conclusions are presented in Section~\ref{sec_summary}.

We assume a standard $\Lambda$CDM cosmology with the following parameters: $H_0$ = 70 km s$^{-1}$ Mpc$^{-1}$, $\Omega_{\Lambda}$ = 0.7, $\Omega_{M}$ = 0.3. The quoted measurement uncertainties refer to the 1$\sigma$ confidence level, unless otherwise specified. 

\section{Early-type galaxy sample and data analysis}\label{sec:2}

Based on the samples of \citet{Beuing_sample:99} and \citet{OSullivan_sample:03}, \citet{Nulsen:09} selected a sample of 104 nearby objects with following criteria: (1) $L_K > $ 10$^{10} L_{\odot}$, (2) absolute magnitude $M_{B} < $ -19, morphological T-type $<$ -2, (3) Virgo-centric flow corrected recession velocity. Only 87 objects have been observed with $Chandra$. To increase our sample we added galaxies from the ATLAS$^{3D}$ (\citealt{Cappellari:11}) and MASSIVE (see \citet{Ma_sample} for sample selection details) samples that were observed by $Chandra$, namely 61 and 39 targets, respectively. In total we analyzed about 150 targets. We selected observations with cleaned exposure times above 10~ks in order to eliminate large uncertainties during spectral analysis. This excludes about half of the targets from the ATLAS$^{3D}$ sample. Our final sample contains 94 objects, which is 1.5 times larger than previous studies of X-ray scaling relations in ETGs. We used LEDA\footnote{Lyon-Meudon Extragalactic Database, \cite{Paturel:97}}, SIMBAD\footnote{http://simbad.u-strasbg.fr/} and NED\footnote{https://ned.ipac.caltech.edu/} databases to classify the objects. The observations have been downloaded from the HEASARC\footnote{http://heasarc.gsfc.nasa.gov/} archive. 

The final ETG sample is shown in Table~\ref{tab1}. The angular and luminosity distances are measured using redshifts from NED. We assumed $D_A$ = 16.5 and $D_L$ = 17.5 Mpc for Virgo galaxies (see \citealt{Cappellari:11}; marked with star in columns 10 and 11).  We include several non-Virgo galaxies whose redshifts are too low to respond reliably to the Hubble flow: NGC1386, NGC3079, NGC4278, NGC4457, and NGC4710. For these we use distances derived from surface brightness fluctuations \citep{Mei:07}. Our sample covers a wide range of distance and includes elliptical, lenticular, SB, BCGs, and cD galaxies. In some cases the morphological type differs between NED and SIMBAD. For example, NGC383, NGC507, NGC3665, NGC4382, NGC4477, and NGC4526 are classified by NED as SAB0 galaxies, while in SIMBAD these objects were classified as S0. We have found about 30 galaxies with discrepant morphological classifications. \citet{Kim:15} claimed that these disparities indicate misclassification due to dust obscuration and/or hidden disks.

The galactic coordinates taken from SIMBAD were derived using data from the 2MASS\footnote{The Two Micron All Sky Survey at near-infrared wavelengths} survey. These coordinates are consistent with the optically-derived coordinates from NED. However, in some objects the location of the peak X-ray emission differs from the optically coordinates. This difference is generally insignificant, so has been neglected. 

\setcounter{table}{0}
\begin{table*}
\centering
\caption{List of early-type galaxy properties.}\label{tab1}
\begin{tabular}{lcccccccccccc}
\hline
Name      & $RA$ & $DEC$ & ObsID & Exposure & Type & BCG & cD &  $z$     &  $D_A$ &  $D_L$  &  $N_{H}$ \\
          &  (J2000)   & (J2000)    &       &    ks        &      &    &     &          &   Mpc  &  Mpc    &  10$^{20}$ cm$^{2}$ \\
          &            &            &       & before/after &      &   &     &           &        &         &    \\
   (1)    &    (2)     &    (3)     &   (4) &     (5)      &  (6)  &  (7)  &  (8)   &   (9)  & (10)   &  (11) & (12) \\
\hline
ESO3060170 & 246.41 & -30.289 & multi  & 28.05/25.96  & E3 & $\surd$ & & 0.035805 & 145.1  & 155.7 & 3.51/13.4  \\
IC1262  & 69.5188 & 32.0738 & multi & 113.68/106.17 & E & $\surd$ & & 0.032649 & 133.0 & 141.8 & 2.47/21.2  \\
IC1459  & 4.6590 & -64.1096 & 2196 & 58.83/45.14 & E3 & && 0.006011&  25.503 & 25.8 & 1.19/1.68  \\
IC1633  & 293.098 & -70.8424 & 4971 & 24.79/22.24 & E1 & $\surd$ & $\surd$ & 0.02425 & 100.0 & 104.9 & 2.01/5.34 \\
IC4296  & 313.5384 & 27.9729 & multi & 48.53/40.05 & E & & & 0.012465 & 52.358 & 53.7 & 4.11/12.4    \\
IC5267  & 350.2369 & -61.801 & 3947 & 54.97/43.71 & SA0 & & & 0.005711 & 24.241 & 24.5 & 1.62/1.91  \\
IC5358  & 25.1415 & -75.8683 & multi & 39.61/38.82 & E4 & $\surd$ & $\surd$ & 0.02884 & 118.1 & 125.0 & 1.54/7.2 \\
NGC315  & 124.5631 & -32.4991 & 4156 & 55.02/39.49 & E &         & $\surd$ & 0.016485 & 68.816 & 71.1 & 5.87/48.2 \\
NGC326  & 124.8444 & -35.9797 & 6830 & 90.83/90.83 & E & $\surd$ & $\surd$ & 0.047400 & 188.8  & 207.1 &  5.81/68.7  \\
NGC383  & 126.8391 & -30.3379 & 2147  & 44.41/41.29  & S0 &         &  & 0.017005 & 70.930 & 73.4 & 5.42/21.5  \\
NGC499  & 130.4977 & -28.9448 & multi & 38.62/38.48 & E5 & & & 0.014673 & 61.423 & 63.2 & 5.26/13.6  \\
NGC507  & 130.6430 & -29.1326 & 317 & 43.63/40.30 & S0 & $\surd$ & & 0.016458 & 68.706 & 71.0 & 5.32/13.7  \\
NGC533  & 140.1457 & -59.9683 & 2880 & 37.61/28.40 & E3 & & $\surd$ & 0.018509 & 77.025 & 79.9 & 3.12/24.7  \\
NGC708  & 136.5695 & -25.0903 & multi & 139.43/137.38 & E & $\surd$ & $\surd$ & 0.016195 & 67.635 & 69.8  &  5.37/11.5 \\
NGC720  & 173.0194 & -70.3572 & multi & 99.21/98.22 & E5 &    &  & 0.005821 & 24.704 & 25.0 & 1.55/13.9  \\
NGC741  & 150.9342 & -53.6764 & 2223 & 30.35/28.14 & E0 &       &  & 0.018549 & 77.186 & 80.1 & 4.47/59.2 \\
NGC821  & 151.5555 & -47.5568 & multi & 188.31/181.64 & E6 &      &   & 0.005787 & 24.561 & 24.8  &  6.34/13.0 \\
NGC1023 & 145.0232 & -19.0892 & multi & 616.18/188.09 & SB0 &     &   & 0.002125 & 16.5* & 17.5* & 7.17/4.26 \\
NGC1265 & 150.1336 & -13.1299 & 3237  & 93.86/93.60  & E   &     & $\surd$   & 0.025137 & 103.6  & 108.8 & 14.3/5.25  \\
NGC1266 & 183.6680 & -47.5077 & 11578 & 153.60/28.63 & SB  &      &   & 0.007238 & 30.649 & 31.1 & 5.39/7.16  \\
NGC1316 & 240.1627 & -56.6898 & 2022  & 29.86/21.21  & S0 &      &   & 0.005871 & 24.914 & 25.2  & 1.92/48.4  \\
NGC1332 & 212.1830 & -54.3661 & multi  & 74.82/20.48  & S0  &    &     & 0.005084 & 21.601 & 21.8  & 2.29/26.8  \\
NGC1386 & 237.6634 & -53.9659 & multi & 80.81/70.47  & SB  &     &    & 0.002895 & 16.5* & 17.5*  & 1.39/181.0  \\
NGC1399 & 236.7164 & -53.6356 & 9530  & 59.35/56.98  & E1  & $\surd$ & & 0.004753 & 20.205 & 20.4  & 1.31/163.0  \\
NGC1404 & 236.9552 & -53.5548 & multi & 319.27/296.7  & E1  &       &  & 0.006494 & 27.531 & 27.9  & 1.35/163.0  \\
NGC1407 & 209.6362 & -50.3838 & 14033 & 54.35/50.26  & E0  &       &  & 0.005934 & 25.179 & 25.5  & 5.41/17.3  \\
NGC1482 & 214.1238 & -47.8035 & 2932  & 28.20/15.36  & SA0 &       &  & 0.006391 & 27.099 & 27.4  & 3.73/10.3  \\
NGC1550 & 190.9760 & -31.8488 & multi  & 89.00/89.00 & SA0   & $\surd$& & 0.012389 & 52.045 & 53.3 & 11.2/5.98  \\
NGC1600 & 200.4164 & -33.2418 & 4371  & 26.75/24.57  & E3  &       &  & 0.015614 & 65.267 & 67.3  & 4.71/4.85  \\ 
NGC1700 & 203.6991 & -27.6137 & 2069  & 42.81/26.79  & E4  &       &  & 0.012972 & 54.445 & 55.9  & 4.80/5.06  \\
NGC2434 & 281.0002 & -21.5444 & 2923  & 52.44/47.31  & E0  &       &  & 0.004637 & 19.716 & 19.9  & 12.1/15.5  \\
NGC2768 & 155.4947 & 40.5634 & 9258   & 153.92/63.56  & E6  &       &  & 0.004513 & 19.191 & 19.4 & 3.89/0.97 \\
NGC3079 & 157.8101 & 48.3598 & 2038  & 33.02/23.51  & SB  &        & & 0.003723 & 16.5* & 17.5* & 0.80/1.36 \\
NGC3091 & 256.7559 & 27.5029 & 3215 & 31.69/27.34 & E3 & $\surd$ & & 0.013222 & 55.473 & 56.9 & 4.75/4.79  \\
NGC3379 & 233.4901 & 57.6328 & multi  & 369.63/319.34 & E1 &        & & 0.003039 & 16.5* & 17.5* & 2.75/1.59 \\
NGC3384 & 233.5221 & 57.7524 & multi & 121.55/115.82 & SB0 &       & & 0.002348 & 16.5* & 17.5* & 2.75/1.75 \\
NGC3557 & 281.5784 & 21.0890  & 4502  & 99.41/93.86  & E3  &   &      & 0.010300 & 43.410 & 44.3 & 7.44/16.7  \\
NGC3585 & 277.2465 & 31.1753 & 9506  & 61.23/57.96  & E6  &     &    & 0.004783 & 20.332 & 20.5  &  5.57/8.49  \\
NGC3607 & 230.5996 & 66.4223  & 2073  & 38.50/32.69  & SA0 &     &    & 0.003142 & 22.2 & 23.5 & 1.52/2.60  \\
NGC3665 & 174.7122 & 68.4932  & 3222  & 17.96/14.59  & SA0 &      &   & 0.006901 & 29.238 & 29.6 & 2.07/1.38  \\
NGC3923 & 287.2759 & 32.2224  & 9507  & 81.00/80.90  & E4  &       &  & 0.005801 & 24.620 & 24.9 & 6.29/11.6  \\
NGC3955 & 286.1398 & 37.8258  & 2955  & 19.71/19.29  & S0  &        & & 0.004973 & 21.133 & 21.3 & 4.86/8.40  \\
NGC4036 & 132.9794 & 54.2466 &  6783  & 21.84/12.20  & S0 &        &  & 0.004620 & 19.643 & 19.9 & 1.90/2.64 \\
NGC4073 & 276.9081 & 62.3697 & 3234 & 29.96/25.76 & E  & $\surd$ & $\surd$ & 0.019584 & 81.364 & 84.6 & 1.90/3.91  \\
NGC4104 & 204.3284 & 80.0306 &  6939 & 35.88/34.86    & S0  & $\surd$ && 0.028196 & 115.6  & 122.2 &  1.68/2.58  \\
NGC4125 & 130.1897 & 51.3391  & 2071   & 64.23/52.97 & E6  &   &      & 0.004523 & 19.234 & 19.4 & 1.86/3.13  \\
NGC4203 & 173.0323 & 80.0788 & 10535  & 42.12/40.61  & SAB0 &   &     & 0.003623 & 17.28 & 17.5 & 1.20/4.08 \\
NGC4261 & 281.8049 & 67.3726 & 9569 & 102.55/102.24  & E2  &     &    & 0.007378 & 31.236 & 31.7 & 1.56/5.50  \\
NGC4278 & 193.7824 & 82.7727  & multi & 470.84/462.52 & E1  &      &    & 0.002068 & 17.11  & 17.4  & 1.75/6.03  \\
NGC4325 & 279.5840 & 72.1969 & 3232 & 30.09/28.30 & E & $\surd$ && 0.025714 & 105.8 & 111.3 & 2.18/5.39  \\
NGC4342 & 283.4824 & 68.8699 & 12955 & 54.54/53.35 & S0 &    &      &  0.002538 & 16.5 & 17.5 & 1.61/5.31 \\
NGC4365 & 283.8070 & 69.1819 &  2015 & 40.43/37.36 & E3 &     &    & 0.004146 & 17.642 & 17.8  &  1.61/5.44  \\
NGC4374 & 278.2045 & 74.4784  & multi & 87.02/79.85  & E1 &    &       & 0.003392 & 16.5* & 17.5* & 2.58/6.02  \\
NGC4382 & 267.7120 & 79.2372  & 2016 & 39.75/29.33  & SA0 &     &     & 0.002432 & 16.5* & 17.5* & 2.51/3.99   \\
NGC4388 & 279.1220 & 74.3355  & 9276 & 170.59/170.59 & SAB &      &    & 0.008419 & 35.586 & 36.2 & 2.58/6.52  \\
NGC4406 & 279.0835 & 74.6369 & 318    & 15.02/13.13 & E3 &        &  & 0.000747 & 16.5* & 17.5* & 2.58/6.36 \\
NGC4457 & 289.1324 & 65.8389  & 3150 & 38.88/32.50  & SAB &        &  & 0.002942 & 16.5* & 17.5* & 1.84/5.53  \\
\hline
\end{tabular}
\end{table*}

\setcounter{table}{0}
\begin{table*}
\centering
\caption{Continued.}
\begin{tabular}{lcccccccccccc}
\hline
Name      & $RA$ & $DEC$ & ObsID & Exposure & Type & BCG & cD  &  $z$     &  $D_A$ &  $D_L$  &  $N_{H}$ \\
          &  (J2000)   & (J2000)    &       &    ks     &      &   &      &          &   Mpc  &  Mpc    &  10$^{20}$ cm$^{2}$ \\
          &            &            &       & before/after &      &   &     &           &        &         &    \\
   (1)    &    (2)     &    (3)     &   (4) &     (5)      &  (6)  &  (7)  &  (8)   &   (9)  & (10)   &  (11) & (12) \\
\hline
NGC4472 & 286.9222 & 70.1961 & 11274 & 39.67/39.67  & E2  &         & & 0.003272 & 16.5* & 17.5*  & 1.65/7.71 \\
NGC4477 & 281.5441 & 75.6119 & 9527   & 38.10/37.68  & SB0 &    &      & 0.004463 & 18.980 & 19.2 & 2.63/6.38 \\
NGC4486 & 283.7777 & 74.4912 & 2707   & 90.03/89.09  & E0  &    & $\surd$ & 0.004283 & 18.220 & 18.4 & 2.52/7.36 \\
NGC4526 & 290.1595 & 70.1385  & 3925  & 43.53/31.73 & SAB &     &     & 0.002058 & 16.5* & 17.5* & 1.66/7.78  \\
NGC4552 & 287.9326 & 74.9668  & multi & 146.96/145.93 & E0  &     &      & 0.001134 & 16.5* & 17.5* & 2.56/6.34  \\
NGC4555 & 221.8117 & 86.4343  & 2884  & 29.97/23.91 & E  &        &   & 0.022292 & 92.231 & 96.4 & 1.37/5.95  \\
NGC4564 & 289.5604 & 73.9207 & 4008  & 16.02/15.84 & E6  &         &  & 0.003809 & 15.8 & 16.3  &  2.27/6.85  \\
NGC4621 & 294.3646 & 74.3621 & 2068  & 24.84/24.84 & E5 &          && 0.001558 & 16.5*  & 17.5*   &  2.22/6.90 \\
NGC4636 & 297.7485 & 65.4729 & multi  & 149.07/133.43  & E0 &         &  & 0.003129 & 14.5 & 15.2  &  1.83/8.12  \\
NGC4649 & 295.8736 & 74.3178 & multi  & 69.88/61.60  & E2 &           && 0.003703 & 16.5* & 17.5*  &  2.13/7.19  \\
NGC4696 & 302.4036 & 21.5580 & 1560  & 84.75/21.20 & E1 & $\surd$ & $\surd$& 0.009867 & 41.613 & 42.4 & 8.07/23.5  \\
NGC4697 & 301.6329 & 57.0637 &  4730 & 40.05/36.98    & E6   &  &  & 0.00414  & 17.616 & 17.8  &  2.12/15.7  \\
NGC4710 & 300.8506 & 78.0300 & 9512  & 29.47/27.72 & SA0 &      &   & 0.003676 & 16.5 & 17.5 & 2.15/8.24 \\
NGC4782 & 304.1379 & 50.2958 & 3220 & 49.33/49.33 & E0 &        &  & 0.015437 & 64.545 & 66.6 & 3.56/31.8  \\
NGC4936 & 306.2037 & 32.2638 & multi  & 28.92/25.14  & E0 &     &      & 0.010397 & 43.812 & 44.7  & 5.91/49.9  \\
NGC5018 & 309.8982 & 43.0614 & 2070  & 30.89/26.54  & E3 &      &     & 0.009393 & 39.643 & 40.4 & 6.98/114.0  \\  
NGC5044 & 311.2340 & 46.0996 & multi & 316.04/302.07 & E0   & $\surd$& & 0.00928  & 39.173 & 39.9  & 5.03/112.0  \\
NGC5171 & 334.8063 & 72.2182 & 3216  & 34.67/30.58 & S0 &    &       & 0.022943 & 94.830 & 99.2 & 1.92/30.8  \\
NGC5353 & 82.6107 & 71.6336 & 14903 & 40.27/37.20 & S0 &      &      & 0.007755 & 32.813 & 33.3 & 0.98/8.39  \\
NGC5532 & 357.9614 & 64.1119 & 3968 & 49.36/44.53 & S0  &      &     & 0.024704 & 101.8 & 106.9 & 1.86/68.2  \\
NGC5813 & 359.1820 & 49.8484 & multi & 488.04/481.82 & E1 &      &  & 0.006525 & 27.662 & 28.0 & 4.23/10.9  \\
NGC5846 & 0.3389   & 48.9043 &  7923 & 90.01/85.25 & E  & $\surd$& & 0.00491  & 20.867 & 21.1  & 4.24/8.73  \\
NGC5866 & 92.0340 & 52.4891 & 2879  & 27.43/25.55  & S0 &    &     & 0.002518 & 14.9 & 15.2 & 1.45/15.4 \\
NGC6098 & 34.9745 & 42.8152 & 10230 & 44.57/43.12 & E &       &$\surd$  & 0.030851 & 126.0 & 133.9 & 4.18/7.56  \\
NGC6107 & 56.2296 & 45.6870 & 8180  & 20.87/19.29 & E & $\surd$ &$\surd$& 0.030658 & 125.2 & 133.0 & 1.46/37.8  \\
NGC6251 & 115.7638 & 31.1958 & 4130 & 45.44/14.06 & E &     &      & 0.024710 & 101.9 & 107.0 & 5.40/4.31  \\
NGC6269 & 49.0135 & 35.9380 & 4972 & 39.64/35.80 & E &       &$\surd$     & 0.034801 & 141.3 & 151.3 & 4.65/12.8  \\
NGC6278 & 43.5694 & 33.945 & 6789  & 15.04/11.79 & S0 &     &      & 0.009447 & 39.865 & 40.6 & 4.91/7.29 \\
NGC6338 & 85.8062 & 35.3991 & 4194 & 47.33/44.52 & S0 &      &     & 0.027303 & 112.1 & 118.3 & 2.55/45.4  \\
NGC6482 & 48.0905 & 22.9122 & 3218 & 19.34/10.03 & E &        &    & 0.013129 & 55.091 & 56.5 & 8.04/9.85  \\
NGC6861 & 350.8772 & -32.2109 & 11752 & 93.50/88.89 & SA0 &    &    & 0.009437 & 39.826 & 40.6 & 4.94/1.22  \\
NGC6868 & 350.9126 & -32.6376 & 11753 & 72.60/69.53 & E2  & $\surd$& & 0.009520 & 40.171 & 40.9 & 4.94/1.27  \\
NGC7176 & 14.9320 & -53.0969 & 905 & 49.53/43.63 & E &     &       & 0.008376 & 35.406 & 36.0 & 1.61/2.42  \\
NGC7196 & 345.3695 & -51.0861 & 10546 & 10.11/10.11 & E &   &      & 0.009750 & 41.127 & 41.9 & 1.84/0.87  \\ 
NGC7618 & 105.5754 & -16.9091 & multi & 235.84/168.14 & E  &  &       & 0.017309 & 72.164 & 74.7  & 11.9/38.1   \\
NGC7626 & 87.8591 & -48.3788 & 2074 & 26.74/23.61 & E &   &    & 0.011358 & 47.790 & 48.9 & 4.94/5.37  \\
UGC408  & 116.977 & -59.40 & 11389 & 93.92/93.80 & SAB0 & & & 0.014723 & 61.628 & 63.5 & 2.80/13.6  \\
\hline
\end{tabular} 
\end{table*}

\subsection{Optical data processing}\label{subsec_opt}

%Previous measurements of effective radius for ETGs were provided for a small number of objects and were performed using slightly different techniques and observations. 
Effective radii, $r_e$, were measured using optical Digitized Sky Survey ($DSS$) images. 10$'$x10$'$ images were downloaded from the $ESO$ web-page\footnote{http://archive.eso.org/dss/dss}. No additional calibrations were performed on these images.  Surface brightness profiles centered on the peak of the optical emission were extracted from each image. The background emission was obtained by fitting a constant to the outskirts of each profile and subtracted from each image. The total flux was determined by numerically integrating the profile including light significantly above background by 5$\sigma$. Uncertainties on $r_e$ were determined using 1000 Monte Carlo realizations of the surface brightness profiles. The 5$r_e$ measurements and their uncertainties are presented in second column of Table~\ref{tab2}. 
%Then, we create the circular regions centered at the maximum of the optical flux and within 5$r_e$ for each galaxy using {\sc ds9\footnote{http://ds9.si.edu/site/Home.html}} software package.  

\subsection{X-ray data processing}

$Chandra$ observations were analyzed using the {\sc ciao} software package version 4.8 and {\sc caldb} version 4.7.1. $Chandra$ data from the Advanced CCD Imaging Spectrometer (ACIS) were analyzed, apart from early observations with CCD temperatures above $-120^{\circ}$C. All galaxies were observed using ACIS chips 3 and 7. Level-2 event files were produced by correcting the level-1 event files for time-dependent gain and charge transfer inefficiency.  Level-2 event files were filtered to delete bad grades.  VFAINT filtering was performed as necessary. Background flares were identified and removed using the {\sc lc\_clean}\footnote{http://cxc.cfa.harvard.edu/contrib/maxim/acisbg/} tool provided by M. Markevitch.  Blank-sky background files were extracted for each observation and processed identically to the target files.  Background files were reprojected to the corresponding position, and then normalized to match the 9.5-12.0 keV flux. Column 5 of Table~\ref{tab1} shows the exposure time before and after corrections were applied.

X-ray images of the ETGs were formed by summing all events within the 0.5 -- 6.0 keV energy range. Point sources were removed using the \texttt{wavdetect} routine with a significance threshold of 10$^{-6}$. Spectra were extracted from circular regions encompassing the central 5$r_e$ of each galaxy. Background spectra were obtained from a nearby region free of sources with the same area as the source region. The local backgrounds are consistent with the blank-sky backgrounds used for creating images. The source and background spectra, the ancillary reference files ($ARF$), and redistribution matrix files ($RMF$), were created using the \texttt{specextract} task in the {\sc ciao} package. Spectra were grouped with one count per energy bin.

\section{Spectral analysis}\label{sec_spec_anal}

\subsection{Multi-component spectral modeling}

Previous studies of the X-ray emission from low-mass systems showed that the unresolved LMXBs and other stellar sources including active binaries (ABs) and cataclysmic variables (CVs) contribute to the total X-ray emission \citep{Pellegrini:94, Revnivtsev:08a}.  Due to their low X-ray luminosities, $\sim$ 10$^{37-38}$ erg/s, LMXBs, ABs, and CVs were often ignored.  However, our sample includes gas-poor galaxies ($L_X < $ 10$^{40}$ erg/s), so we must account for stellar sources. Their fluxes have been measured directly in M31, M32, and the Galactic bulge \citep{Revnivtsev:07a, Boroson:10}. All LMXBs in these galaxies were detected using a combination of thermal and non-thermal (power law) models. \citet{Revnivtsev:08a} measured the temperature of the unresolved stellar sources to be $kT$ = 0.48$\pm$0.07 keV.   The power law slope of the non-thermal component is $\Gamma$ = 1.76$\pm$0.37. These studies showed that power law and thermal models provide good fits to the X-ray emission of both resolved and unresolved LMXBs. Later, \citet{Wong:14} found that a $\Gamma_{LMXBs}$ in the range 1.4 -- 1.8 provided similar results for the hot atmosphere. We applied these and other previous measurements in our spectral fitting.

We use a multi-component model of the form {\sc phabs*(apec+po+mekal+po)} to fit an each spectrum of sampled targets in the {\sc xspec} version 12.9.1 environment \citep{Arnaud:96}. Here {\sc apec} models the thermal emission from the atmosphere, the first {\sc po} is a power law that describes emission from LMXBs, and {\sc mekal+po} describes the thermal ({\sc mekal}) and non-thermal ({\sc po}) contribution from ABs and CVs. The {\sc phabs} model accounts for photoelectric absorption and was fixed to the hydrogen column densities shown in the last column of Table~\ref{tab1} as first value.  They were obtained from \citet{Dickey:90}. The temperature and metallicity of the {\sc apec} model were free parameters. The slopes of two {\sc po} models were fixed to 1.6 and 1.9, respectively.  When metallicity was poorly constrained by the model, it was fixed to 0.5$Z_{\odot}$.  This value was chosen following \citet{Boroson:10} and \citet{Werner:12}. The temperature and metallicity in the {\sc mekal} model were fixed to 0.5 keV and 0.3$Z_{\odot}$, respectively. All frozen parameters used in this spectral fitting were previously tested and applied in previous analyses (e.g. \citealt{Revnivtsev:08a, Boroson:10, Kim:13, Wong:14}). The Cash-statistic\footnote{http://cxc.harvard.edu/sherpa/ahelp/cstat.html} (\citealt{Cash:79}) was applied in spectral fitting. 

The spectra were well-fit by this model ($\chi^2\approx 1$, $\chi^2$ in $C$-statistic is $C-stat$ value divided by degrees of freedom) apart from ten objects: NGC708, NGC1399, NGC3557, NGC4388, NGC4472, NGC4636, NGC4649, NGC5044, NGC5813, and NGC6251. These are among the brightest in the sample. Regrouping the spectra to 50 counts per energy bin, compared to the previous 1 per bin, reduces the $\chi^2$ to about 1 while maintaining the same temperature and flux measurements.

The temperature of ETGs here ranges between 0.20 keV in NGC821 and NGC1023 to 3.34 keV in IC5358. ESO3060170, IC5358, NGC6269, and NGC6278 have been excluded because their temperatures exceed $>$2 keV, presumably because of a larger scale hot atmosphere \citep{Werner:12}. 
Unabsorbed X-ray fluxes, $f_X$, in the 0.5 -- 2.0 keV energy band were measured by adding a {\sc cflux} component to the original model. This energy band was chosen for consistency with previous papers. We  extracted the X-ray fluxes in the $0.5 - 6$ keV range, and found only a $1\%$ to $2\%$ discrepancy with the 0.5 -- 2.0 keV flux. In both cases the spectra were fit over the entire $0.5-6.0$ keV energy range. The corresponding X-ray luminosity was then determined from $L_{X} = 4\pi D_{L}^{2} f_{X}$. Our sample spans a wide range of X-ray luminosities, ($0.02-391$)$\times$10$^{40}$ erg/s. The best-fitting parameters from this spectral analysis are shown in Table~\ref{tab2}.

\setcounter{table}{1}
\begin{table*}
\centering
\caption{The best-fit parameters for spectra extracted from within 5$r_e$. The central velocity dispersion, $\sigma$, was taken from LEDA.}\label{tab2}
\begin{tabular}{lccccccc}
\hline
Name    & 5$r_e$ & $T_{X}$       &   $f_{X}$           & $L_X$             & $Z$     & $\chi^2$ &   $\sigma$            \\
        &   kpc       &   keV         &   0.5-2.0 keV       & 10$^{40}$ erg/s   &  $Z_{\odot}$        &  $C-stat$/d.o.f.   & km/s              \\
        &             &               &   erg/cm$^2$/s      &                   &         &           &                       \\
 (1)    &      (2)    &    (3)         &  (4)               &  (5)              &  (6)     &  (7)     &   (8)   \\   
\hline
ESO3060170 & 89$\pm$7  & 2.40$\pm$0.11 & -11.88$\pm$0.01  & 382.37$\pm$8.80  & 0.69$\pm$0.13 & 1.2  & 271.7$\pm$13.2 \\
IC1262  & 123$\pm$13  & 1.80$\pm$0.05 & -12.02$\pm$0.006 & 229.75$\pm$3.17 & 0.45$\pm$0.05 & 1.2  & 232.5$\pm$9.6\\
IC1459  & 35$\pm$6 & 0.70$\pm$0.01 & -12.52$\pm$0.01 & 2.41$\pm$0.06 & 0.07$\pm$0.01 & 1.1   & 293.6$\pm$6.3\\
IC1633  & 135$\pm$25 & 1.84$\pm$0.14 & -12.40$\pm$0.02 &  52.42$\pm$2.41  & 0.72$\pm$0.21  & 1.0  & 356.6$\pm$12.4\\
IC4296  & 77$\pm$9 & 0.94$\pm$0.01 & -12.28$\pm$0.01 & 18.11$\pm$0.42 & 0.29$\pm$0.05 & 1.1   & 327.1$\pm$5.4 \\
IC5267  & 38$\pm$6 & 0.48$\pm$0.12 & -13.27$\pm$0.03 & 0.39$\pm$0.03 & 0.11$\pm$0.07 & 1.0  & 167.7$\pm$5.2 \\
IC5358  & 250$\pm$32 & 3.34$\pm$0.06 & -11.43$\pm$0.003 & 694.59$\pm$4.80 & 0.66$\pm$0.06 & 1.3  & 214.2$\pm$5.0 \\
NGC315  & 88$\pm$19 & 0.76$\pm$0.01 & -12.35$\pm$0.007 & 27.02$\pm$0.44 & 0.19$\pm$0.02 & 1.5  & 293.3$\pm$0.2\\
NGC326  & 191$\pm$33 & 0.94$\pm$0.10 & -13.80$\pm$0.07 & 8.13$\pm$1.31 & 0.5 & 1.0  & 231.9$\pm$13.1\\
NGC383  & 48$\pm$5 & 0.98$\pm$0.02 & -12.81$\pm$0.01  & 9.98$\pm$0.23  & 0.5 & 1.0  & 271.9$\pm$6.7\\
NGC499  & 57$\pm$8 & 0.79$\pm$0.01 & -11.93$\pm$0.007 & 56.15$\pm$0.91 & 0.48$\pm$0.07 & 1.9  & 253.3$\pm$6.7\\
NGC507  & 107$\pm$12 & 1.25$\pm$0.01 & -11.77$\pm$0.004 & 102.43$\pm$0.94 & 0.56$\pm$0.04 & 1.2  & 291.8$\pm$5.9\\
NGC533  & 112$\pm$15 & 1.04$\pm$0.004 & -11.97$\pm$0.004 & 81.85$\pm$0.75 & 0.41$\pm$0.03 & 1.4  & 271.2$\pm$5.6\\
NGC708  & 110$\pm$14 & 1.56$\pm$0.01 & -11.48$\pm$0.002 & 193.03$\pm$0.89 & 0.65$\pm$0.02 & 3.1  & 222.2$\pm$7.8\\
NGC720  & 32$\pm$3 & 0.62$\pm$0.01 & -12.26$\pm$0.006 & 4.11$\pm$0.06 & 0.26$\pm$0.03 & 1.1  & 235.6$\pm$5.6\\
NGC741  & 88$\pm$7 & 1.02$\pm$0.01 & -12.37$\pm$0.008 & 32.75$\pm$0.60 & 0.25$\pm$0.03 & 1.1  & 286.0$\pm$9.3\\
NGC821  & 30$\pm$7 & 0.20$\pm$0.08 & -13.85$\pm$0.12 & 0.10$\pm$0.03 & 0.5 & 0.9  & 198.4$\pm$2.8\\
NGC1023 & 35$\pm$4 & 0.20$\pm$0.09 & -13.25$\pm$0.03 & 0.21$\pm$0.02 & 0.5 & 1.1  & 197.9$\pm$4.6 \\
NGC1265 & 143$\pm$17 & 0.96$\pm$0.04 & -13.28$\pm$0.03  & 7.43$\pm$0.51  & 0.5 & 1.1  & - \\
NGC1266 & 13$\pm$2  & 0.80$\pm$0.03 & -13.23$\pm$0.03 & 0.67$\pm$0.05 & 0.5 & 0.9  & 94.4$\pm$5.2\\
NGC1316 & 58$\pm$7 & 0.75$\pm$0.01 & -12.07$\pm$0.006  & 6.47$\pm$0.09  & 0.22$\pm$0.02 & 1.2  & 223.7$\pm$3.3\\
NGC1332 & 30$\pm$4 & 0.70$\pm$0.03 & -12.42$\pm$0.007  & 2.16$\pm$0.04  & 0.15$\pm$0.02 & 1.1  & 312.5$\pm$10.7\\
NGC1386 & 16$\pm$4 & 0.32$\pm$0.04 & -13.20$\pm$0.04 & 0.23$\pm$0.02  & 0.5 & 1.1  & 166.2$\pm$18.0\\
NGC1399 & 45$\pm$5 & 1.26$\pm$0.003 & -11.30$\pm$0.002  & 24.96$\pm$0.12  & 0.46$\pm$0.01 & 2.9  & 333.7$\pm$5.3 \\
NGC1404 & 45$\pm$5 & 0.67$\pm$0.004 & -11.54$\pm$0.003 & 26.86$\pm$0.19  & 0.19$\pm$0.007 & 1.6  & 228.1$\pm$3.6\\
NGC1407 & 44$\pm$5 & 1.02$\pm$0.02 & -12.13$\pm$0.01 & 5.77$\pm$0.13 & 0.19$\pm$0.03 & 1.4  & 264.9$\pm$5.1\\
NGC1482 & 69$\pm$7 & 0.80$\pm$0.01 & -12.64$\pm$0.01  & 2.06$\pm$0.05  & 0.5 & 1.6  & 108.5$\pm$7.4 \\
NGC1550 & 54$\pm$5 & 1.27$\pm$0.005 & -11.51$\pm$0.003 & 105.04$\pm$0.73 & 0.48$\pm$0.02 & 1.4  & 300.3$\pm$5.3\\
NGC1600 & 97$\pm$10 & 1.24$\pm$0.02 & -12.49$\pm$0.01 & 17.14$\pm$0.40  & 0.32$\pm$0.05 & 1.1  & 331.4$\pm$7.0\\ 
NGC1700 & 45$\pm$6 & 0.51$\pm$0.02 & -12.68$\pm$0.01  & 7.81$\pm$0.18  & 0.13$\pm$0.03 & 1.1  & 233.1$\pm$3.9\\
NGC2434 & 25$\pm$3 & 0.59$\pm$0.03 & -12.78$\pm$0.02  & 0.79$\pm$0.04 & 0.5 & 1.2  & 183.7$\pm$5.3\\
NGC2768 & 31$\pm$4 & 0.35$\pm$0.02 & -12.7848$\pm$0.01 & 0.74$\pm$0.02 & 0.5 & 1.1  & 184.2$\pm$2.8 \\
NGC3079 & 25$\pm$4 & 0.78$\pm$0.01 & -12.58$\pm$0.01 & 0.97$\pm$0.02 & 0.5 & 1.7  & 175.1$\pm$12.3 \\
NGC3091 & 47$\pm$6 & 0.88$\pm$0.01 & -12.14$\pm$0.006 & 28.06$\pm$0.39 & 0.57$\pm$0.08 & 1.0  & 310.2$\pm$7.6\\
NGC3379 & 29$\pm$5 & 0.24$\pm$0.08 & -13.31$\pm$0.05 & 0.18$\pm$0.02 & 0.11$\pm$0.08 & 1.4  & 203.7$\pm$1.8\\
NGC3384 & 23$\pm$3 & 0.31$\pm$0.04 & -13.59$\pm$0.05 & 0.09$\pm$0.01 & 0.5 & 1.0  & 145.7$\pm$2.5 \\
NGC3557 & 60$\pm$7 & 0.43$\pm$0.10  & -12.68$\pm$0.06  & 4.91$\pm$0.67  & 0.5 & 2.2  & 264.1$\pm$7.2\\
NGC3585 & 35$\pm$4 & 0.32$\pm$0.07 & -13.25$\pm$0.04  & 0.28$\pm$0.03  & 0.5 & 1.1  & 210.9$\pm$6.2\\
NGC3607 & 28$\pm$3 & 0.59$\pm$0.11  & -12.75$\pm$0.02  & 0.73$\pm$0.03  & 0.5 & 1.1  & 220.8$\pm$4.2\\
NGC3665 & 38$\pm$5 & 0.30$\pm$0.05  & -12.85$\pm$0.03  & 1.48$\pm$0.10  & 0.5 & 0.9  & 214.7$\pm$8.6\\
NGC3923 & 50$\pm$6 & 0.58$\pm$0.01  & -12.17$\pm$0.005  & 5.02$\pm$0.06  & 0.18$\pm$0.02 & 1.5  & 246.6$\pm$5.6\\
NGC3955 & 11$\pm$1 & 0.31$\pm$0.03  & -13.41$\pm$0.04  & 0.21$\pm$0.02  & 0.5 & 0.9  & 94.4$\pm$5.3 \\
NGC4036 & 18$\pm$2 & 0.46$\pm$0.07 & -13.46$\pm$0.06 & 0.16$\pm$0.01 & 0.5 & 0.7  & 197.9$\pm$6.3 \\
NGC4073 & 104$\pm$16 & 1.88$\pm$0.02 & -11.58$\pm$0.003 & 225.24$\pm$1.56 & 1.67$\pm$0.10 &  1.5  & 267.0$\pm$6.3\\
NGC4104 & 132$\pm$14 & 1.43$\pm$0.04 &  -12.38$\pm$0.01 & 74.48$\pm$1.71 & 0.30$\pm$0.04 &  1.3  & 291.0$\pm$5.5\\
NGC4125 & 32$\pm$3 & 0.49$\pm$0.01  & -12.45$\pm$0.007   & 1.60$\pm$0.03 & 0.18$\pm$0.02 & 1.0  & 238.2$\pm$7.0\\
NGC4203 & 21$\pm$3 & 0.28$\pm$0.03 & -12.92$\pm$0.02 & 0.44$\pm$0.02 & 0.5 & 1.0  & 160.8$\pm$5.5 \\
NGC4261 & 45$\pm$4 & 0.80$\pm$0.006 & -12.26$\pm$0.005 & 6.61$\pm$0.08 & 0.19$\pm$0.01 & 1.5  & 296.4$\pm$4.3\\
NGC4278 & 25$\pm$3 & 0.33$\pm$0.02  & -13.02$\pm$0.01 & 0.35$\pm$0.01 & 0.5 & 1.0  & 234.3$\pm$4.5\\
NGC4325 & 16$\pm$1 & 0.93$\pm$0.006 & -11.82$\pm$0.004 & 224.34$\pm$2.07 & 0.51$\pm$0.04 & 0.9 & - \\
NGC4342 & 11$\pm$1 & 0.64$\pm$0.02 & -13.06$\pm$0.01 & 0.32$\pm$0.01 & 0.16$\pm$0.04 & 1.0  & 240.7$\pm$5.7 \\
NGC4365 & 28$\pm$2 & 0.44$\pm$0.02 &  -12.97$\pm$0.02 & 0.41$\pm$0.02 & 0.5 & 1.1  & 246.9$\pm$2.6\\
NGC4374 & 32.51$\pm$3 & 0.81$\pm$0.005  & -11.85$\pm$0.004 & 5.18$\pm$0.05 & 0.16$\pm$0.007 & 1.4  & 274.9$\pm$2.4 \\
NGC4382 & 34$\pm$3 & 0.44$\pm$0.03  & -12.51$\pm$0.01 & 1.13$\pm$0.03 & 0.23$\pm$0.05 & 1.1  & 175.3$\pm$3.5 \\
NGC4388 & 39$\pm$4 & 0.98$\pm$0.05  & -12.63$\pm$0.03 & 3.68$\pm$0.25 & 0.5 & 4.9  & 98.9$\pm$9.4\\
NGC4406 & 35$\pm$3 & 0.88$\pm$0.01 & -11.55$\pm$0.01 & 10.28$\pm$0.24 & 0.34$\pm$0.03 & 1.1  & 230.0$\pm$2.6 \\
NGC4457 & 18$\pm$2 & 0.59$\pm$0.02  & -12.83$\pm$0.02 & 0.54$\pm$0.03 & 0.17$\pm$0.05 & 1.1  & 113.3$\pm$9.8\\
\hline
\end{tabular}
\end{table*}

\setcounter{table}{1}
\begin{table*}
\centering
\caption{Continued.}
\begin{tabular}{lccccccc}
\hline
Name    & $5 r_e$ & $T_{X}$       &   $f_{X}$           & $L_X$             & $Z$     & $\chi^2$   &  $\sigma$            \\
        &   kpc       &   keV         &   0.5-2.0 keV       & 10$^{40}$ erg/s   &  $Z_{\odot}$       & $C-stat$/d.o.f. &    km/s           \\
        &             &               &   erg/cm$^2$/s      &                   &         &             &               \\
 (1)    &      (2)    &    (3)         &  (4)               &  (5)              &  (6)     &  (7)     &   (8)   \\   
\hline
NGC4472 & 36$\pm$3 & 1.06$\pm$0.002 & -11.30$\pm$0.002 & 18.36$\pm$0.08 & 0.55$\pm$0.02 & 3.0 & 280.8$\pm$2.9\\
NGC4477 & 21$\pm$3 & 0.34$\pm$0.02 & -12.63$\pm$0.01 & 1.04$\pm$0.03 & 0.5 & 1.1  & 172.2$\pm$6.2 \\
NGC4486 & 35$\pm$4 & 1.85$\pm$0.002 & -10.19$\pm$0.0003 & 262.93$\pm$0.18 & 0.75$\pm$0.004 & 4.3  & 321.7$\pm$4.3 \\
NGC4526 & 25$\pm$3 & 0.37$\pm$0.02  & -13.06$\pm$0.02  & 0.32$\pm$0.02 & 0.5 & 1.2  & 224.4$\pm$9.4\\
NGC4552 & 24$\pm$2 & 0.64$\pm$0.01  & -12.17$\pm$0.007 & 2.48$\pm$0.04 & 0.16$\pm$0.02 & 1.2  & 249.7$\pm$2.9\\
NGC4555 & 63$\pm$7 & 1.05$\pm$0.02  & -12.82$\pm$0.01  & 16.83$\pm$0.33 & 0.33$\pm$0.07 & 1.3  & 344.0$\pm$28.5\\
NGC4564 & 13$\pm$3 & 0.38$\pm$0.15 & -13.61$\pm$0.11  & 0.09$\pm$0.02 & 0.5 &  0.9  & 155.9$\pm$2.2\\
NGC4621 & 24$\pm$3 & 0.26$\pm$0.07 & -13.65$\pm$0.09  & 0.02$\pm$0.004 & 0.5 & 1.1  & 227.7$\pm$3.8\\
NGC4636 & 34$\pm$3 & 0.75$\pm$0.003 & -11.27$\pm$0.002  & 19.68$\pm$0.09  & 0.42$\pm$0.02 & 1.7  & 199.7$\pm$2.7 \\
NGC4649 & 41$\pm$4 & 0.94$\pm$0.003 & -11.48$\pm$0.002  & 12.13$\pm$0.06 & 0.58$\pm$0.03 & 2.0  & 329.1$\pm$4.6\\
NGC4696 & 64$\pm$5 & 1.88$\pm$0.02 & -10.74$\pm$0.003  & 391.42$\pm$2.70 & 0.86$\pm$0.04 &  2.2  & 243.8$\pm$6.5\\
NGC4697 & 35$\pm$3 & 0.31$\pm$0.01 & -12.77$\pm$0.02 & 0.64$\pm$0.03 & 0.5 & 1.1  & 166.6$\pm$1.6\\
NGC4710 & 17$\pm$2 & 0.32$\pm$0.05 & -13.56$\pm$0.06 & 0.10$\pm$0.01 & 0.5 & 1.1  & 116.5$\pm$6.4 \\
NGC4782 & 64$\pm$3 & 1.02$\pm$0.01 & -12.65$\pm$0.01 & 11.88$\pm$0.28 & 0.5 & 1.8  & 308.5$\pm$11.2 \\
NGC4936 & 69$\pm$4 & 0.91$\pm$0.04 & -12.31$\pm$0.02  & 11.71$\pm$0.54  & 0.22$\pm$0.09 & 0.8  & 278.2$\pm$14.8\\
NGC5018 & 39$\pm$4 & 0.53$\pm$0.07 & -13.11$\pm$0.03  & 1.52$\pm$0.10  & 0.5 & 1.1  & 206.5$\pm$4.5\\  
NGC5044 & 51$\pm$3 & 0.95$\pm$0.002 & -10.97$\pm$0.002 & 204.11$\pm$0.47 & 0.35$\pm$0.007 & 3.5 & 225.7$\pm$9.2 \\
NGC5171 & 88$\pm$7 & 0.81$\pm$0.05 & -13.59$\pm$0.05  & 3.02$\pm$0.35 & 0.5 & 0.7  & -\\
NGC5353 & 32$\pm$3 & 0.74$\pm$0.02 & -12.48$\pm$0.01 & 4.39$\pm$0.10 & 0.17$\pm$0.03 & 1.1  & 283.5$\pm$4.8\\
NGC5532 & 73$\pm$6 & 0.97$\pm$0.02 & -12.88$\pm$0.01 & 18.02$\pm$0.41 & 0.20$\pm$0.04 & 1.0  & 277.8$\pm$18.6\\
NGC5813 & 32$\pm$2 & 0.71$\pm$0.002 & -11.33$\pm$0.001 & 43.88$\pm$0.10 & 0.45$\pm$0.01 & 2.5  & 235.4$\pm$3.4 \\
NGC5846 & 34$\pm$6 & 0.79$\pm$0.003 & -11.53$\pm$0.003 & 15.72$\pm$0.11 & 0.33$\pm$0.01 & 1.2  & 237.1$\pm$3.5 \\
NGC5866 & 26$\pm$3 & 0.41$\pm$0.08 & -13.05$\pm$0.02 & 0.33$\pm$0.02 & 0.5 & 1.1  & 161.6$\pm$4.8 \\
NGC6098 & 108$\pm$12 & 1.60$\pm$0.11 & -13.04$\pm$0.03 & 19.56$\pm$1.35 & 0.5 & 0.8  & 275.3$\pm$25.2\\
NGC6107 & 156$\pm$17 & 1.61$\pm$0.06 & -12.66$\pm$0.02  & 46.30$\pm$2.13 & 0.58$\pm$0.14 &  1.0  & 240.9$\pm$28.1 \\
NGC6251 & 130$\pm$23 & 0.83$\pm$0.03 & -12.46$\pm$0.01 & 47.50$\pm$1.09 & 0.5 & 2.5  & 311.5$\pm$18.6\\
NGC6269 & 91$\pm$9 & 2.40$\pm$0.16 & -12.51$\pm$0.01 & 84.64$\pm$1.95 & 1.01$\pm$0.27 & 1.1  & 317.9$\pm$22.4\\
NGC6278 & 27$\pm$3 & 2.05$\pm$0.30 & -13.17$\pm$0.04 & 1.33$\pm$0.12 & 0.5 & 0.7  & 193.2$\pm$12.8 \\
NGC6338 & 124$\pm$24 & 1.84$\pm$0.03 & -11.80$\pm$0.004 & 264.49$\pm$2.44 & 0.84$\pm$0.07 & 1.2  & 348.4$\pm$40.2\\
NGC6482 & 37$\pm$4 & 0.82$\pm$0.007 & -11.78$\pm$0.006 & 63.39$\pm$0.88 & 0.38$\pm$0.04 & 1.1  & 316.8$\pm$9.8\\
NGC6861 & 41$\pm$4 & 1.24$\pm$0.03 & -12.50$\pm$0.01 & 6.24$\pm$0.14 & 0.17$\pm$0.03 & 1.1  & 406.9$\pm$19.6\\
NGC6868 & 62$\pm$5 & 0.75$\pm$0.02 & -12.37$\pm$0.01 & 8.54$\pm$0.20 & 0.18$\pm$0.03 &  1.2  & 250.1$\pm$3.7 \\
NGC7176 & 60$\pm$5 & 0.77$\pm$0.03 & -12.79$\pm$0.02 & 2.51$\pm$0.12 & 0.5 & 1.0 & 245.9$\pm$5.7\\
NGC7196 & 39$\pm$3 & 0.64$\pm$0.04 & -12.64$\pm$0.03 & 4.81$\pm$0.33 & 0.16$\pm$0.09 & 0.9  & 277.9$\pm$37.5\\ 
NGC7618 & 79$\pm$6 & 0.93$\pm$0.006 & -12.05$\pm$0.004 & 59.50$\pm$0.55 & 0.25$\pm$0.02 & 1.3  & 292.8$\pm$30.3 \\
NGC7626 & 52$\pm$3 & 0.93$\pm$0.02 & -12.60$\pm$0.02 & 7.19$\pm$0.33 & 0.5 & 1.1  & 267.0$\pm$3.7\\
UGC408  & 57$\pm$4 & 0.82$\pm$0.01 & -12.80$\pm$0.01 & 7.65$\pm$0.18 & 0.18$\pm$0.02 & 1.0  & 197.6$\pm$4.8\\
\hline
\end{tabular}
\end{table*}

The spectral analysis was also performed using {\sc mekal} instead of {\sc apec} to model the thermal emission from the hot cluster gas. This approach permits a comparison to previous measurements, and to test systematic differences between the two thermal models. We found that {\sc apec} temperatures are 10-20\% higher than those measured using {\sc mekal}. In contrast, the X-ray flux provided by both these models are equal within uncertainties. The temperature differences are likely due to out-of-date atomic libraries in the {\sc mekal} model. We have used {\sc apec} version 3.0.7, which contains the most up-to-date atomic libraries, as well as photoionization and recombination rates.

An {\sc apec} model was used to fit the thermal emission from stellar sources to check the {\sc mekal} results. The previous temperature and flux measurements of the hot gas were than compared with the new spectral model (i.e., {\sc phabs*(apec+po+apec+po)}). We found that using {\sc apec} to model the thermal component of stellar sources, with all other parameters fixed to their previous values, provides essentially the same temperatures and fluxes for the thermal component of the hot gas. Thus, we adopt our primary model for consistency with previous results.

\begin{figure*}
\includegraphics[width=0.49\textwidth]{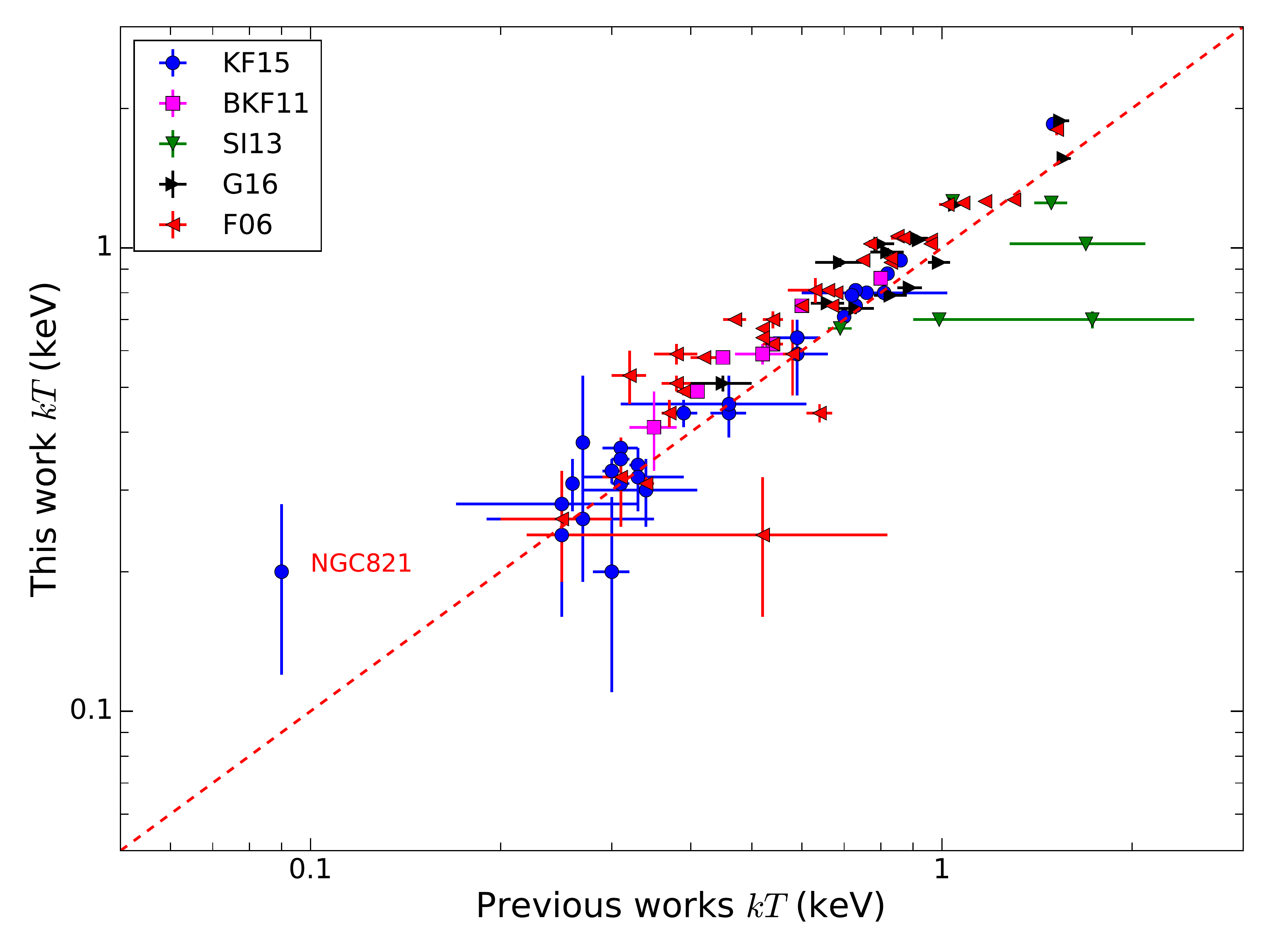}
\includegraphics[width=0.49\textwidth]{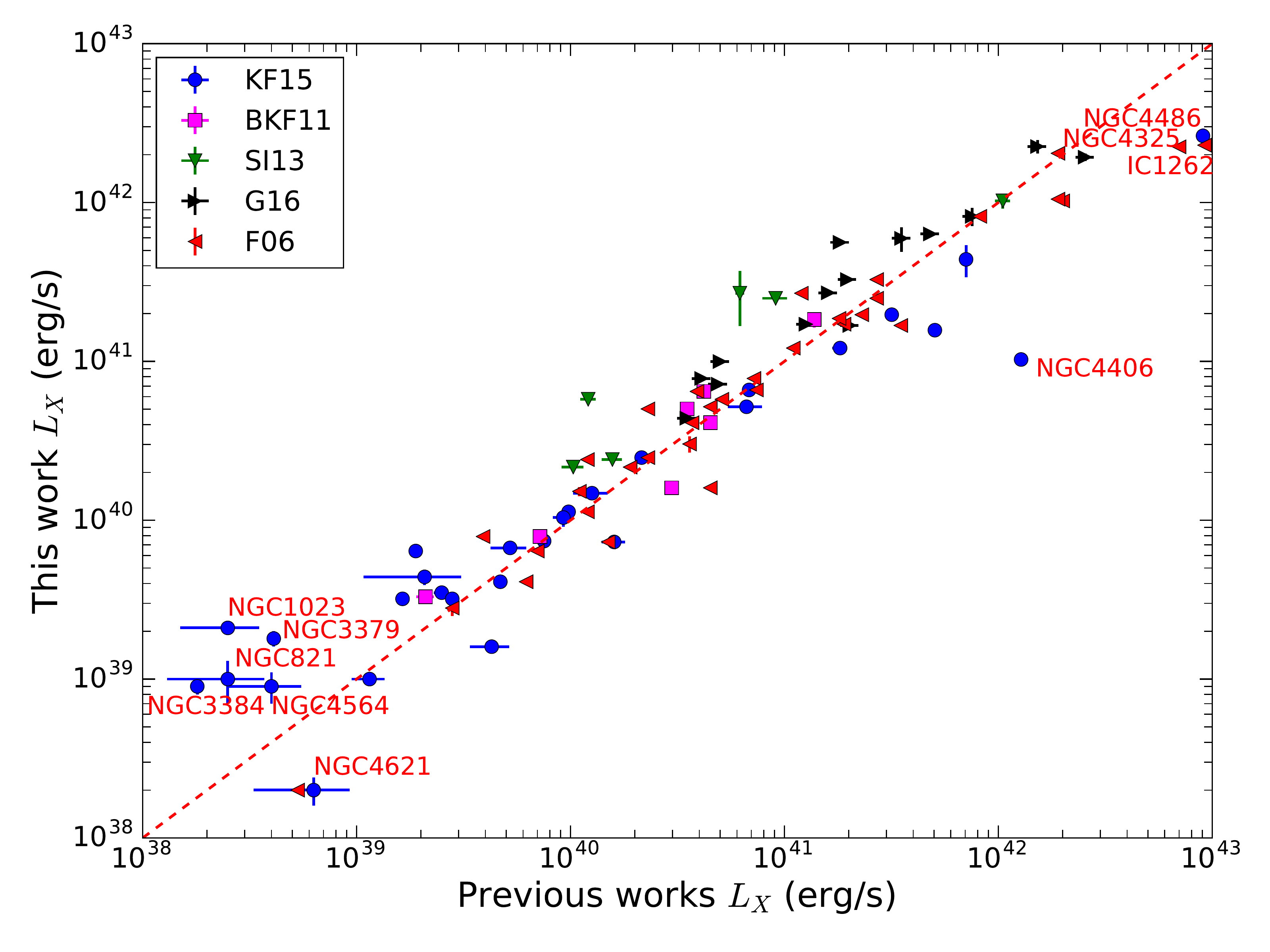}
\caption{Comparisons of temperature (left panel) and X-ray luminosity (right panel) with previous works. The dashed line indicates the line of equality.}
\label{fig_comp}
\end{figure*}

\subsection{Comparison with previous results}

Of our sample of 94, 56 objects have previous temperature and luminosity measurements. These values are taken from \citet{Fukazawa:06, Boroson:10, Su:13, Kim:15, Goulding:16}, referred to as F06, BKF11, SI13, KF15, G16, respectively, in Fig.~\ref{fig_comp}. Our measurements generally agree well with earlier studies. However, our the temperatures are slightly higher than previous results.  This small discrepancy might be due to several factors including,: (1) use of the $C$-statistic instead of $\chi^2$ in our spectral analysis; (2) adopting different fixed parameters in our multi-component model; (3) using the {\sc mekal} model instead {\sc apec}; or (4) using different energy ranges during spectral fitting. 

To investigate the impact of fixing the foreground extinction, $N_H$, to the \citet{Dickey:90} values,  we performed additional spectral fitting allowing $N_H$  to vary.  The best-fitting $N_H$ values are systematically 1.5-2 times higher than those obtained by \citet{Dickey:90}.  Nevertheless, the choice of fixing $N_H$ or allowing it to vary affected the slope, zero point, and scatter on the derived temperatures and luminosities insignificantly.  The fixed and free values for the zero point are 0.08$\pm$0.03, 0.06$\pm$0.03, slope, 1.23$\pm$0.13, 1.28$\pm$0.14, and scatter, 0.44$\pm$0.04 and 0.42$\pm$0.04, respectively.

The \citet{Dickey:90} values of $N_H$ have been surpassed by the Leiden/Argentine/Bonn radio survey by \citet{Kalberla:05}. In addition, molecular hydrogen and dust contribute to the absorption at higher $N_H$ \citep{Willingale:13}. Thus, we also study the impact of \citet{Kalberla:05} $N_H$ values (shown as second value in the last column of Table~\ref{tab1}) on the best-fit spectral parameters. We found no significant influence of these new column density values on the derived temperatures and luminosities. The values for zero point, slope, and scatter are 0.05$\pm$0.03, 1.25$\pm$0.19, and 0.43$\pm$0.07, respectively. Moreover, we found that high $N_H$ values produce higher uncertainties on the derived spectral parameters.

A small discrepancy was found between our X-ray luminosities and those from F06 and KF15.  This discrepancy likely originates from differences in luminosity distances, $D_L$. Adopting the distances quoted in KF15, we obtain consistent luminosities for NGC1023, NGC3379, NGC3384, NGC4564, and NGC4621. However, the luminosity differences for NGC821, NGC4406, and NGC4486 are apparently not caused by differing luminosity distances. When the $D_L$ quoted in KF15 are using to determined $L_X$ from our measured fluxes, we obtain significantly different results. For instance, using the KF15 distances for NGC4406 gives a luminosity of 9.5$\times$10$^{40}$ erg/s, while KF15 quote 12.7$\times$10$^{40}$ erg/s. For this object, G16 and SI13 measured luminosities of 9.98 and 10.4 $\times$10$^{40}$ erg/s, which is in agreement with our measurements.

\section{Mass profiles}\label{sec_mass}

In this section we describe the derivation of the total and gas mass profiles. Studies have shown that the hot atmospheres of galaxies, groups, and clusters rest in hydrostatic equilibrium \citep{Moore:94, Churazov:08, Churazov:10, Navarro:10, Babyk:14, Babyk:16}. Here we use a simple $\beta$-model (\citealt{Cavaliere:78}) to describe the X-ray surface brightness profiles and to calculate total mass. 
%It is known that stellar component become a significant fraction of the baryonic and total mass of galaxies, especially at the central part where stars become dominate fraction \citep{Capelo:10}.

\subsection{Surface brightness profile}

The X-ray surface brightness profiles were extracted from the 0.5 -- 6.0 keV X-ray images. Each profile contained 100 annular regions of uniform width, each centered on the X-ray peak. The radius of the outermost annulus, $r_X$, is distinct from the aperture used for the optical analysis. The X-ray SBPs were then fit with a single $\beta$-model:
\begin{equation}
S(r) = S_{0} \left( 1 + \left( \frac{r}{r_{c}}\right)^{2}\right)^{-3\beta + 1/2} + C,
\label{sr}
\end{equation}
where $S(r)$ is the X-ray surface brightness as a function of projected radius. $S_{0}$, $r_{c}$, $\beta$, and $C$ are free parameters in the model. We find that the SBPs of early-type galaxies are characterized by $\beta \approx$ 0.45 -- 0.50, which is smaller than the $\sim$ 0.67 typical of galaxy clusters.

\subsection{Total mass estimates}
We derive the total gravitating mass for each galaxy assuming spherical symmetry and that the hot gas is in hydrostatic equilibrium. For gas in hydrostatic equilibrium,
\begin{equation}
\frac{dP}{dr} = -\frac{G M(r)}{r^{2}} \rho_{\rm g}(r),
\label{dpdr}
\end{equation}
where $P$ is the gas pressure, $G$ is the gravitational constant, $\rho_{\rm g}$ is the gas density, and $M$ is the total mass inside a sphere of radius $r$. The gas pressure is related to the gas density and temperature through the ideal gas law, $P = \frac{\rho_{\rm g}kT(r)}{\mu m_{p}}$. The total gravitating mass can then be written as
\begin{equation}
M(r) = -\frac{kT(r)r}{G \mu m_{p}} \left( \frac{d \ln \rho_{\rm g} }{d \ln r} + \frac{d \ln T}{d \ln r}\right).
\label{mtr}
\end{equation}
Here $\mu$ = 0.62 is the mean molecular weight of the hot gas and $m_p$ is the mass of a proton. Assuming that the gas is isothermal with mean temperature $T$, equation (\ref{mtr}) becomes
\begin{equation}
M = -\frac{k T r}{G \mu m_{p}} \left( \frac{d \ln \rho_{g} }{d \ln r} \right).
\label{mtot}
\end{equation}

The gas density profiles were determined using $\beta$-model fits to the X-ray surface brightness profiles. The gas density formulation for the $\beta$-model is
\begin{equation}
\rho_{\rm g}(r) = \rho_{0} \left( 1 + \left( \frac{r}{r_{c}}\right)^{2}\right)^{-3\beta/2},
\label{rog}
\end{equation}
where $r_{c}$ is the core radius and $\rho_{0} = 2.21 \mu m_p n_0$ is the central gas density. The central concentration, $n_0$, can be calculated from emissivity, $\epsilon$, as 
\begin{equation}
 n_0 = \left[\frac{S_0}{r_c \epsilon B(3\beta-0.5, 0.5) } \right]^{0.5},
\end{equation}
where $B(a,b)$ is the validity of the beta function (see \citet{Ettori:00} for details). The total mass sampled range from 10$^{12}$ to 10$^{13} M_{\odot}$.

\subsection{Gas mass estimates}
The hot gas mass within a radius $r_X$ was determined by integrating the gas density profiles (see Eq.~\ref{rog}): 
\begin{equation}
M_{\rm g} = 4\pi \rho_{0} \int_{0}^{r_X} r^{2} \left( 1 + \left( \frac{r}{r_{c}}\right)^{2}\right)^{-3\beta/2} dr.
\label{mg1}
\end{equation}
The gas masses of our sampled elliptical and lenticular galaxies spans 10$^{9}$$-$10$^{11} M_{\odot}$, while the gas masses of groups, BCGs, and cDs are an order of magnitude higher.
%The result obtained by integrating the density profile over a defined volume of the galaxy depends on the cosmology since the projected radius of the galaxy is calculated using the galaxy's angular distance $D_{A}$. 

\subsection{Summary of the calculations}

\setcounter{table}{2}
\begin{table*}
\centering
\caption{The best-fit parameters for an isothermal $\beta$-model.}\label{tab3}
\begin{tabular}{lcccccc}
\hline
Name   & $\beta$       & $r_c$         & $\rho_0$   & $\chi^2$ &  $M_{g}$              & $M$                   \\
       &               & kpc           & 10$^{-24}$ &          &  10$^{11}$ $M_{\odot}$ & 10$^{13}$ $M_{\odot}$ \\
       &               &               & g/cm$^3$   &          &                       &                       \\
  (1)  &         (2)   &  (3)          &  (4)       & (5)      &   (6)                 & (7)                   \\ 
\hline
ESO3060170 & 0.41$\pm$0.03 & 3.23$\pm$0.68 & 1.28$\pm$0.14  & 93  & 17.09$\pm$1.33 & 1.06$\pm$0.18$\pm$0.35 \\
IC1262  & 0.57$\pm$0.04 & 27.3$\pm$2.27 & 0.25$\pm$0.05 & 217 & 38.6$\pm$3.7 & 1.40$\pm$0.20$\pm$0.42 \\
IC1459  & 0.50$\pm$0.01 & 0.20$\pm$0.02 & 12.8$\pm$1.60 & 220 & 0.27$\pm$0.11 & 0.14$\pm$0.01$\pm$0.03    \\
IC1633  & 0.63$\pm$0.02 & 1.88$\pm$0.15 & 3.37$\pm$0.40 & 243 & 4.63$\pm$0.45 & 1.70$\pm$0.08$\pm$0.21   \\
IC4296  & 0.65$\pm$0.02 & 0.69$\pm$0.04 & 10.1$\pm$1.25 & 120 & 0.61$\pm$0.16 & 0.55$\pm$0.08$\pm$0.12     \\
IC5267  & 0.37$\pm$0.02 & 0.18$\pm$0.01 & 2.71$\pm$0.30 & 120 & 0.44$\pm$0.12 & 0.08$\pm$0.01$\pm$0.03   \\
IC5358  & 0.19$\pm$0.01 & 7.00$\pm$0.26 & 0.48$\pm$0.12 & 200 & 15.65$\pm$2.11 & 1.77$\pm$0.11$\pm$0.20  \\
NGC315  & 0.53$\pm$0.01 & 0.79$\pm$0.04 & 6.77$\pm$0.72 & 108 & 2.72$\pm$0.65 & 0.38$\pm$0.02$\pm$0.05   \\
NGC326  & 0.67$\pm$0.13 & 3.28$\pm$0.90 & 0.61$\pm$0.17 & 54 & 1.74$\pm$0.61 & 1.31$\pm$0.24$\pm$0.51   \\
NGC383  & 0.90$\pm$0.13 & 2.88$\pm$0.07 & 1.74$\pm$0.15  & 160  & 0.18$\pm$0.03 & 0.52$\pm$0.02$\pm$0.04   \\
NGC499  & 0.34$\pm$0.01 & 2.86$\pm$0.44 & 0.73$\pm$0.12 & 141 & 4.78$\pm$0.63 & 0.16$\pm$0.01$\pm$0.02   \\
NGC507  & 0.35$\pm$0.01 & 0.47$\pm$0.07 & 3.57$\pm$0.39 & 178 & 6.96$\pm$0.54 & 0.52$\pm$0.01$\pm$0.02   \\
NGC533  & 0.53$\pm$0.01 & 2.61$\pm$0.11 & 2.49$\pm$0.24 & 114 & 9.28$\pm$1.12 & 0.65$\pm$0.02$\pm$0.04   \\
NGC708  & 0.35$\pm$0.01 & 2.69$\pm$0.06 & 1.70$\pm$0.21 & 462 & 18.10$\pm$2.64 & 0.49$\pm$0.01$\pm$0.02    \\
NGC720  & 0.40$\pm$0.01 & 0.65$\pm$0.07 & 1.58$\pm$0.17 & 112 & 0.54$\pm$0.05 & 0.09$\pm$0.01$\pm$0.02   \\
NGC741  & 0.46$\pm$0.01 & 1.09$\pm$0.09 & 0.33$\pm$0.05 & 161 & 5.96$\pm$0.73 & 0.43$\pm$0.02$\pm$0.04   \\
NGC821  & 0.48$\pm$0.01 & 0.04$\pm$0.004 & 4.55$\pm$0.51 & 470 & 0.10$\pm$0.02 & 0.03$\pm$0.02$\pm$0.02   \\
NGC1023 & 0.35$\pm$0.01 & 0.01$\pm$0.003 & 5.02$\pm$0.68 & 124 & 0.22$\pm$0.05 & 0.03$\pm$0.01$\pm$0.02 \\
NGC1265 & 0.90$\pm$0.03 & 3.78$\pm$0.12 & 0.82$\pm$0.11 & 114  & 0.30$\pm$0.03 & 1.47$\pm$0.01$\pm$0.02   \\
NGC1266 & 0.48$\pm$0.02 & 0.09$\pm$0.03 & 2.08$\pm$1.08 & 301 & 0.05$\pm$0.01 & 0.05$\pm$0.01$\pm$0.02 \\
NGC1316 & 0.43$\pm$0.01 & 0.21$\pm$0.04 & 12.6$\pm$1.69  & 376 & 1.21$\pm$0.22 & 0.21$\pm$0.02$\pm$0.05   \\
NGC1332 & 0.84$\pm$0.04 & 0.64$\pm$0.04 & 9.11$\pm$1.53  & 358  & 0.05$\pm$0.01  & 0.19$\pm$0.03$\pm$0.05  \\
NGC1386 & 0.38$\pm$0.01 & 0.004$\pm$0.002 & 1.39$\pm$1.04 & 223  & 0.06$\pm$0.01  & 0.02$\pm$0.003$\pm$0.005 \\
NGC1399 & 0.48$\pm$0.01 & 0.32$\pm$0.01 & 12.5$\pm$1.42  & 254  & 0.71$\pm$0.11  & 0.30$\pm$0.02$\pm$0.04   \\
NGC1404 & 0.46$\pm$0.02 & 0.35$\pm$0.01 & 8.78$\pm$1.16 & 159  & 1.12$\pm$0.21 & 0.14$\pm$0.02$\pm$0.04    \\
NGC1407 & 0.40$\pm$0.01 & 0.47$\pm$0.05 & 2.21$\pm$0.21 & 136  & 0.74$\pm$0.11  & 0.21$\pm$0.02$\pm$0.04   \\
NGC1482 & 0.67$\pm$0.04 & 0.91$\pm$0.09 & 3.71$\pm$0.39  & 275 & 0.06$\pm$0.02 & 0.39$\pm$0.02$\pm$0.05   \\
NGC1550 & 0.40$\pm$0.01 & 2.07$\pm$0.07 & 2.09$\pm$0.23 & 323 & 5.48$\pm$0.55 & 0.29$\pm$0.02$\pm$0.03 \\
NGC1600 & 0.59$\pm$0.03 & 2.68$\pm$0.26 & 1.28$\pm$0.13  & 184  & 2.48$\pm$0.32  & 0.74$\pm$0.08$\pm$0.13   \\ 
NGC1700 & 0.55$\pm$0.03 & 1.67$\pm$0.17 & 1.33$\pm$0.14  & 245  & 0.74$\pm$0.12  & 0.13$\pm$0.02$\pm$0.04    \\
NGC2434 & 0.33$\pm$0.03 & 0.22$\pm$0.17 & 1.33$\pm$0.11  & 159 & 0.17$\pm$0.02  & 0.05$\pm$0.004$\pm$0.01  \\
NGC2768 & 0.29$\pm$0.02 & 0.27$\pm$0.12 & 0.81$\pm$0.12 & 128 & 0.30$\pm$0.05 & 0.04$\pm$0.003$\pm$0.01 \\
NGC3079 & 0.40$\pm$0.02 & 0.15$\pm$0.13 & 7.01$\pm$1.12 & 212 & 0.33$\pm$0.04 & 0.09$\pm$0.005$\pm$0.01 \\
NGC3091 & 0.37$\pm$0.01 & 0.46$\pm$0.05 & 4.36$\pm$0.51 & 181 & 3.01$\pm$0.35 & 0.16$\pm$0.015$\pm$0.02   \\
NGC3379 & 0.40$\pm$0.02 & 0.01$\pm$0.003 & 7.59$\pm$1.41 & 126 & 0.11$\pm$0.04 & 0.03$\pm$0.002$\pm$0.005 \\
NGC3384 & 0.42$\pm$0.04 & 0.01$\pm$0.005 & 7.47$\pm$1.65 & 256 & 0.04$\pm$0.01 & 0.03$\pm$0.003$\pm$0.008 \\
NGC3557 & 0.29$\pm$0.01 & 0.64$\pm$0.19  & 0.50$\pm$0.06  & 205  & 2.01$\pm$0.33  & 0.07$\pm$0.01$\pm$0.02  \\
NGC3585 & 0.48$\pm$0.06 & 0.02$\pm$0.01 & 7.36$\pm$1.15 & 70 & 0.03$\pm$0.01 & 0.05$\pm$0.01$\pm$0.02    \\
NGC3607 & 0.43$\pm$0.01 & 0.01$\pm$0.003  & 1.24$\pm$0.15  & 123  & 0.08$\pm$0.01 & 0.05$\pm$0.003$\pm$0.007   \\
NGC3665 & 0.65$\pm$0.15 & 1.76$\pm$0.74  & 0.41$\pm$0.05  & 109  & 0.11$\pm$0.01 & 0.09$\pm$0.01$\pm$0.03   \\
NGC3923 & 0.53$\pm$0.01 & 0.55$\pm$0.03 & 3.43$\pm$0.36  & 135  & 0.60$\pm$0.08  & 0.15$\pm$0.02$\pm$0.04   \\
NGC3955 & 0.33$\pm$0.05 & 0.56$\pm$0.35 & 0.49$\pm$0.06 & 124 & 0.03$\pm$0.01 & 0.01$\pm$0.002$\pm$0.006   \\
NGC4036 & 0.26$\pm$0.01 & 0.02$\pm$0.002 & 11.3$\pm$1.57 & 106 & 0.30$\pm$0.05 & 0.02$\pm$0.003$\pm$0.008 \\
NGC4073 & 0.41$\pm$0.01 & 2.16$\pm$0.12 & 3.02$\pm$0.31 & 150 & 8.34$\pm$1.43  & 0.81$\pm$0.05$\pm$0.08   \\
NGC4104 & 0.55$\pm$0.01 & 1.52$\pm$0.10 & 4.55$\pm$0.51 & 57  & 6.91$\pm$0.42  & 1.14$\pm$0.05$\pm$0.08   \\
NGC4125 & 0.33$\pm$0.01 & 0.38$\pm$0.06  & 1.58$\pm$0.16 & 174 & 0.33$\pm$0.05 & 0.06$\pm$0.01$\pm$0.02    \\
NGC4203 & 0.57$\pm$0.01 & 0.06$\pm$0.002 & 4.55$\pm$0.58 & 134 & 0.03$\pm$0.01 & 0.04$\pm$0.002$\pm$0.005 \\
NGC4261 & 0.56$\pm$0.01 & 0.43$\pm$0.01 & 8.48$\pm$1.17 & 376 & 0.37$\pm$0.04 & 0.21$\pm$0.01$\pm$0.02    \\
NGC4278 & 0.58$\pm$0.01 & 0.08$\pm$0.01 & 2.43$\pm$0.25 & 198 & 0.03$\pm$0.01  & 0.05$\pm$0.002$\pm$0.005   \\
NGC4325 & 0.67$\pm$0.03 & 13.1$\pm$0.71 & 0.78$\pm$0.11 & 230 & 2.27$\pm$0.14 & 0.12$\pm$0.01$\pm$0.02     \\
NGC4342 & 0.31$\pm$0.01 & 0.01$\pm$0.002 & 5.09$\pm$0.63 & 141 & 0.05$\pm$0.01 & 0.02$\pm$0.002$\pm$0.006 \\
NGC4365 & 0.36$\pm$0.02 & 0.22$\pm$0.09 & 1.69$\pm$0.15 & 107 & 0.23$\pm$0.03 & 0.04$\pm$0.002$\pm$0.005   \\
NGC4374 & 0.48$\pm$0.01 & 0.49$\pm$0.03 & 3.52$\pm$0.37 & 339  & 0.37$\pm$0.05 & 0.14$\pm$0.01$\pm$0.02     \\
NGC4382 & 0.31$\pm$0.01 & 0.01$\pm$0.003  & 2.72$\pm$0.31 & 125 & 0.48$\pm$0.07 & 0.04$\pm$0.003$\pm$0.008    \\
NGC4388 & 0.44$\pm$0.01 & 0.10$\pm$0.02 & 1.83$\pm$0.22 & 276 & 0.48$\pm$0.06 & 0.19$\pm$0.02$\pm$0.03     \\
NGC4406 & 0.34$\pm$0.01 & 0.51$\pm$0.08 & 2.36$\pm$0.27 & 171 & 1.05$\pm$0.16 & 0.12$\pm$0.02$\pm$0.03 \\
NGC4457 & 0.43$\pm$0.01 & 0.02$\pm$0.001  & 9.59$\pm$1.72 & 181 & 0.11$\pm$0.02 & 0.05$\pm$0.001$\pm$0.002    \\
\hline
\end{tabular}
\end{table*}

\setcounter{table}{2}
\begin{table*}
\centering
\caption{Continued.}
\begin{tabular}{lcccccc}
\hline
Name   & $\beta$       & $r_c$         & $\rho_0$   & $\chi^2$ &  $M_{g}$              & $M$                    \\
       &               & kpc           & 10$^{-24}$ &          &  10$^{11}$ $M_{\odot}$ & 10$^{13}$ $M_{\odot}$  \\
       &               &               & g/cm$^3$   &          &                       &                        \\
  (1)  &         (2)   &  (3)          &  (4)       & (5)      &   (6)                 & (7)                   \\
\hline
NGC4472 & 0.43$\pm$0.01 & 0.32$\pm$0.01 & 6.66$\pm$0.74 & 243 & 0.74$\pm$0.08 & 0.20$\pm$0.01$\pm$0.02  \\
NGC4477 & 0.65$\pm$0.13 & 2.73$\pm$0.67 & 0.33$\pm$0.05 & 116 & 0.10$\pm$0.02 & 0.05$\pm$0.006$\pm$0.01 \\
NGC4486 & 0.50$\pm$0.01 & 2.33$\pm$0.02 & 4.70$\pm$0.53 &  339 & 2.63$\pm$0.12 & 0.41$\pm$0.03$\pm$0.05 \\
NGC4526 & 0.52$\pm$0.04 & 0.59$\pm$0.11  & 1.39$\pm$0.15 & 138 & 0.10$\pm$0.02 & 0.05$\pm$0.003$\pm$0.005   \\
NGC4552 & 0.50$\pm$0.01 & 0.18$\pm$0.01  & 2.70$\pm$0.41 & 395 & 0.24$\pm$0.04 & 0.07$\pm$0.006$\pm$0.01   \\
NGC4555 & 0.55$\pm$0.02 & 1.30$\pm$0.14  & 3.14$\pm$0.27  & 85 & 1.96$\pm$0.10  & 0.41$\pm$0.02$\pm$0.03   \\
NGC4564 & 0.90$\pm$0.73 & 0.47$\pm$0.10 & 1.25$\pm$0.11 & 74 & 0.01$\pm$0.003 & 0.05$\pm$0.002$\pm$0.004    \\
NGC4621 & 0.28$\pm$0.03 & 0.04$\pm$0.01 & 2.54$\pm$0.26 & 86 & 0.13$\pm$0.02 & 0.03$\pm$0.002$\pm$0.003   \\
NGC4636 & 0.33$\pm$0.01 & 0.32$\pm$0.02 & 3.03$\pm$0.31 & 464  & 1.18$\pm$0.12 & 0.10$\pm$0.01$\pm$0.02   \\
NGC4649 & 0.49$\pm$0.01 & 0.30$\pm$0.01 & 12.6$\pm$1.41 & 596 & 0.66$\pm$0.08 & 0.20$\pm$0.02$\pm$0.03    \\
NGC4696 & 0.31$\pm$0.01 & 0.96$\pm$0.10 & 3.55$\pm$0.37 & 280 & 6.46$\pm$0.83 & 0.42$\pm$0.03$\pm$0.05  \\
NGC4697 & 0.26$\pm$0.03 & 0.26$\pm$0.20 & 0.58$\pm$0.09 & 121 & 0.35$\pm$0.06 & 0.03$\pm$0.004$\pm$0.005   \\
NGC4710 & 0.90$\pm$0.67 & 0.99$\pm$0.11 & 0.76$\pm$0.06 & 77 & 0.01$\pm$0.01 & 0.07$\pm$0.01$\pm$0.02 \\
NGC4782 & 0.33$\pm$0.02 & 1.62$\pm$0.50 & 0.56$\pm$0.07 & 322 & 4.11$\pm$0.52 & 0.24$\pm$0.02$\pm$0.03    \\
NGC4936 & 0.48$\pm$0.05 & 1.10$\pm$0.57 & 1.11$\pm$0.09  & 89 & 1.16$\pm$0.14 & 0.23$\pm$0.04$\pm$0.06  \\
NGC5018 & 0.57$\pm$0.05 & 0.86$\pm$0.17 & 1.58$\pm$0.17  & 102  & 0.16$\pm$0.02 & 0.13$\pm$0.02$\pm$0.03  \\  
NGC5044 & 0.27$\pm$0.01 & 2.24$\pm$0.27 & 1.06$\pm$0.11 & 284 & 2.43$\pm$0.26 & 0.14$\pm$0.02$\pm$0.03  \\
NGC5171 & 0.90$\pm$0.05 & 5.07$\pm$0.40 & 2.22$\pm$0.27  & 177 & 1.82$\pm$0.15 & 0.81$\pm$0.01$\pm$0.02   \\
NGC5353 & 0.44$\pm$0.01 & 0.34$\pm$0.04 & 4.09$\pm$0.42 & 133 & 0.33$\pm$0.04 & 0.13$\pm$0.01$\pm$0.02    \\
NGC5532 & 0.66$\pm$0.02 & 1.43$\pm$0.10 & 3.85$\pm$0.32 & 73 & 1.35$\pm$0.12  & 0.55$\pm$0.03$\pm$0.05   \\
NGC5813 & 0.27$\pm$0.01 & 0.34$\pm$0.04 & 3.17$\pm$0.32 & 114 & 1.68$\pm$0.19 & 0.07$\pm$0.01$\pm$0.02   \\
NGC5846 & 0.38$\pm$0.01 & 0.73$\pm$0.04 & 1.73$\pm$0.27 & 293 & 0.66$\pm$0.08  & 0.11$\pm$0.01$\pm$0.02   \\
NGC5866 & 0.37$\pm$0.02 & 0.39$\pm$0.11 & 1.19$\pm$0.08 & 182 & 0.22$\pm$0.04 & 0.05$\pm$0.01$\pm$0.02 \\
NGC6098 & 0.42$\pm$0.01 & 0.20$\pm$0.10 & 9.41$\pm$1.64 & 277 & 5.53$\pm$0.61 & 0.81$\pm$0.05$\pm$0.08   \\
NGC6107 & 0.36$\pm$0.01 & 0.46$\pm$0.17 & 4.89$\pm$0.55 & 46 & 13.47$\pm$1.58 & 0.91$\pm$0.06$\pm$0.10   \\
NGC6251 & 0.47$\pm$0.004 & 0.38$\pm$0.03 & 3.00$\pm$0.41 & 196 & 14.51$\pm$1.82 & 0.52$\pm$0.03$\pm$0.05   \\
NGC6269 & 0.37$\pm$0.01 & 0.08$\pm$0.01 & 4.37$\pm$0.71 & 194 & 13.62$\pm$2.19 & 0.85$\pm$0.02$\pm$0.03   \\
NGC6278 & 0.42$\pm$0.07 & 6.53$\pm$1.52 & 0.31$\pm$0.06 & 413 & 0.95$\pm$0.11 & 0.24$\pm$0.03$\pm$0.05 \\
NGC6338 & 0.82$\pm$0.04 & 9.45$\pm$0.50 & 1.16$\pm$0.12 & 264 & 6.21$\pm$0.51 & 1.99$\pm$0.15$\pm$0.21  \\
NGC6482 & 0.49$\pm$0.01 & 1.44$\pm$0.07 & 3.96$\pm$0.44 & 165 & 2.11$\pm$0.37 & 0.16$\pm$0.02$\pm$0.03   \\
NGC6861 & 0.55$\pm$0.01 & 0.74$\pm$0.05 & 2.91$\pm$0.25 & 153 & 0.39$\pm$0.04 & 0.31$\pm$0.03$\pm$0.05   \\
NGC6868 & 0.36$\pm$0.02 & 0.55$\pm$0.10 & 1.07$\pm$0.08 & 311 & 0.73$\pm$0.15 & 0.23$\pm$0.02$\pm$0.03    \\
NGC7176 & 0.47$\pm$0.04 & 0.70$\pm$0.15 & 0.83$\pm$0.12 & 171 & 0.35$\pm$0.02 & 0.26$\pm$0.03$\pm$0.05   \\
NGC7196 & 0.71$\pm$0.11 & 2.30$\pm$0.46 & 0.87$\pm$0.11 & 139 & 0.22$\pm$0.03 & 0.21$\pm$0.04$\pm$0.06    \\ 
NGC7618 & 0.40$\pm$0.01 & 2.19$\pm$0.15 & 1.02$\pm$0.08 & 334 & 5.53$\pm$0.29 & 0.35$\pm$0.02$\pm$0.03   \\
NGC7626 & 0.48$\pm$0.02 & 0.81$\pm$0.10 & 1.88$\pm$0.17 & 190 & 0.64$\pm$0.05 & 0.24$\pm$0.03$\pm$0.05   \\
UGC408  & 0.90$\pm$0.68 & 1.54$\pm$0.13 & 1.11$\pm$0.10 & 64 & 0.03$\pm$0.01 & 0.51$\pm$0.02$\pm$0.03  \\
\hline
\end{tabular}
\end{table*}

The best-fitting parameters from the density formulation of the $\beta$-model are shown in Table \ref{tab3}. We present the gas and total masses measured within 5$r_e$. When the X-ray surface brightness profile does not reach 5$r_e$, we extrapolated the total and gas mass profiles out to 5$r_e$ using the linear slope of the last 20 points in log-log space.

The main source of error in the gas mass are the modeled parameters $\beta$ and $r_c$.  The $\chi^2$ values presented in Table~\ref{tab3} have not been divided by the degrees of freedom, 96. For most observations the $\beta$-model provides an accurate fit to the X-ray profile. We apply two methods to define the total mass uncertainties. First, we use the best-fitting parameters of $\beta$-model and Monte Carlo simulations. Due to the small uncertainties on $\beta$ and $r_c$, the density and mass profiles have small statistical uncertainties. Second, we estimate the total mass uncertainties by propagating the errors through Eq.~\ref{mtot}. The total mass uncertainties obtained from the best-fitting parameters are given as a second value in column 7 of Tab.~\ref{tab3}, while propagated errors are given as a third value in the same column. We found that uncertainties obtained by the propagation method are higher by a factor of $1.5-2$ than errors estimated from the best-fitting parameters.

\subsection{Comparison with previous results}\label{subs_comp}

\begin{figure}
\includegraphics[width=0.49\textwidth]{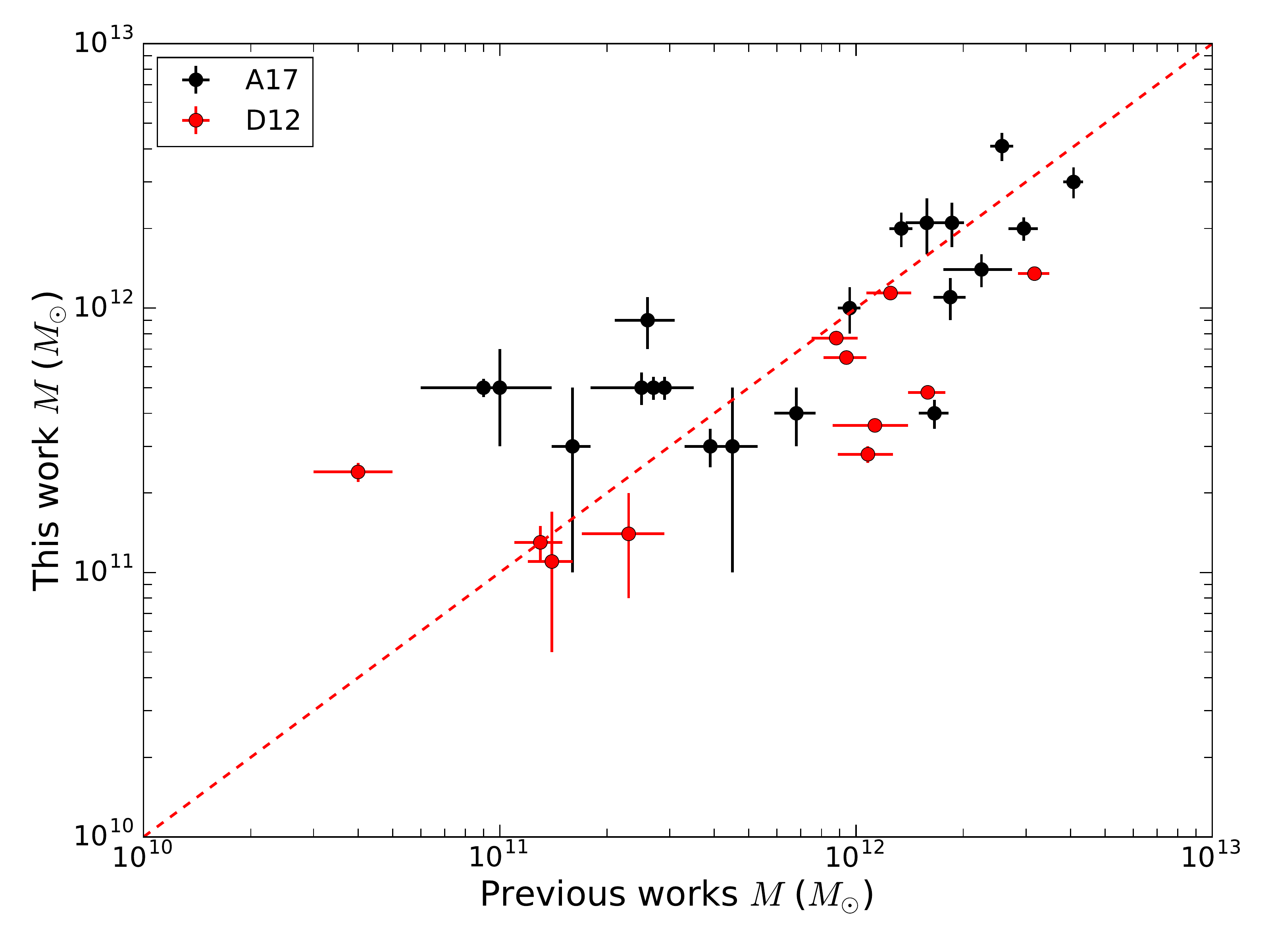}
\caption{Comparison of total mass within 5$r_e$ to previous works.}
\label{fig_mc}
\end{figure}

In Figure~\ref{fig_mc} total mass measured within 5$r_e$ is plotted against the total mass derived from stellar velocity dispersions.  Velocity dispersions were obtained from \citet{Deason:12} and \citet{Alabi:17} (D12 and A17 in Figure~\ref{fig_mc}, respectively). The dynamical measurements of total mass were also restricted to the central 5$r_e$. D12 and A17 define their total early-type galaxy masses within 5$r_e$ using the velocities of planetary nebulae and globular clusters. In general our masses agree with those from D12 and A17, although with large scatter for systems lying below 10$^{12} M_{\odot}$. The principal source of error lies in the measurement of effective radius. Our effective radii are derived from optical data, while D12 and A17 use the near-IR. We explore this bias below.

Measurements of galaxy size are difficult to standardize because of their dependence on wavelength and background noise. As such, it is difficult to find a consensus on galaxy sizes in the literature, particularly at high stellar masses. For example, the SLUGGS sample \citep{Alabi:16} used galaxy sizes taken from ATLAS$^{3D}$, which is based on $2MASS$ and $RC3$ estimates. These measurements underestimate galaxy sizes at high stellar masses by up to a factor 3 relative to the $Spitzer$ masses from \citet{Forbes:16}. The ATLAS$^{3D}$ collaboration acknowledged this issue using the size-stellar mass relation.  They found that the galaxies with high stellar masses are significantly smaller than expected (see \citealt{Cappellari:11}). Variations in effective radius naturally propagate into differences in mass measured at a fixed multiple of $r_e$.  Deeper observations are better able to trace light in the outskirts of galaxies, enabling higher fidelity measurements of $r_e$. The $Spitzer$ data are 3 magnitudes deeper than $2MASS$, so are better suited for determining $r_e$. Additionally, the near-IR light observed by $Spitzer$ traces old stars, where age and metallicity degeneracies are unimportant and effects of dust are minimized.

\begin{figure}
\includegraphics[width=0.49\textwidth]{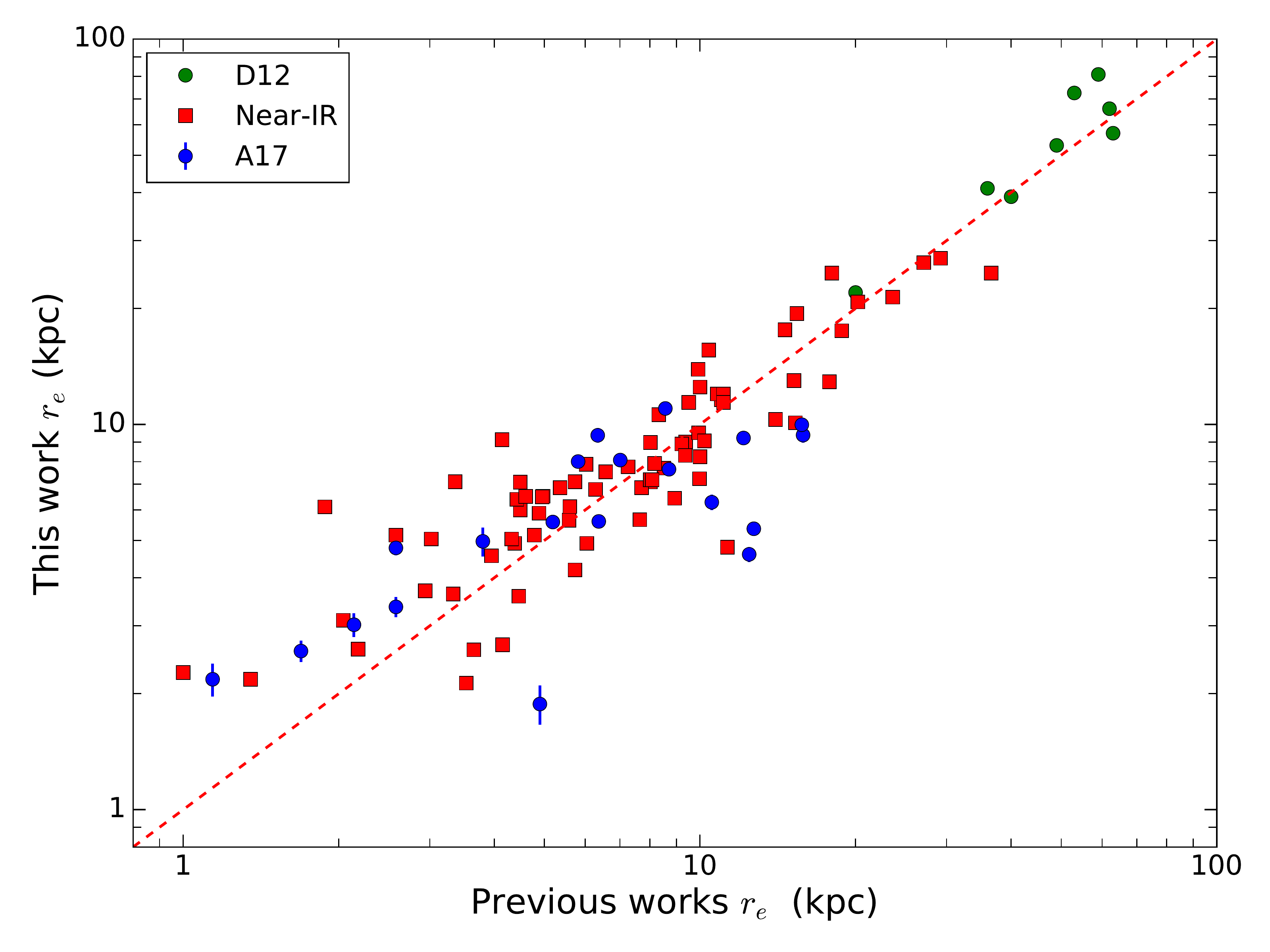}
\caption{Comparison of effective radius estimates using optical $DSS$ and near-IR $Spitzer$ images to the radii defined in D12 and A17.}
\label{fig_rc}
\end{figure}

In Figure~\ref{fig_rc} we compare our optical effective radii to the near-IR radii measured by D12 and A17.  Only 30 objects overlap.  We therefore have repeated our measurement of $r_e$ (see Sec.~\ref{subsec_opt}) using data from the $Spitzer$ database, which includes 75 of our galaxies. We find general agreement between the optical and near-IR measurements for $r_e\geq10$~kpc. At smaller $r_e$ our optical radii are systematically larger than the near-IR values. We conclude that the large scatter seen in Fig.~\ref{fig_mc} for low masses is related to issues in the effective radius measurement.

\section{Scaling relation results}\label{sec_rel}
\begin{table*}
\centering
\caption{Scaling relations of the form log($y$) = $a$ + $b$ log($x$). 
%Luminosities are expressed in units of 10$^{40}$~erg/s and masses in 10$^{12}M_{\odot}$.
}\label{tab_res}
\begin{tabular}{llcccccccc}
\hline
Sample & $N$ &  $C_{cor}$ & $a$ & $b$ & $p$-Pearson & $p$-Spearman & $p$-AD & $p$-SW & rms scatter \\
\hline
       &  & &   & $L_X-T$  \\
\hline
Full (BCES) & 90 & 0.86 & 41.37$\pm$0.07 & 4.42$\pm$0.19 & $>>$0.0001 & $>>$0.0001 & 0.54 & 0.11 & 0.99 \\
Full (Kelly's) & 90 & 0.86 & 41.26$\pm$0.07 & 4.53$\pm$0.40 & $>>$0.0001 & $>>$0.0001 & 0.65 & 0.10 & 0.99 \\

$L_X > 10^{40}$ (BCES) & 65 & 0.75 & 43.15$\pm$13.80 & 4.24$\pm$0.16 & $>>$0.0001 & $>>$0.0001 &0.55 &0.28& 0.67 \\
$L_X > 10^{40}$ (Kelly's) & 65 & 0.75 & 41.33$\pm$0.06 & 3.78$\pm$0.48 & $>>$0.0001 & $>>$0.0001 & 0.46 & 0.32 & 0.67 \\

$L_X < 10^{40}$ (BCES) & 25 & 0.60 & 40.95$\pm$0.67 & 0.85$\pm$1.24 & 0.0016 & 0.0016 & 0.74 & 0.08 & 0.38 \\
$L_X < 10^{40}$ (Kelly's) & 25 & 0.60 & 39.92$\pm$0.26 & 1.09$\pm$0.75 & 0.0016 & 0.0016 & 0.76 & 0.03 & 0.38 \\
\hline
 &&&& $M-T$ &&\\
\hline
Full (BCES) & 90 & 0.85 & 12.56$\pm$0.04 & 2.43$\pm$0.19 & $>>$0.0001 & $>>$0.0001 & 0.35 & 0.29 & 0.52 \\
Full (Kelly's) & 90 & 0.85 & 12.47$\pm$0.04 & 2.43$\pm$0.25 & $>>$0.0001 & $>>$0.0001 & 0.38 & 0.43 & 0.52 \\
\hline
&&&&$L_X-M$ &&\\
\hline
Full (BCES) & 90 & 0.78 & 13.35$\pm$2.34 & 2.78$\pm$0.33 & $>>$0.0001 & $>>$0.0001 & 0.74 & 0.12 & 0.99 \\
Full (Kelly's) & 90 & 0.78 & 20.40$\pm$1.95 & 2.65$\pm$0.15 & $>>$0.0001 & $>>$0.0001 & 0.72 & 0.10 & 0.99 \\
\hline
&&&&$M-Y_X$ &&\\
\hline
Full (BCES) & 90 & 0.76 & 7.41$\pm$0.41 & 0.45$\pm$0.04 & $>>$0.0001 & $>>$0.0001 & 0.75 & 0.09 & 0.52\\
Full (Kelly's) & 90 & 0.76 & 8.14$\pm$0.49 & 0.38$\pm$0.05 & $>>$0.0001 & $>>$0.0001 & 0.77 & 0.03 & 0.52\\
\hline
\end{tabular}
\end{table*}

In this section we explore four scaling relations for our sampled ETGs. In addition to the $L_X-T$, $M-T$, and $L_X-M$ relations, we derive the relation between total mass and $Y_X = T \times M_g$. The $Y_X$ indicator has been studied primarily in galaxy clusters but not lower mass systems (\citealt{Kravtsov:06, Nagai:07}). 

To determine the form of the scaling relation, we performed (1) linear fits in log space using the bivariate correlated error and intrinsic scatter (BCES) algorithm \citep{Akritas:96} as well as (2) likelihood-based approach of \citet{Kelly:07}. The orthogonal BCES algorithm performs a linear least-squares regression that minimizes the orthogonal distance to the best-fit relation. Parameter uncertainties were determined using 10,000 Monte Carlo bootstrap re-samplings. Although the BCES method de-biases least squares linear regression for measurement errors, it is not perfect. Kelly's regression is better, both in bias removal and in improved confidence intervals. It is a Bayesian method based on deriving a likelihood function. This method is implemented in Linmix package\footnote{Python version - https://github.com/jmeyers314/linmix.} and takes intrinsic scatter into account. Parameter uncertainties for Kelly's methods were obtained by running $\sim$ 15,000 steps of a Markov Chain Monte Carlo.

We find that both methods provide similar results. Since both these methods assume residuals are normally distributed, we perform the Anderson-Darling (AD) and Shapiro-Wilks (SW) tests to check their residuals for normality. We obtain $p >$ 0.5 in our scaling relations, indicating that the residuals are normally distributed.

We also used the Pearson and Spearman correlation tests to determine the significance of linear relationship between two datasets. Finally we define a root mean square scatter (rms scatter) for each relation as
\begin{equation}
{\rm rms} = \sqrt{\frac{\sum(f_{i}-< f >)^{2}}{N}},
\label{rms}
\end{equation}
where $<f>$ is the fitted relation. This sample includes both gas-rich and gas-poor objects.  We have therefore subdivided the sample based on X-ray luminosity. For consistency to previous works (e.g., \citet{Kim:13, Kim:15}), we set the $L_X$ threshold at 10$^{40}$ erg/s. The resulting best fits, their uncertainties, correlation coefficients, $p$-values for null-hypothesis and normality as well as rms scatters are shown in Table~\ref{tab_res}. 

%This sample includes both gas-rich and gas-poor objects.  We have therefore subdivided the sample based on several physical parameters and determined the power law scaling relations for each subsample. The subsamples were defined based on X-ray luminosity, metallicity, stellar velocity dispersion, and a discernible X-ray core. 

\subsection{$L_X-T$}\label{sec_LT}

Our measured $L_X-T$ relation is shown on the left side of Figure~\ref{fig_lt_mt}. From top to bottom the relation is color-coded by galaxy type, metallicity, and stellar velocity dispersion. The contribution of LMXBs (green dashed line) and other faint stellar sources (magenta dashed line) to the total X-ray emission is shown in the upper left panel. These components were modeled simultaneously with the thermal emission described in Section~\ref{sec_spec_anal}. Our X-ray luminosity measurements for LMXBs and other stellar sources agree with previous estimates \citep{Irwin:96, Irwin:98, Revnivtsev:07a, Revnivtsev:07b, Boroson:10}. In addition, the X-ray luminosities of these components are consistent with expectations when scaling from stellar mass \citep{Revnivtsev:08a, Revnivtsev:08b}. Using ROSAT observations, \citet{Irwin:98} measured the X-ray luminosities of these stellar components to be in the range 10$^{36}$-10$^{39}$ erg/s. \citet{Revnivtsev:08a} found that the low-mass X-ray binaries are characterized by X-ray luminosities of 10$^{37}$-10$^{39}$ erg/s. In the 0.5-2.0 keV energy band, the unresolved X-ray emission is characterized by $L_X/M_{*} \sim $ 8.2$\times$10$^{27}$ erg/s/$M_{\odot}$ (emissivity per unit stellar mass). It is consistent with measurements of dwarf ellipticals, spiral bulges, and the Milky Way. Such consistency suggests that the bulk of the unresolved emission is produced by an old stellar population that can be characterized by a universal emissivity per unit stellar mass \citep[see][for more details]{Revnivtsev:08a}.

The best-fitting $L_X-T$ relations over our entire sample are $L_X \propto T^{4.42\pm0.19}$ and $L_X \propto T^{4.53\pm0.40}$ using BCES and Kelly's regression methods, respectively. For gas-rich objects with X-ray luminosities $>10^{40}$~erg/s, the slopes are slightly shallower, 4.24$\pm$0.16 and 3.78$\pm$0.48, respectively. Gas-poor objects below this value reveal no clear correlation. The metallicity and velocity dispersion subsamples are discussed further in the next section.

Our relation holds over a wide range of X-ray luminosity ($\sim10^{38}-5\times 10^{42}$~erg/s) and temperature ($\sim0.1-2$~keV). The most massive and luminous objects in our sample are BCGs and cDs, which occupy top-right corner of the plot. The red dash-dotted line indicates the self-similar scaling. Our observed $L_X-T$ relation is significantly steeper than the self-similar prediction of $L_X \propto T^2$.  This steepening indicates baryonic physics on both small and large scales. The $L_X-T$ relation is steeper in ETGs than in clusters, indicating that non-gravitational processes are more efficient in low-mass systems. 

\subsection{$M-T$}\label{sec_MT}

\begin{figure*}
\begin{minipage}{0.495\textwidth}
\includegraphics[width=1.0\textwidth]{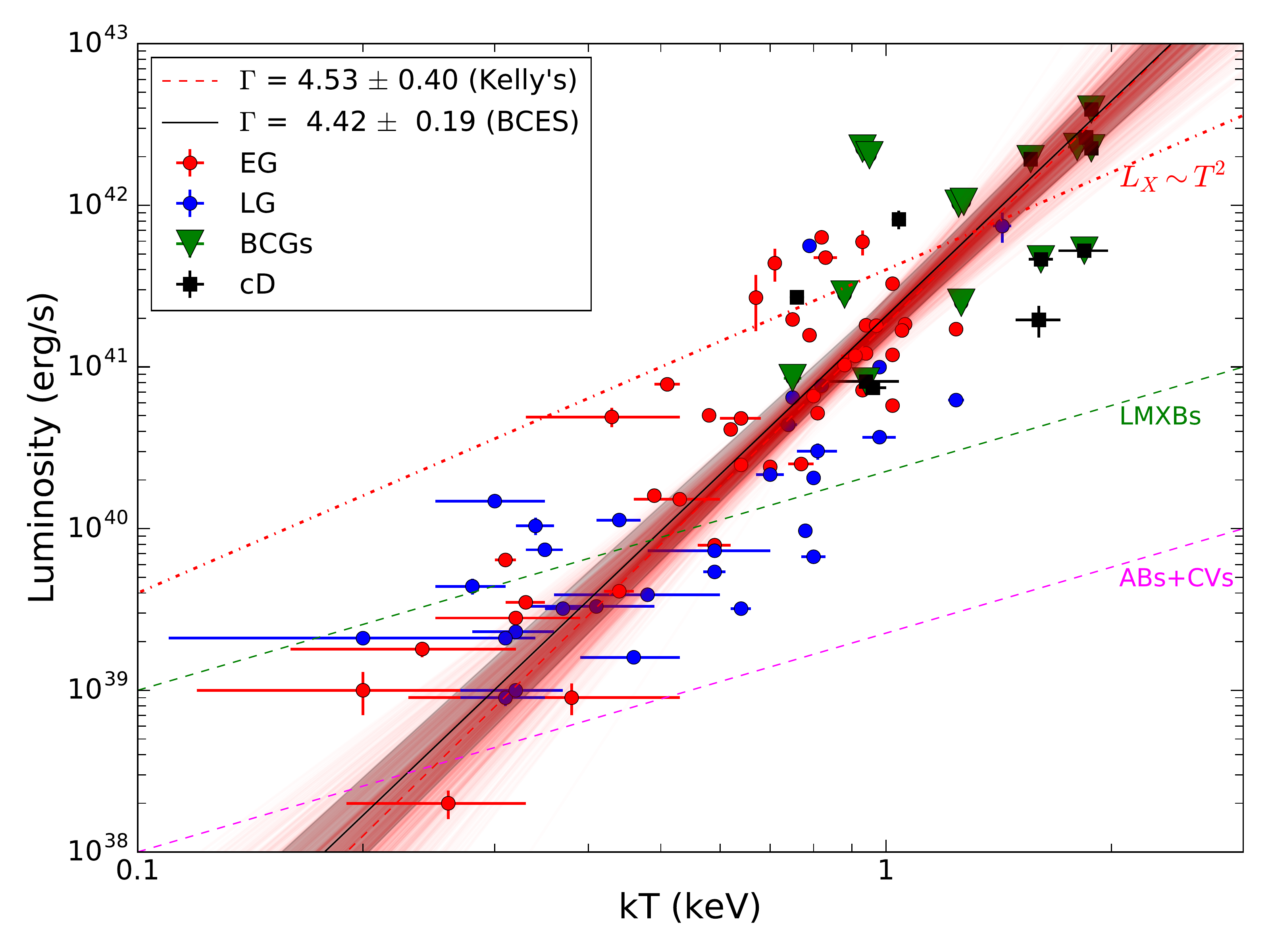}
\includegraphics[width=1.0\textwidth]{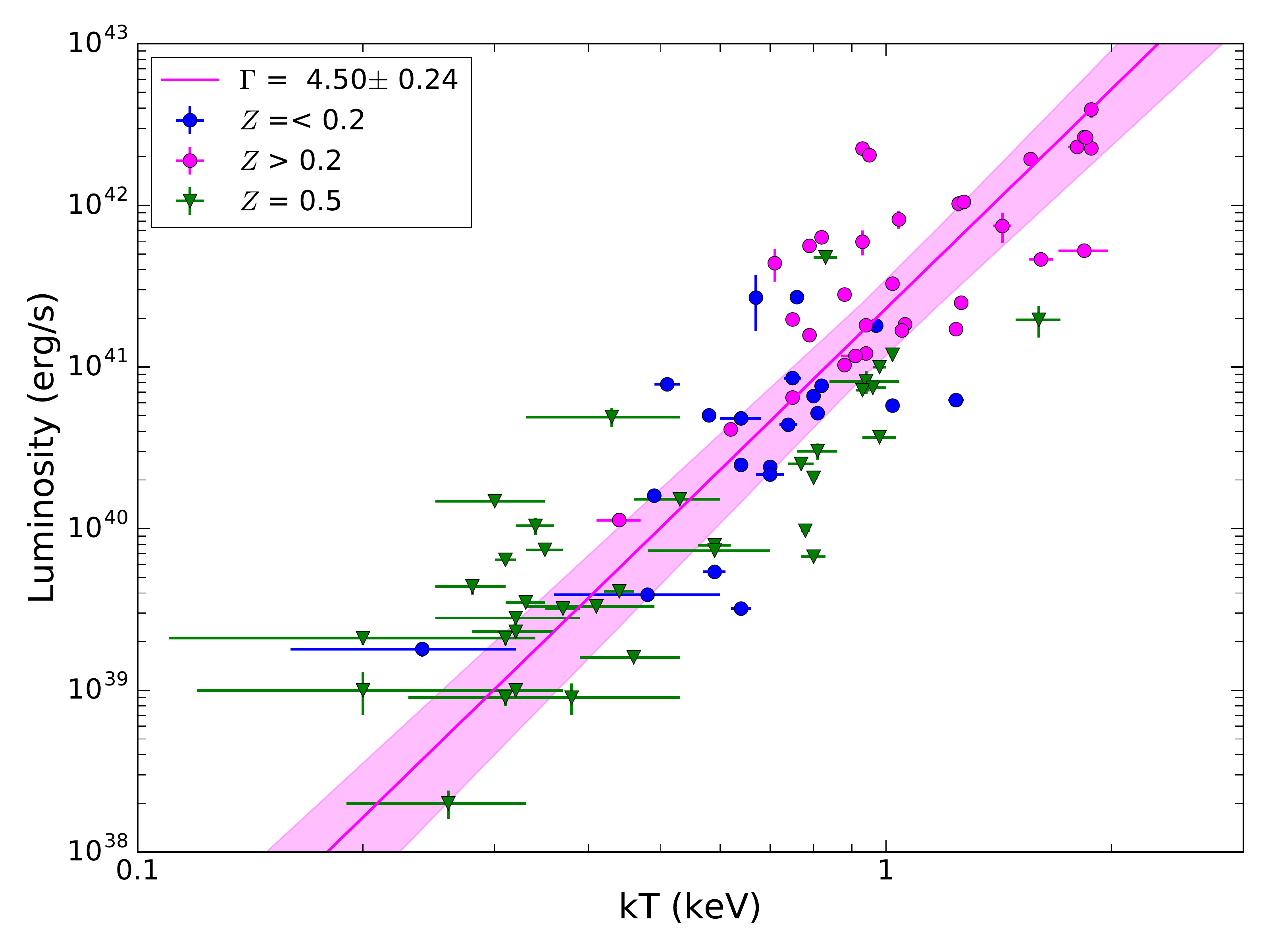}
\includegraphics[width=1.0\textwidth]{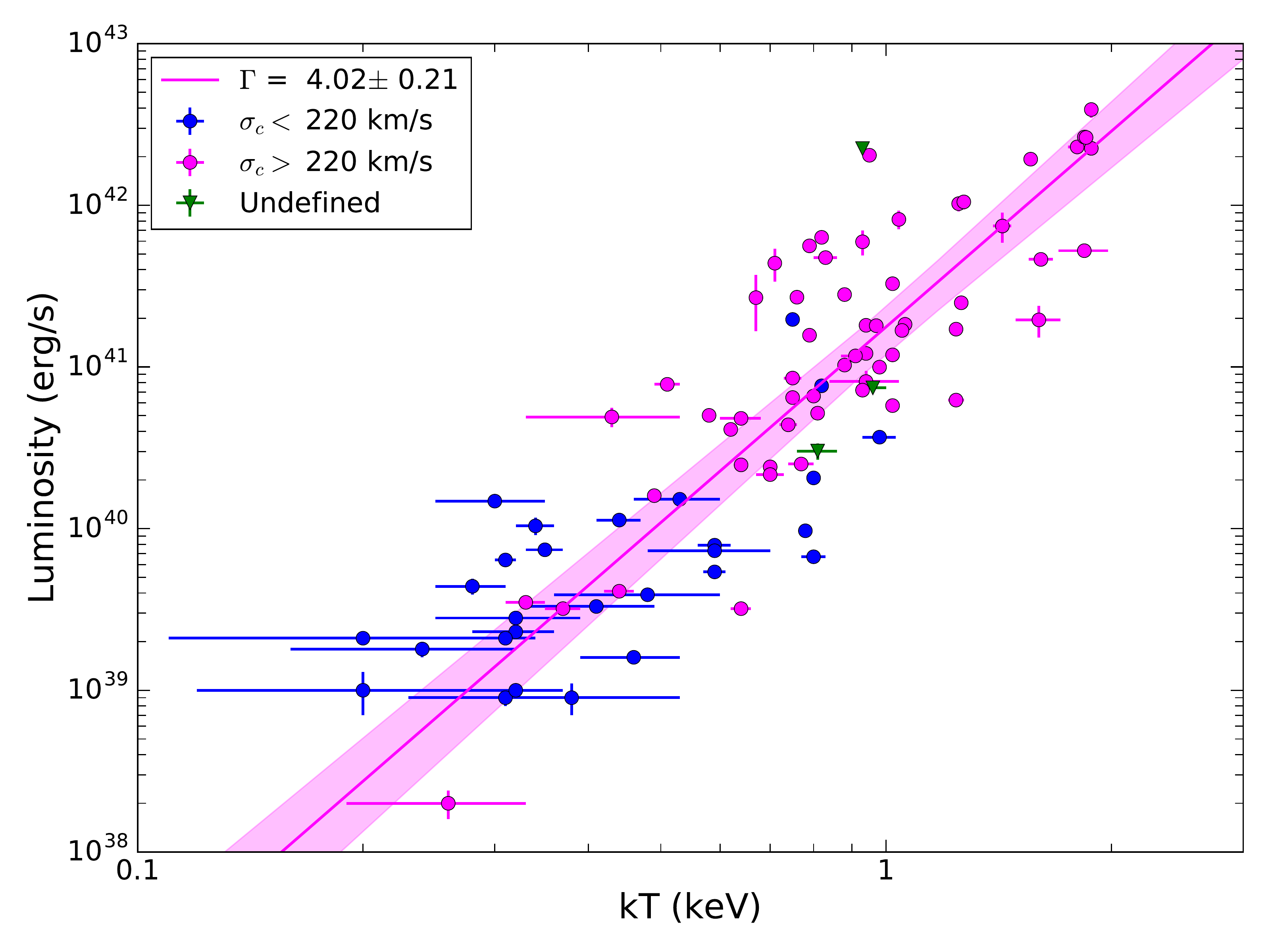}
\end{minipage}
\begin{minipage}{0.495\textwidth}
\includegraphics[width=1.0\textwidth]{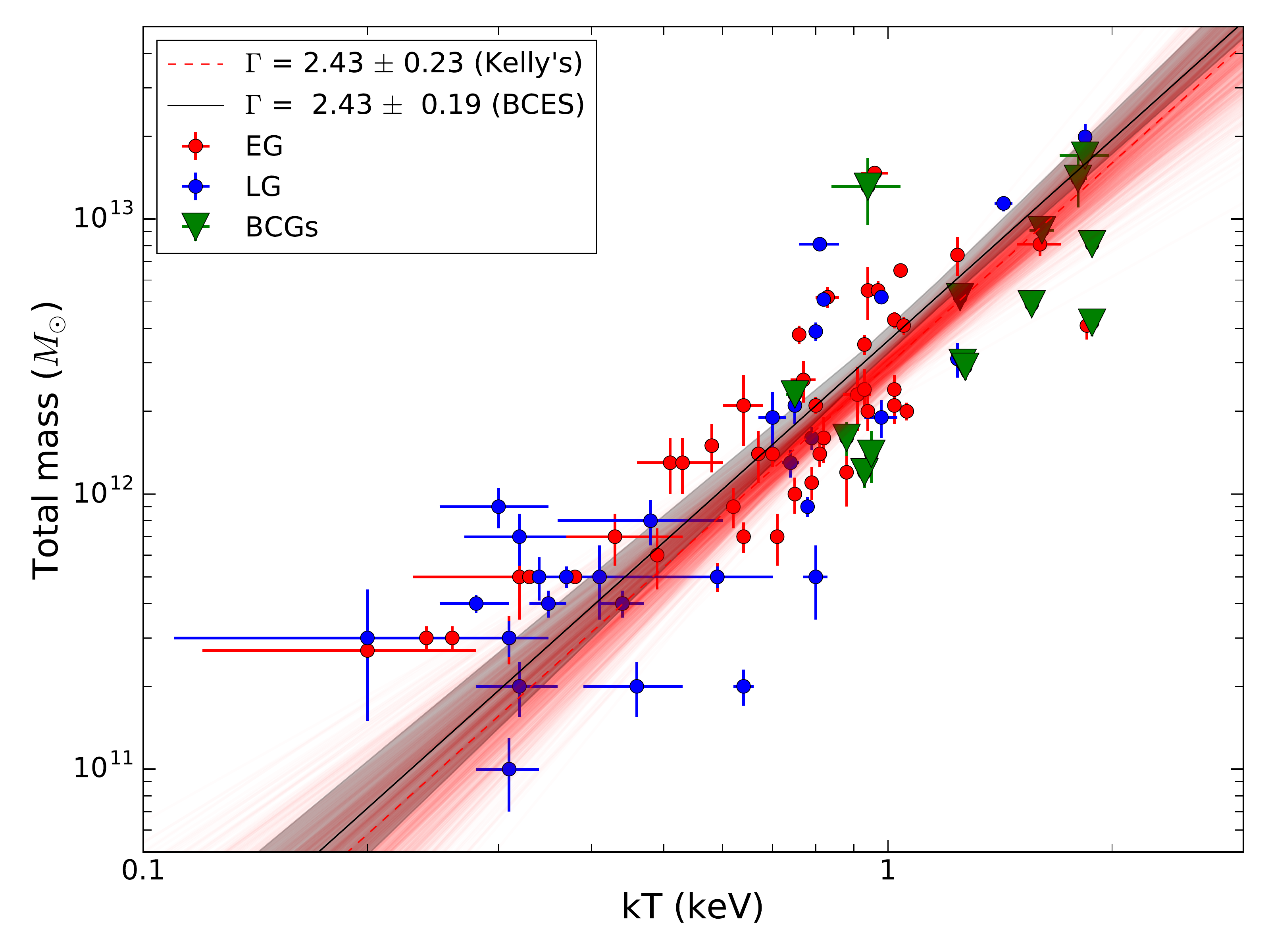}
\includegraphics[width=1.0\textwidth]{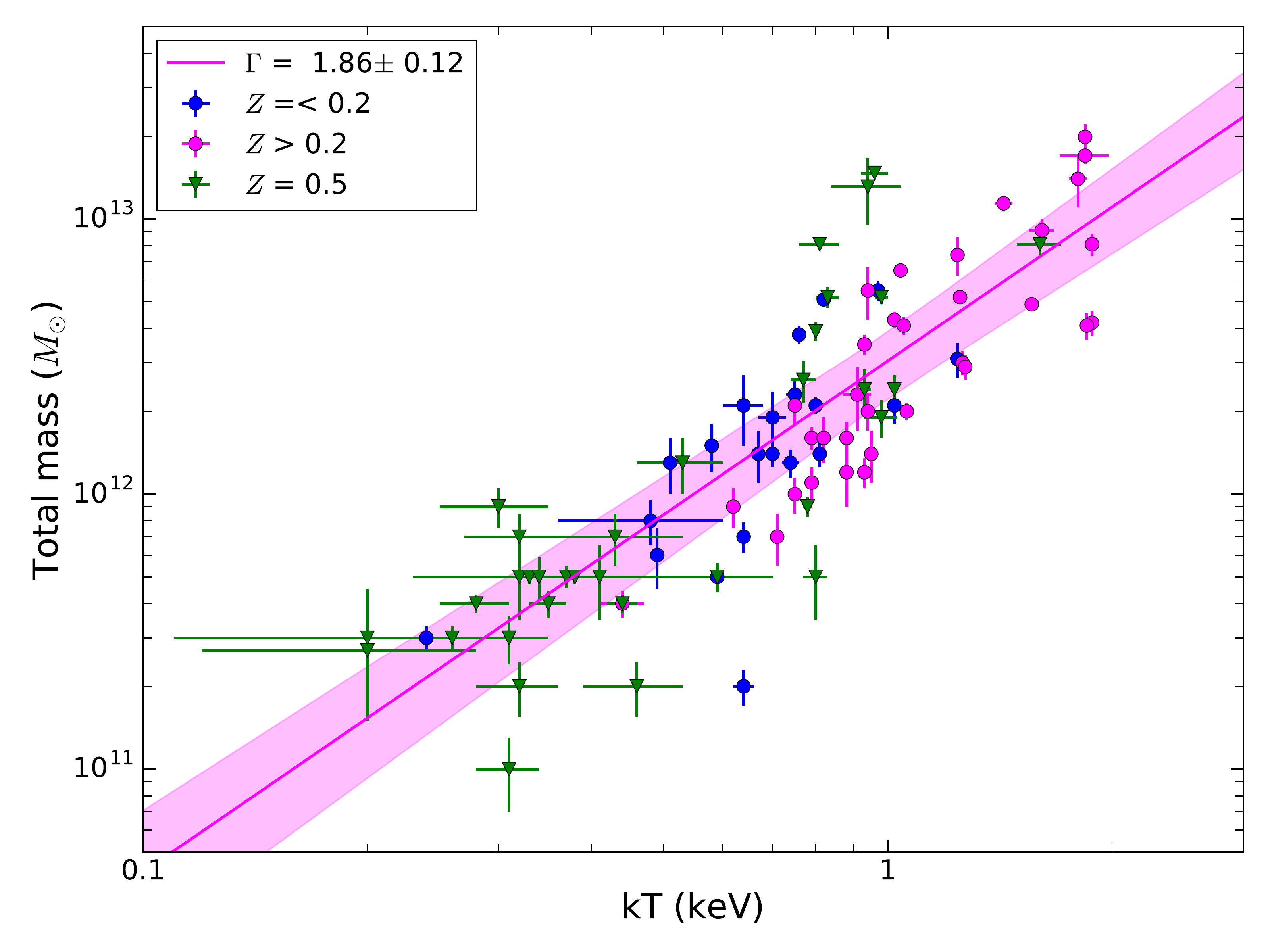}
\includegraphics[width=1.0\textwidth]{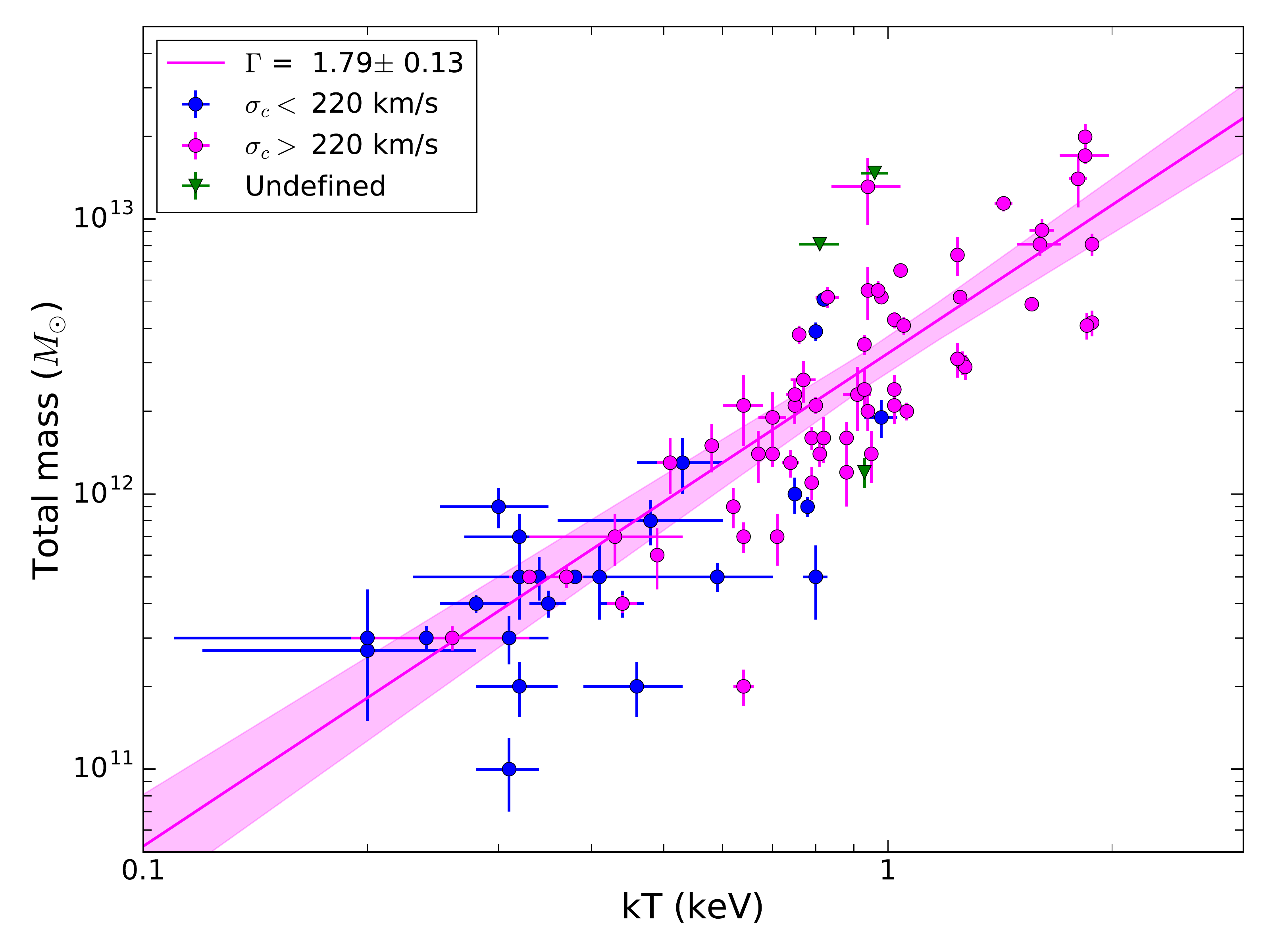}
\end{minipage}
\caption{The relations between X-ray luminosity (left) and total mass (right) with temperature, each derived within 5$r_e$. The solid black and dashed red lines indicate the best-fitting relation for the entire ETG sample using BCES and Kelly's regression methods. In the upper left plot the red dashed-dotted line indicates the self-similar $L_{X} \sim T^2$ relation, and the green and magenta dashed lines show the contributions to the total X-ray luminosity from LMXBs and other stellar sources. The shaded regions in all panels indicate the 1$\sigma$ confidence levels for the fitted scaling relations. The subsamples in each row are color-coded by morphological type (top), metallicity (middle), and $\sigma_c$-fast/slow (bottom). The power laws shown in magenta were BCES fit with the gas-rich (middle) and $\sigma_c$-fast (bottom) galaxies only.}
\label{fig_lt_mt}
\end{figure*}

We have investigated the $M-T$ relation of ETGs in a sample of 90 systems. The right-hand plots in Figure~\ref{fig_lt_mt} show the $M-T$ relation, with color coding the same as in Section~\ref{sec_LT}. The best-fit results are $M \propto T^{2.43\pm0.19}$ for BCES and $M \propto T^{2.43\pm0.25}$ for Kelly's, respectively, with an rms deviation of 0.52 dex. According to the self-similar model, the total mass should scale with temperature as $M \propto T^{3/2}$. Thus, this relation is also significantly steeper than is predicted by self-similarity. The main contributors to this steepening are galaxies with temperatures below 0.7 keV. Fitting only galaxies with $kT < 0.7$ keV we find a steeper relation, with $M \propto T^{3.2\pm0.4}$.

Most of the X-ray flux emerges from the centers of galaxies.  Therefore,  the $L_X-T$ scaling relation shows larger scatter compared to the $M-T$ relation, consistent with previous results for galaxy clusters \citep{Markevitch:98, Arnaud:99, Pratt:09, Maughan:12}. Thus, the total mass-temperature relation is less sensitive to the non-gravitational processes. 

\subsection{$L_X-M$}\label{sec_LM}

\begin{figure*}
\begin{minipage}{0.495\textwidth}
\includegraphics[width=1.0\textwidth]{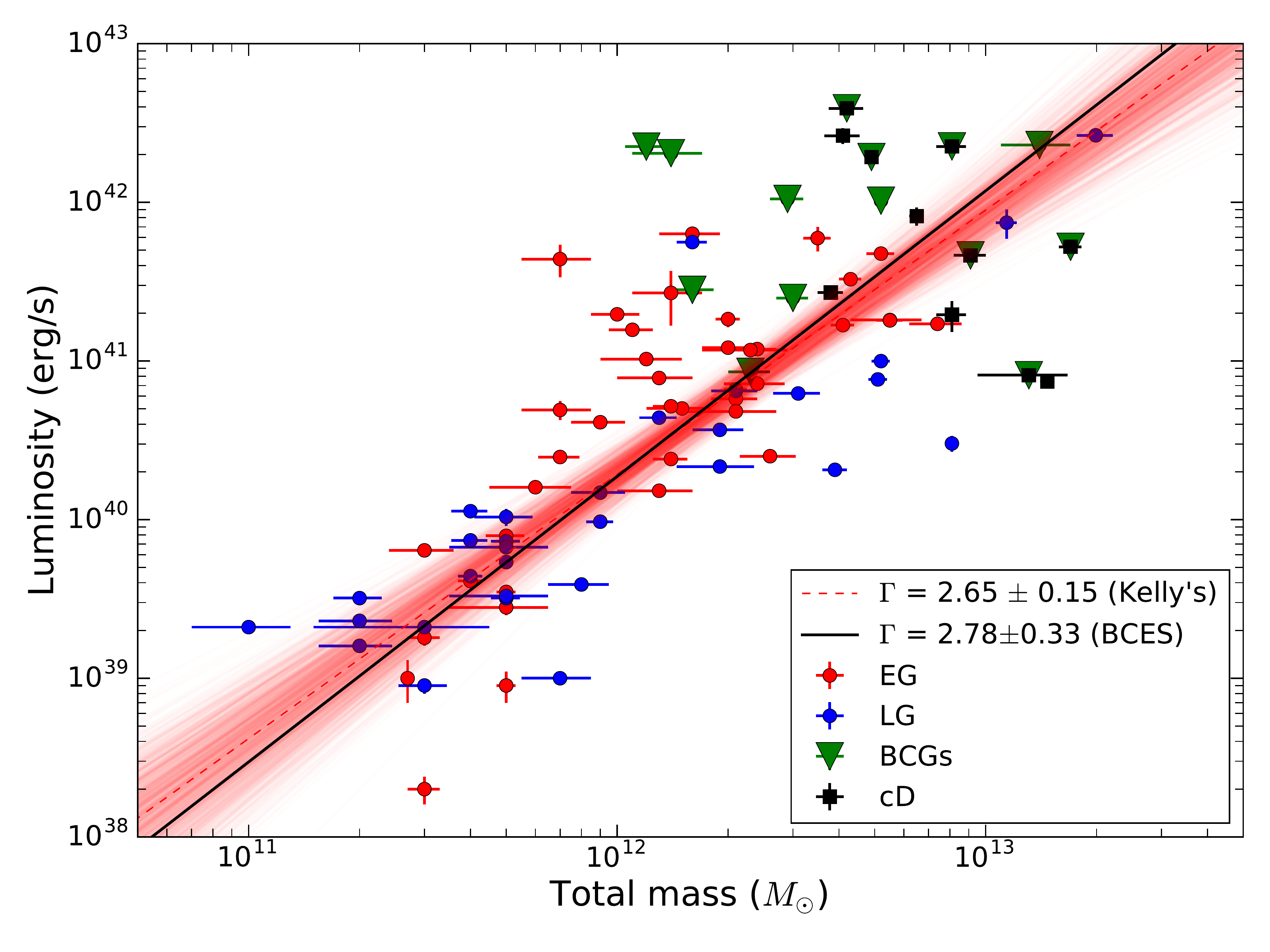}
\includegraphics[width=1.0\textwidth]{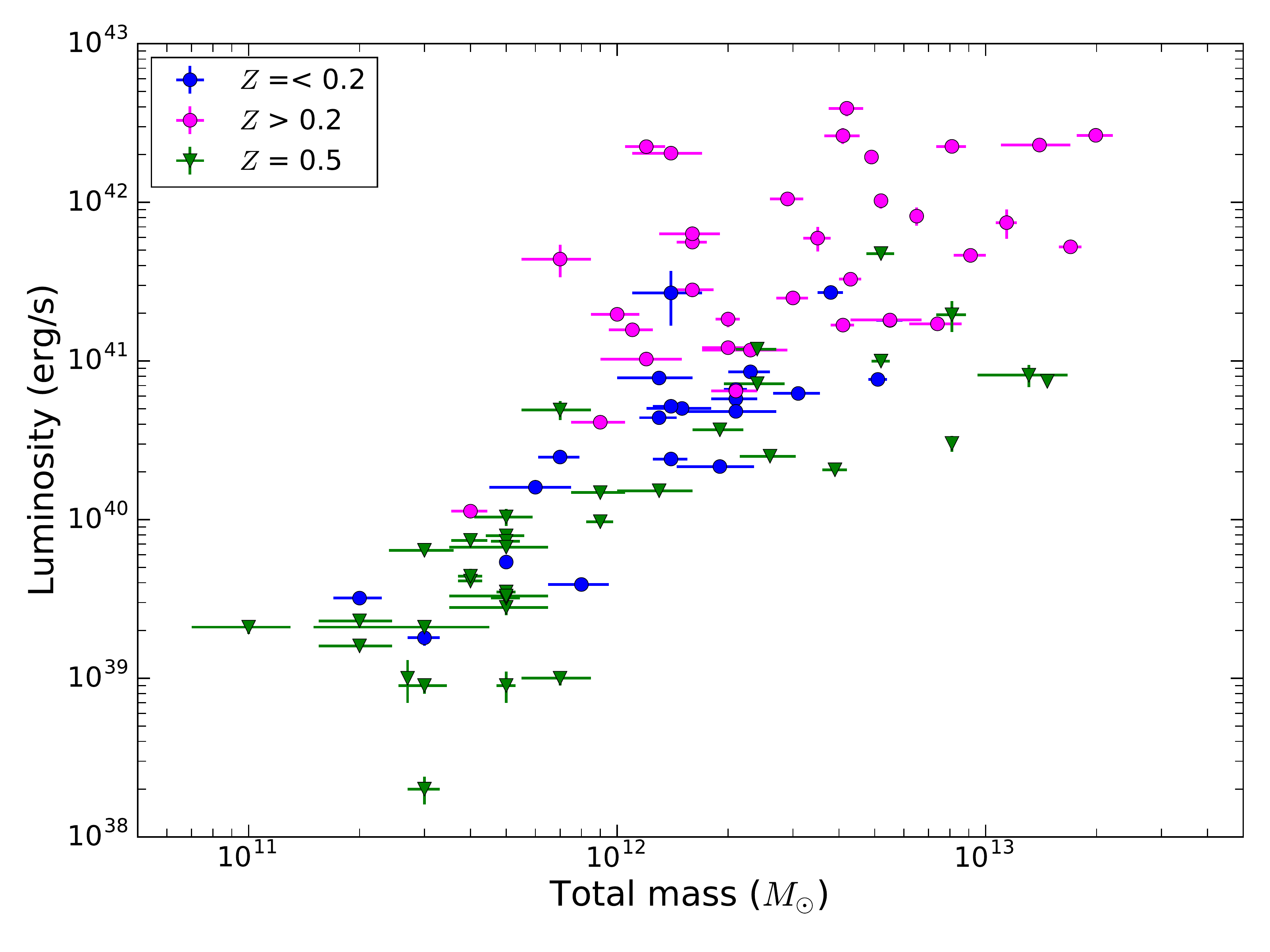}
\includegraphics[width=1.0\textwidth]{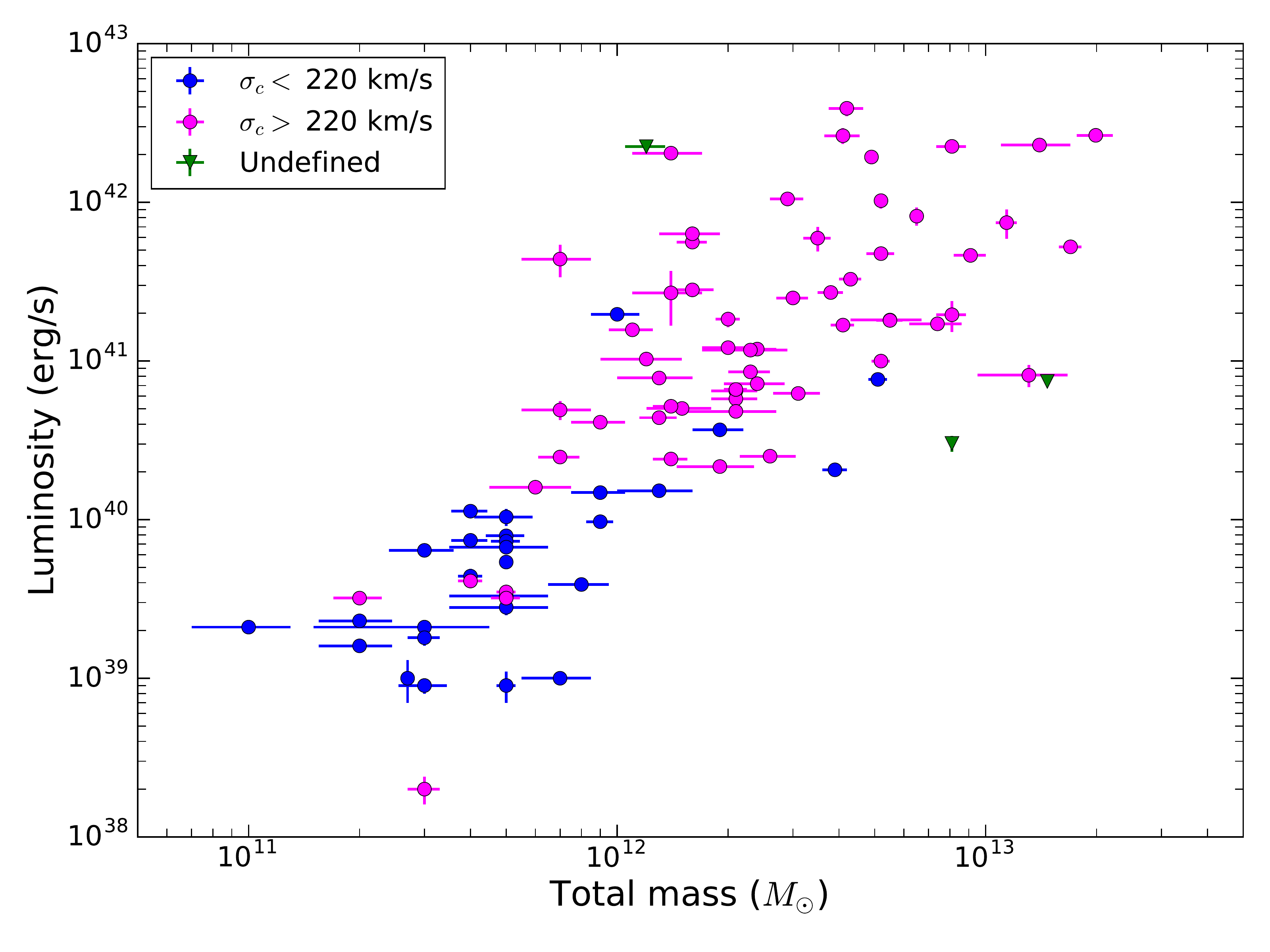}
\end{minipage}
\begin{minipage}{0.495\textwidth}
\includegraphics[width=1.0\textwidth]{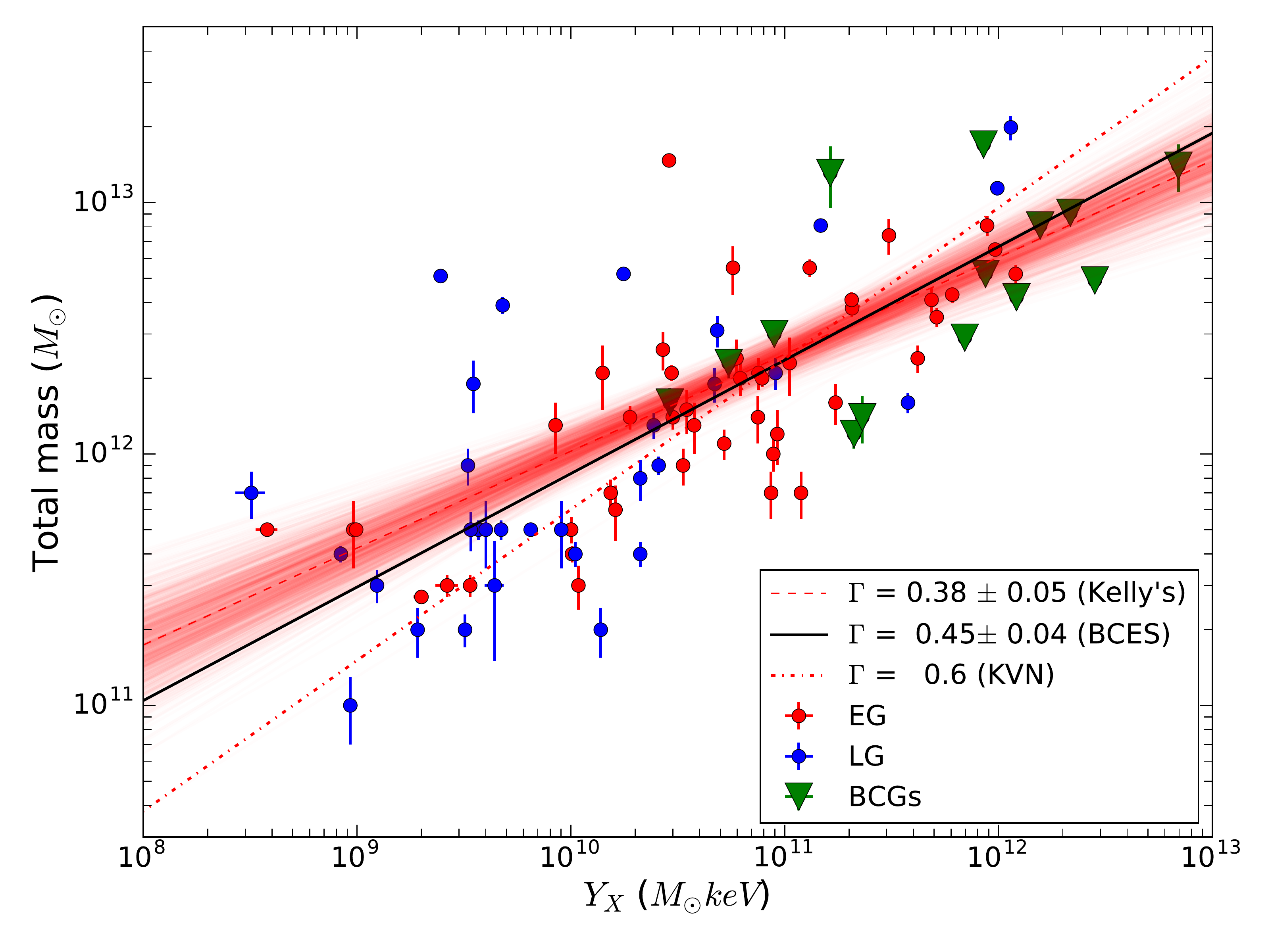}
\includegraphics[width=1.0\textwidth]{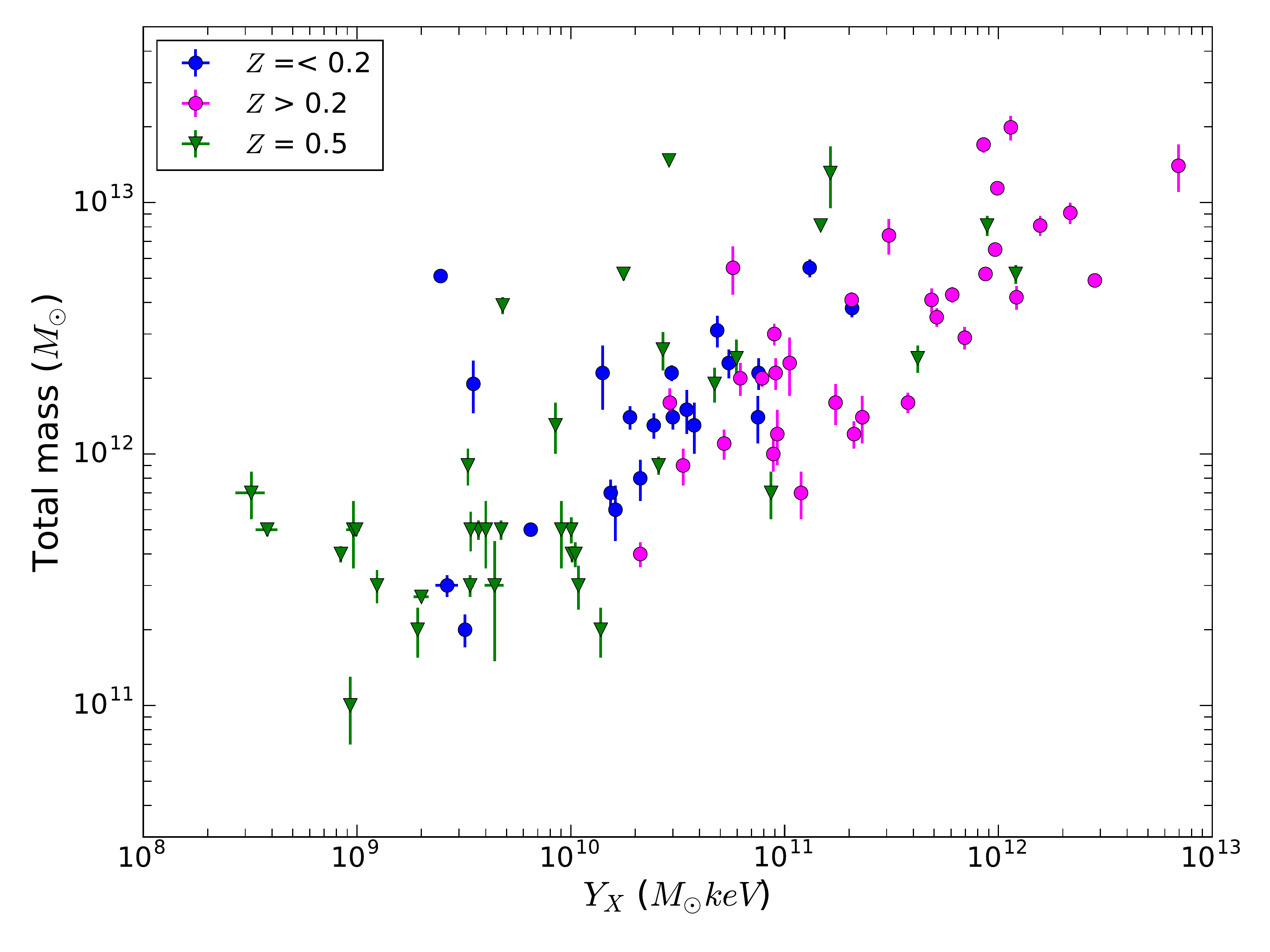}
\includegraphics[width=1.0\textwidth]{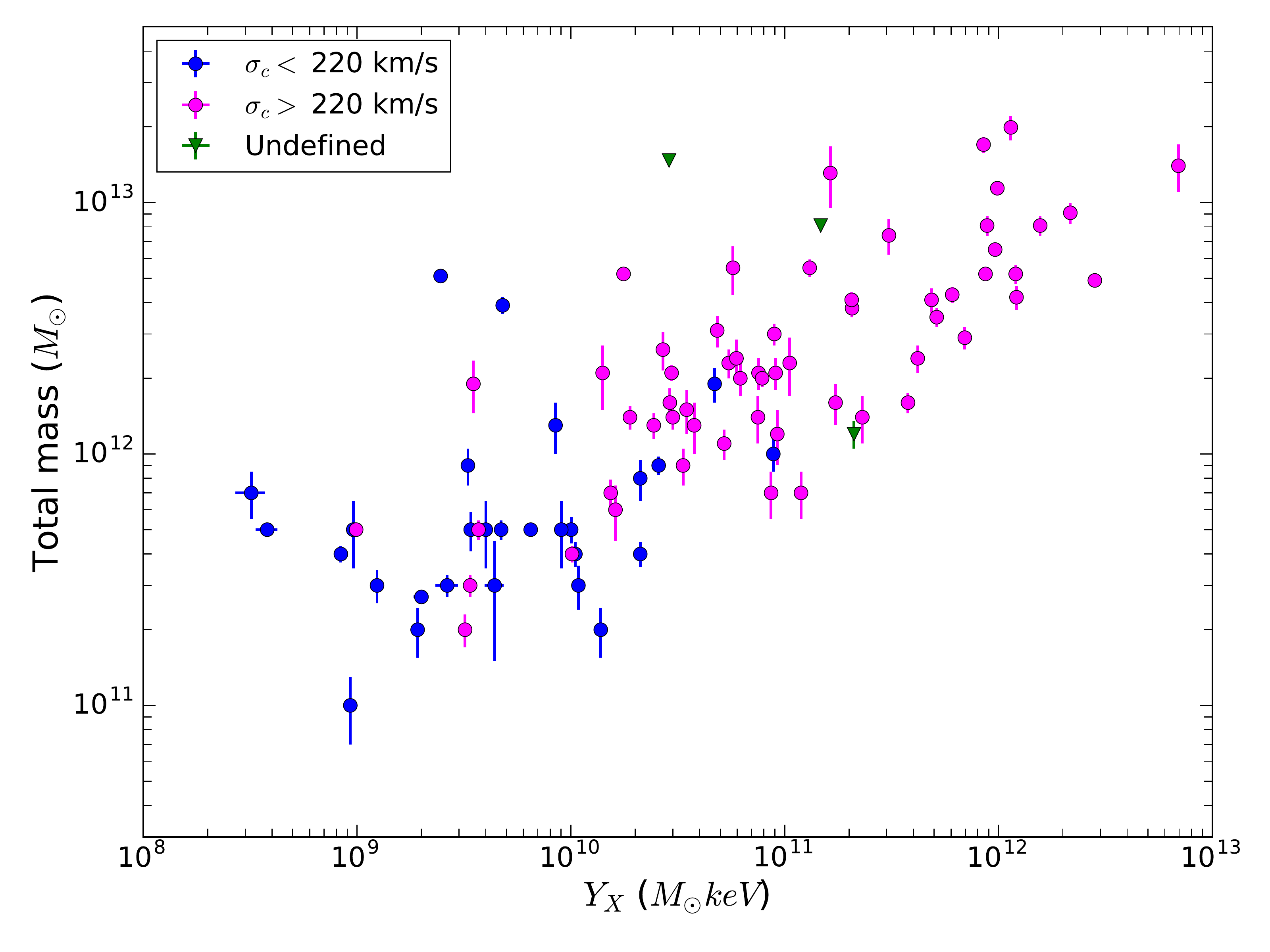}
\end{minipage}
\caption{The X-ray luminosity -- total mass (left) and total mass -- $Y_X$ (right) relations for the entire ETG sample. The solid black and dashed red lines indicate the best-fitting relation for the entire ETG sample using BCES and Kelly's regression methods. The red dash-dotted line shows the best-fitting result from \citet{Kravtsov:06} obtained for galaxy clusters. The color-coding in each row is the same as in Fig.~\ref{fig_lt_mt}.}
\label{fig_lm_ym}
\end{figure*}

The scaling relations between X-ray luminosity and total mass are shown on the left-hand side of Figure~\ref{fig_lm_ym}. For the entire sample we measured $L_X \propto M^{2.78\pm0.23}$ and $L_X \propto M^{2.65\pm0.15}$, for BCES and Kelly's methods respectively. Due to the high scatter in high-mass systems (mostly BCGs and cDs), the slope has relatively large uncertainty.

Furthermore, Malmquist bias is present in both the $L_X-T$ and $L_X-M$ scaling relations \citep{Stanek:06, Vikhlinin:09, Main:17}. This bias can be quantified by $\delta \ln L_X = 3/2\sigma^{2}_{i}$, where $\sigma_i$ is an intrinsic scatter in the log-normal value of luminosity for a given $T$. We correct for this bias to our $L_X-T$ and $L_X-M$ relations, modifying the normalization of our fit but not the power law slope involving $L_X$.

\subsection{$M-Y_X$}
The total mass -- $Y_X$ scaling relation was initially investigated in galaxy clusters by \citet{Kravtsov:06}. Here we explore for the first time this relation in ETGs. The $M-Y_X$ relation is shown on the right-hand side of Figure~\ref{fig_lm_ym}. We find a strong positive correlation with best-fitting power law scaling $M \propto Y_X^{0.45\pm0.04}$ (BCES) and rms deviation of 0.52 dex. Using Kelly's regression method we find a slightly shallower scaling relation, $M \propto Y_X^{0.38\pm0.05}$.  Both slopes are shallower than the slope of 3/5 measured in clusters alone \citep[][referred to as KVN]{Kravtsov:06}. The dashed red line in the top-right plot of Fig.~\ref{fig_lm_ym} shows the scaling measured by KVN for their sample of relaxed galaxy clusters.

\begin{table}
\centering
\caption{Comparison with previous X-ray scaling relation measurements.}\label{tab_comp_scal}
\begin{tabular}{cccccc}
\hline
Our results & Literature & Ref \\
\hline
 & $L_X-T$ & \\
\hline
 & $L_X \propto T^{4.8\pm0.7}$ & \citet{OSullivan_sample:03} \\
$L_X \propto T^{4.4\pm0.2}$ (BCES) & $L_X \propto T^{4.6\pm0.7}$ & \citet{Boroson:10} \\
$L_X \propto T^{4.5\pm0.4}$ (Kelly's) & $L_X \propto T^{4.5\pm0.3}$ & \citet{Kim:15} \\
 & $L_X \propto T^{4.7\pm0.4}$ & \citet{Goulding:16} \\
 \hline
 & $L_X-M$ & \\
\hline
$L \propto M^{2.8\pm0.3}$ (BCES) & $L_X \propto M^{2.7\pm0.3}$ & \citet{Kim:13} \\
$L \propto M^{2.7\pm0.2}$ (Kelly's) & $L_X \propto M^{3.13\pm0.32}$ & \citet{Forbes:16} \\
\hline
\end{tabular}

\end{table}

\subsection{Comparison with previous results}

In Table~\ref{tab_comp_scal} we compare our results to previously published X-ray scaling relations for ETGs \citep{OSullivan_sample:03, Boroson:10, Kim:13, Kim:15, Goulding:16, Forbes:16}. Our results are consistent with but have lower uncertainties than these studies. \citet{OSullivan_sample:03} performed their $L_X-T$ scaling relation analysis for a sample of cD galaxies. Later, \citet{Kim:15} found a significantly steeper $L_X-T$ relation ($L_X \propto T^{5.4\pm0.6}$) for their sample, which included cD galaxies as well. \citet{Boroson:10} found that the gas-poor ETGs follow the power law fit $L_X \propto T^{4.5\pm0.55}$. However, \citet{David:06} found no clear correlation between X-ray luminosity and temperature for a sample of 18 gas-poor ETGs. 

The X-ray $L_X-M$ relation of 14 ETGs has been investigated by \citet{Kim:13}. They found the scatter in gas-poor objects ($L_X \lessapprox 3\times 10^{39}$~erg/s) to be larger than in gas-rich galaxies. \citet{Forbes:16} presented a strong correlation between X-ray luminosity and galaxy dynamical mass within 5$r_e$ for a sample of 29 massive ETGs obtained using the SLUGGS survey. 

%{\bf Although we obtain smaller uncertainties on the best-fit scaling relation slopes of different scaling relations, we do not adopt regression methods used in these previous studies. Thus, we conclude that care should be taken when comparing regression coefficients from different studies.}

\section{Discussion}\label{sec_disc}

Here we use X-ray scaling relations to explore the structural and dynamical properties of ETGs. We first subdivided our sample on metallicity, stellar velocity dispersion, and X-ray core size. The stellar velocity dispersion threshold was chosen to be 220 km/s for consistency with previous results. For metallicity and X-ray core size we have built histograms of metallicity and core radius. The peaks were chosen as our thresholds. Due to the similarity of best-fitting results for scaling relations analyzed in the previous section with BCES and Kelly's methods, further fitting has been performed with the BCES method only. The best-fitting relations for each subsample are shown in Table~\ref{tab_res_subs}. We found no clear correlations in the $M - Y_X$ scaling relation for the subdivided samples.

We find that more massive galaxies are characterized by a higher metallicity, higher central velocity dispersions, and larger X-ray cores. We also explore the gas-to-total mass fraction and the $L_X-\sigma_c$ relation. Finally, we explore whether AGN feedback causes the X-ray scaling relations to deviate from self-similarity.

\begin{table*}
\centering
\caption{Scaling relations of the form log($y$) = $a$ + $b$ log($x$). Luminosities are expressed in units of 10$^{40}$~erg/s and masses in 10$^{12}M_{\odot}$.}\label{tab_res_subs}
\begin{tabular}{llcccccc}
\hline
Sample & $N$ &  $C_{cor}$ & $a$ & $b$ & $p$-Pearson & $p$-Spearman & rms scatter \\
\hline
       &  & &   & $L_X-T$  \\
\hline
X-ray core & 42 & 0.84 & 56.08$\pm$22.27 & 4.46$\pm$0.23 & $>>$0.0001 & $>>$0.0001 & 0.90 \\
X-ray coreless & 35 & 0.83 & 8.93$\pm$3.82 & 2.87$\pm$0.55 & $>>$0.0001 & $>>$0.0001 & 0.92 \\
$Z$-rich & 33 & 0.75 & 76.82$\pm$36.46 &  4.50$\pm$0.24 & $>>$0.0001 & $>>$0.0001 & 0.59 \\
$Z$-poor & 21 & 0.64 & 8.57$\pm$4.59 & 1.68$\pm$0.44 & 0.0016 & 0.0033 & 0.59 \\
$\sigma_c$-fast & 59 & 0.82 & 35.42$\pm$11.07 & 4.02$\pm$0.21 & $>>$0.0001 & $>>$0.0001 & 0.83 \\
$\sigma_c$-slow & 28 & 0.70 & 3.42$\pm$2.06 & 1.79$\pm$0.89 & $>>$0.0001 & 0.00025 & 0.57 \\
\hline
 &&&& $M-T$ &&\\
\hline
X-ray core & 42 & 0.78 & 9.52$\pm$1.50 & 1.86$\pm$0.12 & $>>$0.0001 & $>>$0.0001 & 0.43 \\
X-ray coreless & 35 & 0.85 & 2.52$\pm$0.36 & 1.99$\pm$0.82 & $>>$0.0001 & $>>$0.0001 & 0.47 \\
$Z$-rich & 33 & 0.86 & 3.05$\pm$0.81 & 1.86$\pm$0.20 & $>>$0.0001 & $>>$0.0001 & 0.41 \\
$Z$-poor & 21 & 0.69 & 3.34$\pm$1.25 & 2.64$\pm$0.40 & 0.0005 & $>>$0.0001 & 0.36 \\
$\sigma_c$-fast & 59 & 0.83 & 3.28$\pm$0.44 & 1.79$\pm$0.13 & $>>$0.0001 & $>>$0.0001 & 0.43 \\
$\sigma_c$-slow & 28 & 0.68 & 1.97$\pm$0.53 & 1.61$\pm$0.43 & $>>$0.0001 & $>>$0.0001 & 0.36 \\
\hline
&&&&$L_X-M$ &&\\
\hline
X-ray core & 42 & 0.90 & 5.37$\pm$0.17 & 0.06$\pm$0.33 & $>>$0.0001 & 0.0005 & 0.92 \\
X-ray coreless & 35 & 0.86 & 3.28$\pm$0.13 & 1.63$\pm$0.46 & $>>$0.0001 & $>>$0.0001 & 0.92 \\
$Z$-rich & 33 & 0.53 & 3.53$\pm$0.13 & 0.43$\pm$0.42 & 0.0014 & 0.047 & 0.59 \\
$Z$-poor & 21 & 0.83 & 2.05$\pm$0.77 & 1.14$\pm$0.45 & $>>$0.0001 & $>>$0.0001 & 0.59 \\
$\sigma_c$-fast & 59 & 0.72 & 4.49$\pm$1.15 & 0.78$\pm$0.30 & $>>$0.0001 & $>>$0.0001 & 0.83 \\
$\sigma_c$-slow & 28 & 0.70 & 1.04$\pm$0.64 & 1.03$\pm$0.65 & $>>$0.0001 & $>>$0.0001 & 0.56 \\
\hline
% &&&&$M-Y_X$ &&\\
% \hline
%Core & 42 & 0.63 & 0.14$\pm$0.03 & 0.00$\pm$0.19 & 2.17$\times$10$^{-6}$ & 2.63$\times$10$^{-5}$ & 0.43 \\
%Cusp & 35 & 0.83 & 8.33$\pm$1.95 & 0.32$\pm$0.23 & 1.57$\times$10$^{-10}$ & 2.90$\times$10$^{-8}$ & 0.47 \\
%$Z$-rich & 33 & 0.67 & 0.002$\pm$0.0005 & 0.24$\pm$0.57 & 1.72$\times$10$^{-5}$ & 6.70$\times$10$^{-5}$ & 0.33 \\
%$Z$-poor & 21 & 0.70 & 0.22$\pm$0.09 & 0.37$\pm$0.48 & 3.21$\times$10$^{-5}$ & 0.0019 & 0.35 \\
%Fast & 59 & 0.76 & 0.26$\pm$0.05 & 0.17$\pm$0.09 & 2.87$\times$10$^{-12}$ & 3.11$\times$10$^{-10}$ & 0.36 \\
%Slow & 28 & 0.41 & 0.005$\pm$0.002 & 0.33$\pm$0.17 & 0.03 & 0.016 & 0.27 \\ 
\end{tabular}
\end{table*}

% --- rich/poor on metallicity description
\subsection{Metallicity in ETGs}

Atmospheric metallicity is sensitive to several physical processes, including star formation history and outflows. The stars in massive galaxies are metal-rich compared to lower mass galaxies. Due to their shallower gravitational potential wells, lower mass galaxies are more easily stripped of enriched gas \citep{Faber:73, Babyk:13}. 

Using the metallicities measured from spectral fitting in Table~\ref{tab2}, we subdivided our sample into metal-rich and metal-poor systems. A metallicity of $0.2~Z_{\odot}$ divided the sample. The middle panels in Figures.~\ref{fig_lt_mt} and \ref{fig_lm_ym} show the metal-rich and metal-poor subsamples in each scaling relation. We find a strong, steep $L_X-T$ correlation for metal-rich ($Z > 0.2~Z_{\odot}$) galaxies with an rms deviation of 0.59. Metal-poor galaxies ($Z < 0.2~Z_{\odot}$), on the other hand, show a weak $L_X-T$ relation that is shallower than the metal-rich $L_X-T$ relation. The $M-T$ relation is strongly correlated for metal-rich galaxies, while the $L_X-M$ and $M-Y_X$ relations have significantly higher scatter.

%An increasing of metallicity of ETGs with luminosity and total mass as well was firstly claimed by \citet{Faber:73}. Using $(U-V)-m_V$ color-magnitude relation of early-type galaxies, \citet{Faber:73} concluded that due to weaker potential
%Note that green points show objects were metallicity was fixed at 0.5 value since during spectral fitting we got too low value of metallicity (usually 0.01 or lower). Thus, assuming that green points have about the same low value of metallicity as blue points (below 0.2$Z_{\odot}$) we are able to see a clear distribution of metal-poor and metal-rich galaxies in the bottom left and upper right corners respectively. 

Galaxies with temperatures lying between 0.5 and 1.0 keV are usually degenerate in metallicity, temperature and normalization of the thermal model \citep{Werner:12}. \citet{Werner:12} found that overestimating metallicity by a factor of two will underestimate the normalization by a factor of $\sim1.35$. To explore the effect on our scaling relations, we fit the most metal-rich and metal-poor objects with a fixed metallicity that varied between 0.1 and $1.0~Z_{\odot}$ in steps of $0.1~Z_{\odot}$. We found that objects with temperatures $\lesssim$1.2 keV were not affected by metallicity variations. The sources with temperature $\gtrsim$1.2 keV correlate with metallicity. 
%{\bf Put the objects tested at multiple metallicities into a Table.}
%For example, we got $T$ of 0.30 keV at fixed 0.1$Z_{\odot}$ and 0.33 keV at fixed 1.0$Z_{\odot}$ for NGC1386, 1.00 and 1.04 for NGC533, 0.30 and 0.36 for NGC2768, whilst we got 1.37 keV at 0.1$Z_{\odot}$ and 1.70 at 1.0$Z_{\odot}$ for IC1262, 1.02 and 1.43 for NGC1399, 0.96 and 1.55 for NGC1550. 
A mis-estimate of metallicity of this size has a relatively small impact on X-ray luminosity. 

%\citet{Su:13} have obtained higher values of metallicity for the faint ETGs using multi-component model with two-temperature models, i.e., {\sc phabs*(vapec+vapec+po+mekal+po)}. They got higher metallicities with bigger uncertainties as well. So high statistical uncertainties on the metallicity come from the spectra with low number of counts. They also concluded that the low count rate do not bias calculation of the hot gas metallicity. In contrast, \citet{Werner:12} argued that in the case of too low count rate, the multi-temperature model fits are infeasible. 

\subsection{$\sigma_c$-fast/slow galaxies}
% ---- slow/fast description
We investigated the gross dynamical properties of ETGs using their central velocity dispersions. We divided our sample into $\sigma_c$-fast and slow objects using a threshold of $\sigma_c = 220$~$\rm km~s^{-1}$, following \citet{Kim:15}. These subsamples are highlighted in the bottom panels of Figures~\ref{fig_lt_mt} and \ref{fig_lm_ym}. The $L_X-T$ and $M-T$ relations both show strong correlations for $\sigma_c$-fast galaxies.  Galaxies with higher central velocity dispersions are hotter, more luminous, and are more massive (in both total and gas mass) than those with lower velocities.

\begin{figure*}
\begin{minipage}{0.495\textwidth}
\includegraphics[width=1.0\textwidth]{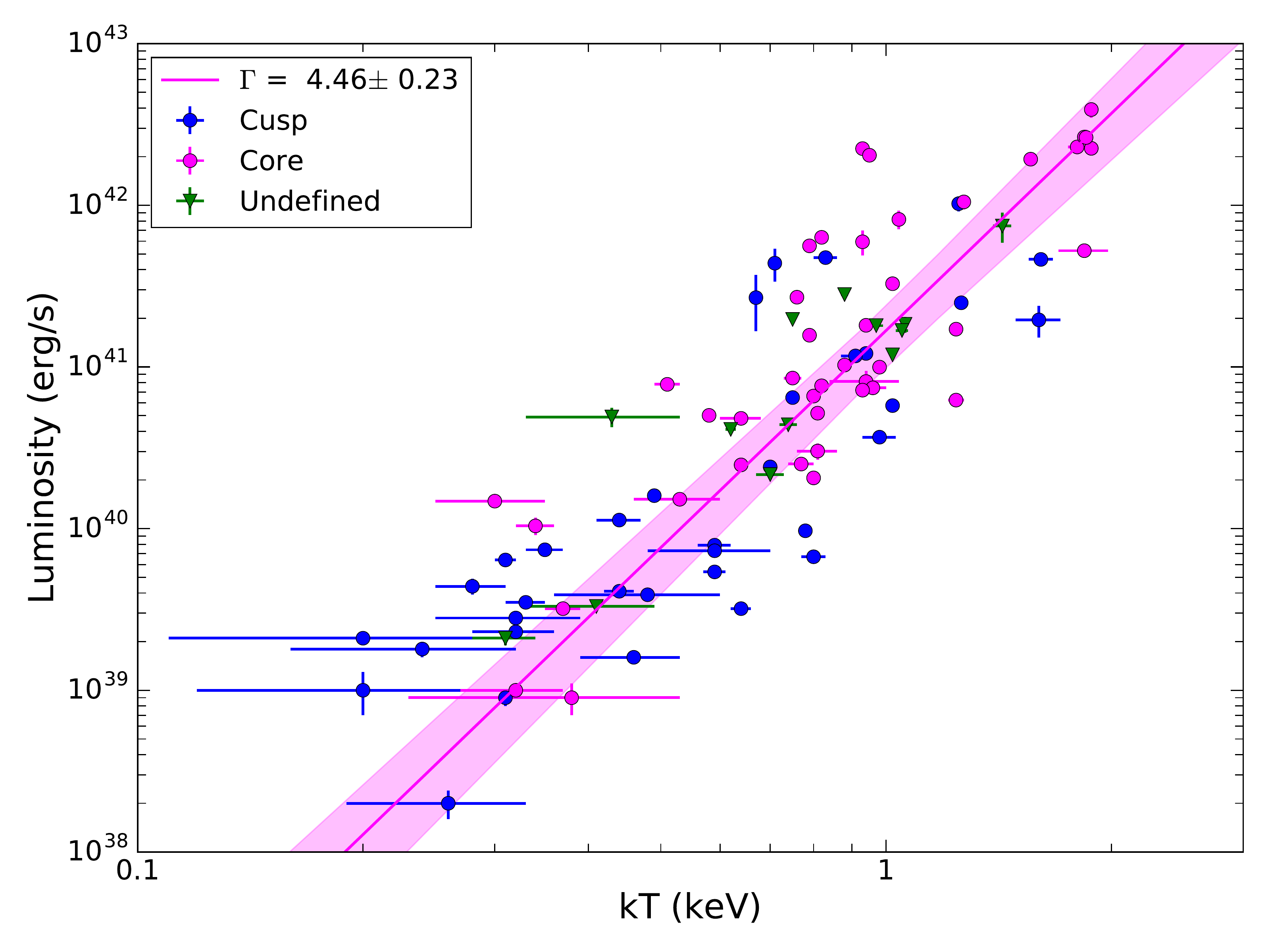}
\includegraphics[width=1.0\textwidth]{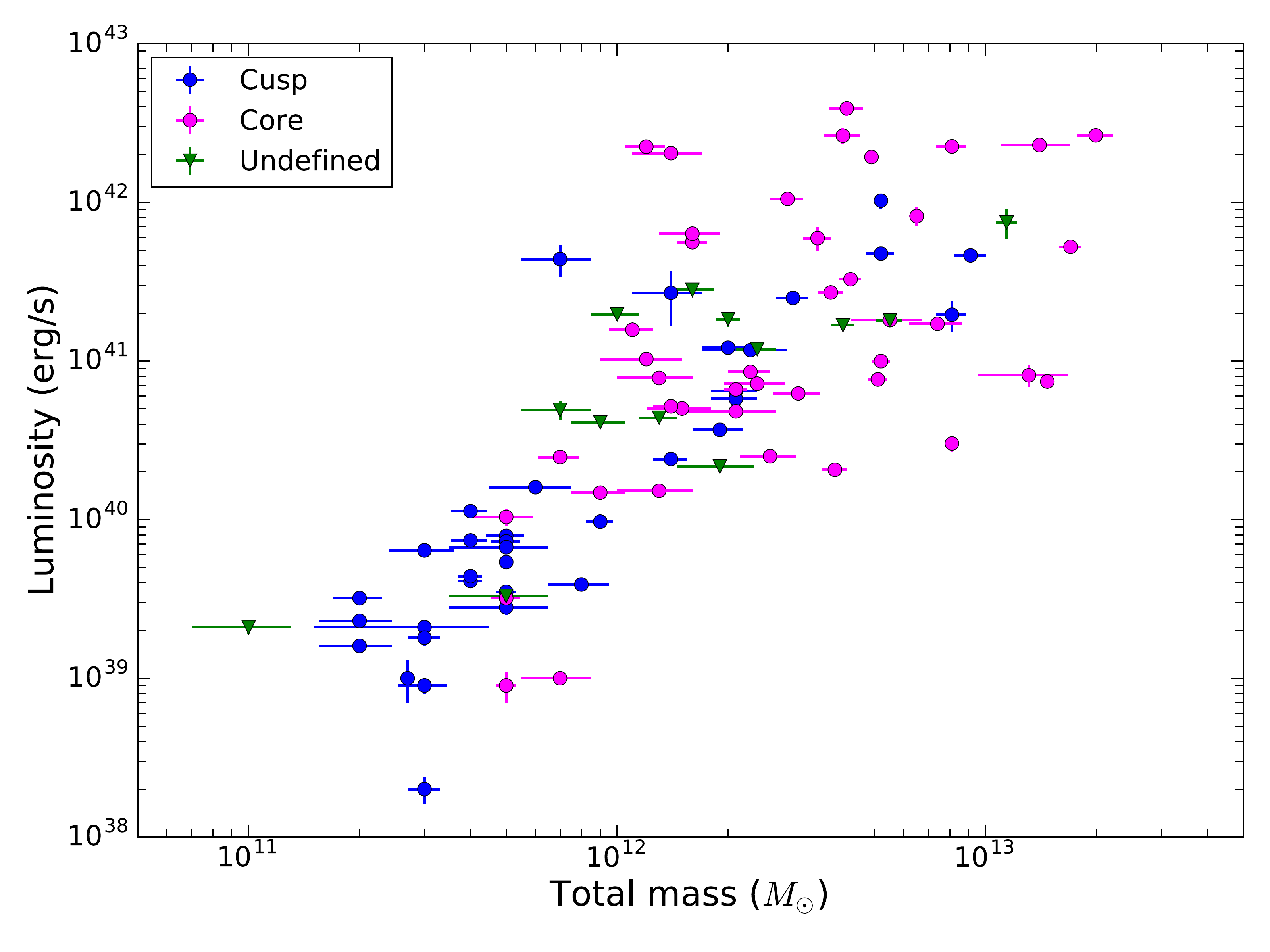}
\end{minipage}
\begin{minipage}{0.495\textwidth}
\includegraphics[width=1.0\textwidth]{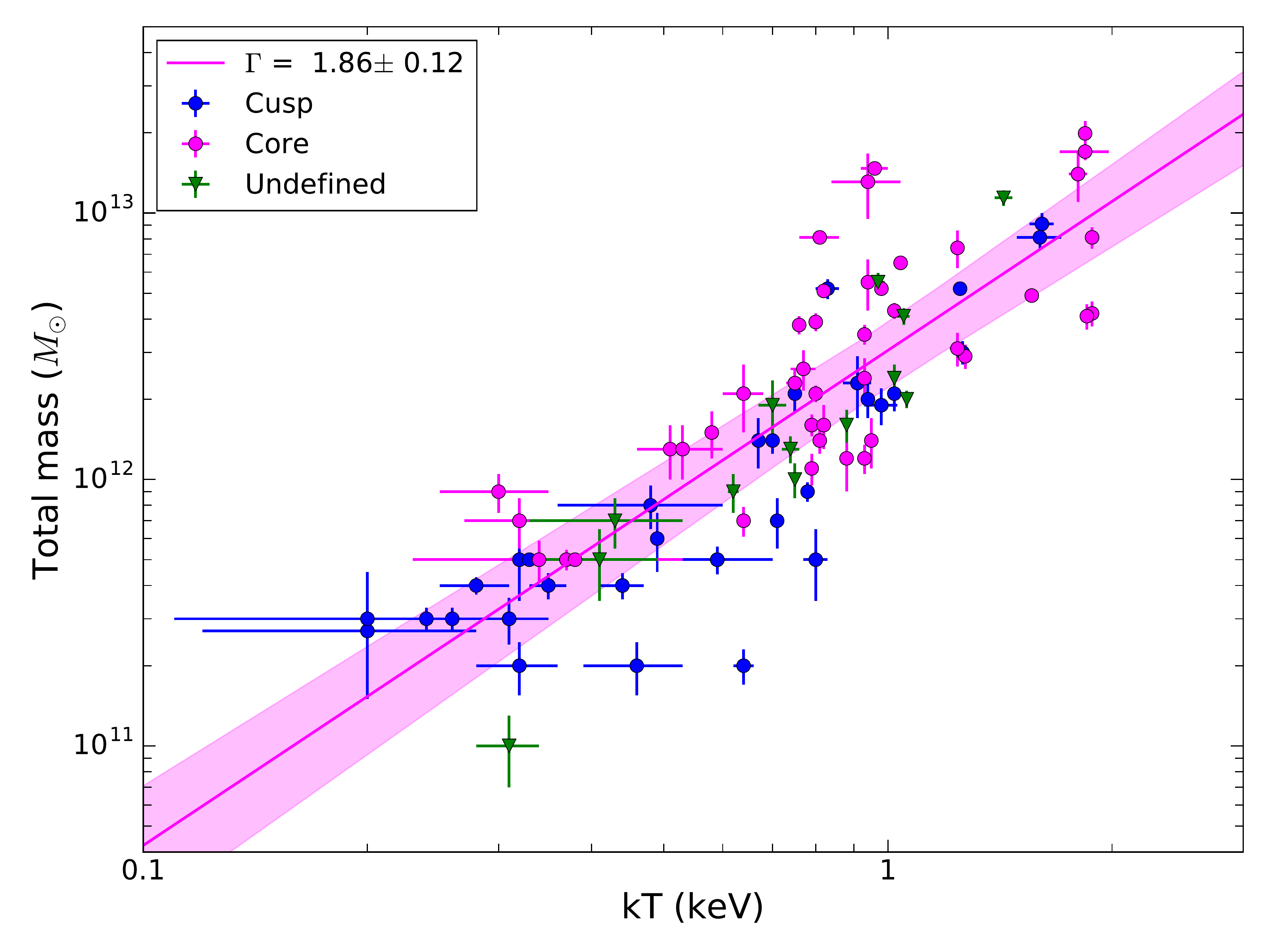}
\includegraphics[width=1.0\textwidth]{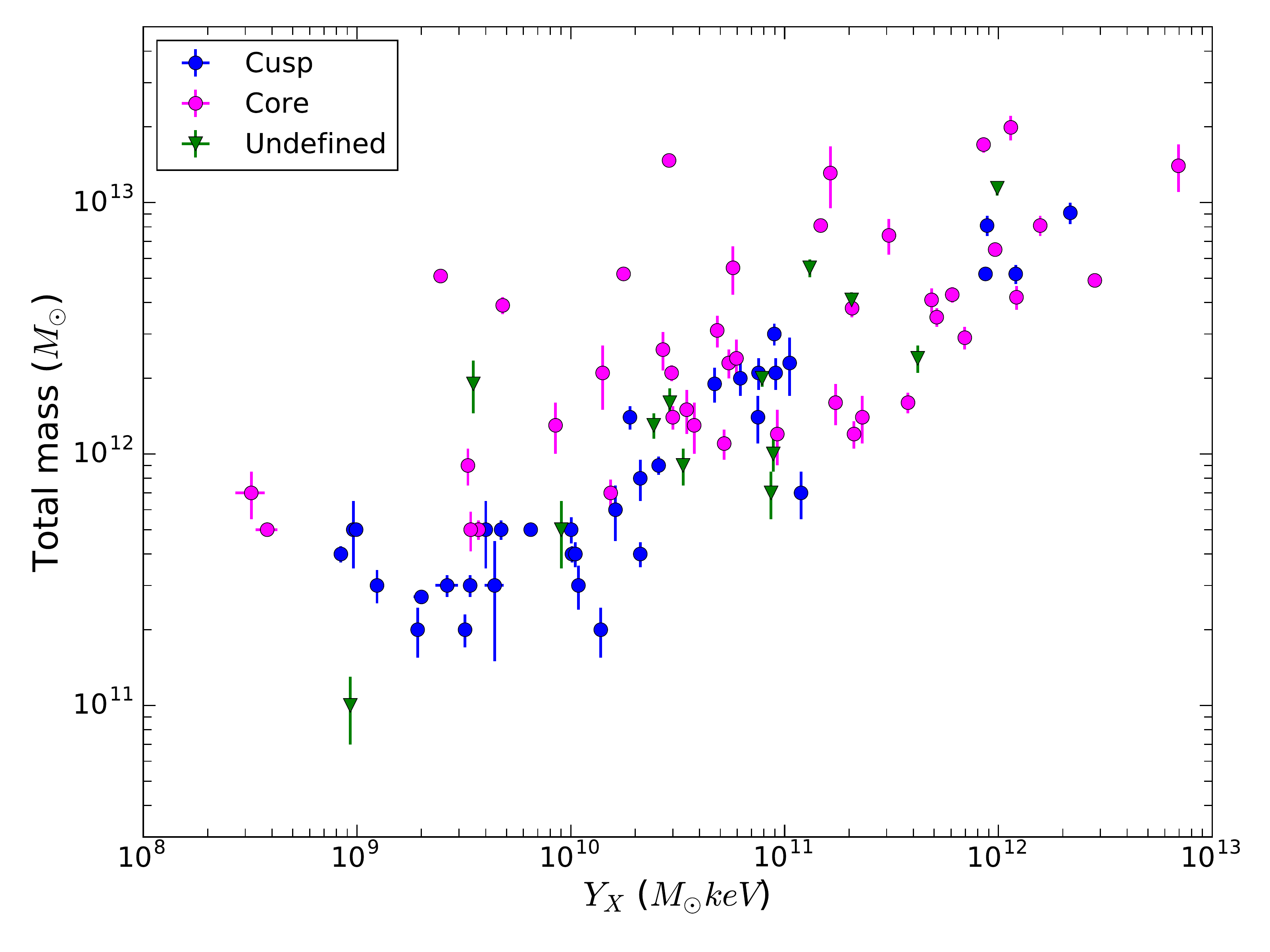}
\end{minipage}
\caption{The X-ray scaling relations for galaxies with an X-ray core/cusp. The fits in magenta were obtained from X-ray core galaxies only.}
\label{fig_core}
\end{figure*}

%---- core/cusp description
\subsection{X-ray core/coreless galaxies}

To explore the relationship between stellar cores and atmospheric cores we subdivide our sample using the core radius $r_c$ obtained from the $\beta$-model fit to the X-ray surface brightness profiles. A significant X-ray core corresponds to systems with $r_c > 0.5$~kpc, while coreless galaxies have $r_c < 0.5$~kpc.  All four scaling relations, color-coded by the presence or not of an X-ray core, are shown in Figure~\ref{fig_core}. The power law fits to the subsamples are given in Table~\ref{tab_res_subs}. No clear separation between X-ray core and coreless galaxies in temperature, luminosity, and mass is found. 
% However, we find that a big fraction of} galaxies with X-ray cores are hotter, more luminous, and more massive than the coreless galaxies. Moreover, the presence of an X-ray core correlates with the presence of optically-defined cores (\citealt{Krajnovic:13, Lauer:07, Hopkins:09, Kormendy:09}). This indicates that the properties of the hot gas are coupled to the stellar content of the galaxy.

\subsection{The $M_g-M$ relation}

\begin{figure}
\includegraphics[width=0.49\textwidth]{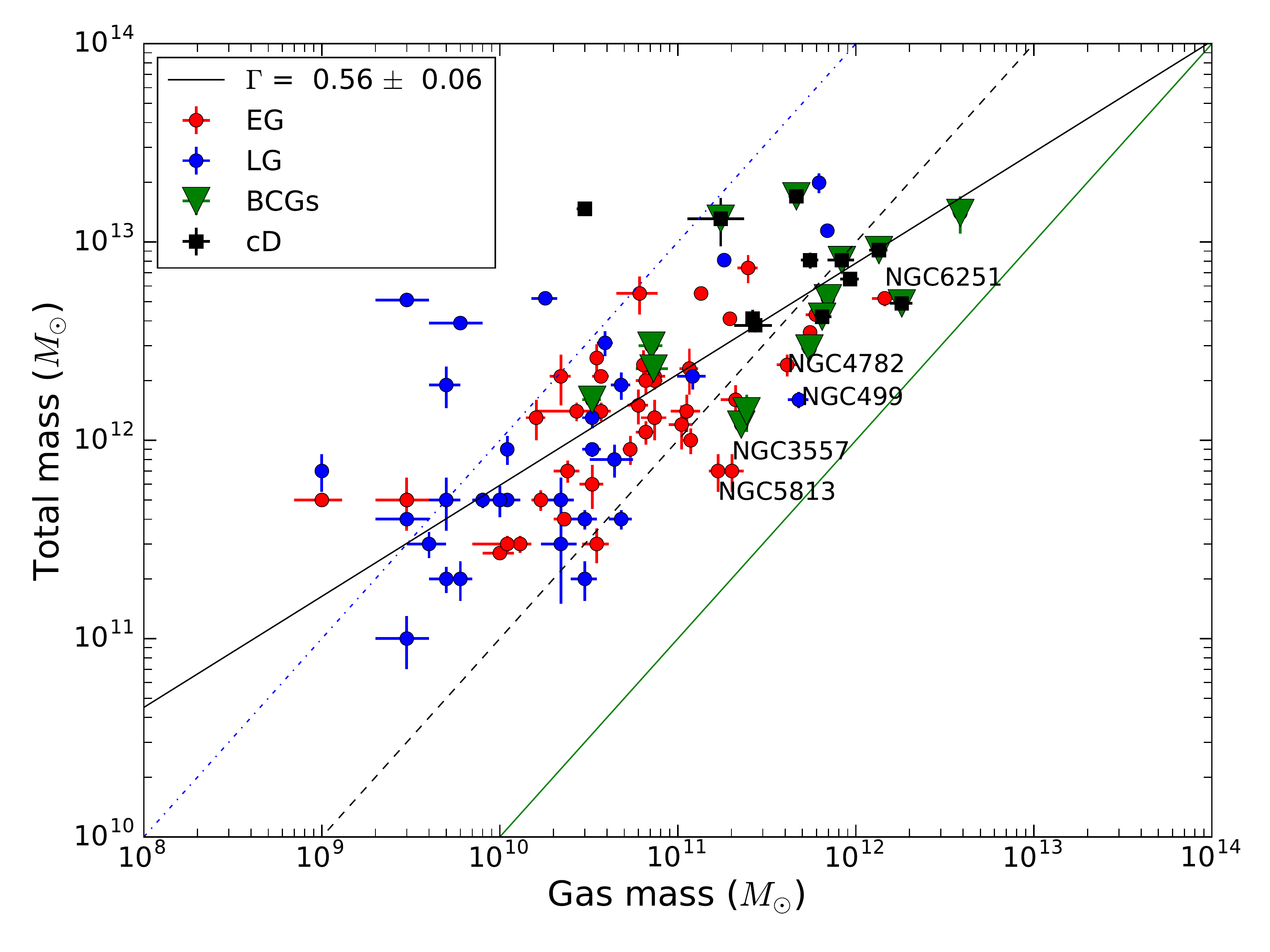}
\caption{The relation between total mass and gas mass, both derived within 5$r_e$. The solid, dashed, and dash-dotted lines correspond to gas fractions of 1, 0.1, and 0.01, respectively.}
\label{fig_mm}
\end{figure}

The relationship between total mass and atmospheric mass is shown in Figure~\ref{fig_mm}. The dash-dotted, dashed and solid lines correspond to constant gas-to-total mass fractions of 0.01, 0.1, and 1. We observe a weak correlation, $M \propto M_g^{0.56\pm0.06}$, spanning 4 decades in atmospheric mass and $\sim$2 decades in total mass.  The total masses of BCGs and cD galaxies are similar at roughly 10$^{12}~M_{\odot}$. The gas fraction in lenticular galaxies is lower than that in elliptical galaxies.  The gas fractions in elliptical and lenticular galaxies lie below $\sim 0.1$. Only 5 ETGs (those labeled in Figure~\ref{fig_mm}), apart from BCGs and cDs, have gas fractions above $0.1$. 

\subsection{The $L_X-\sigma_c$ relation}

\begin{figure}
\includegraphics[width=0.49\textwidth]{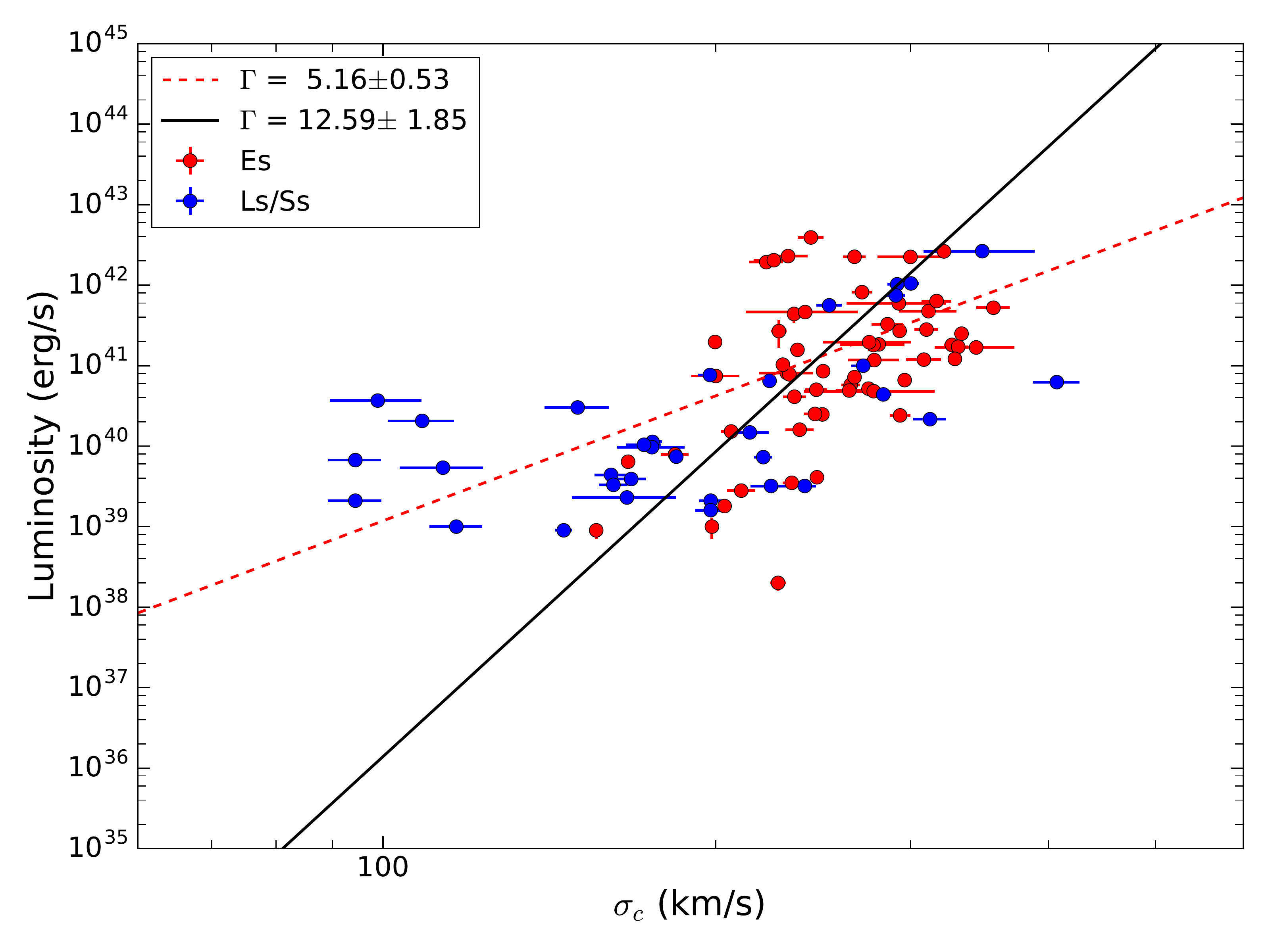}
\caption{The relation between X-ray luminosity and central velocity dispersion.}
\label{fig_lx_vel}
\end{figure}

The X-ray luminosity of a virialized, self-similar atmosphere should scale with stellar velocity dispersion as $L_X \propto \sigma_c^4$. Recent observations of galaxy clusters give steeper slopes, e.g., $L_X \propto \sigma_c^{5.2\pm0.3}$ (\citealt{Wu:99}).  Steep $L_X-\sigma_c$ relations for elliptical galaxies $L_X \propto \sigma_c^{8-11}$, have been found (\citealt{Diehl:07, Goulding:16}). We find $L_X \propto \sigma_c^{5.16\pm0.53}$ for the entire sample, Figure~\ref{fig_lx_vel}.  Including ellipticals only steepens the relationship to $L_X \propto \sigma_c^{12.6\pm1.9}$, consistent with observational results of \citet{Goulding:16} and cosmological simulations of \citet{Dave:02}. 

\subsection{Impact of AGN feedback}\label{subsec_feedback}

The slopes of the X-ray scaling relations are steeper than self-similar scaling. Departures from self-similarity indicate extra thermal processes acting on the atmospheres of early-type galaxies \citep{Giodini:13}. The effects increase significantly from clusters to the lower mass atmospheres of groups and galaxies.  Heat sources capable of driving the scaling away from self-similarity include supernovae feedback, stellar mass loss, thermal conduction, cosmic rays, and AGN feedback.  Each of these contributes at a level that varies with the mass of the system.  However, the impact of any individual process on a hot atmosphere is poorly known. AGN feedback is the largest heat source that is prevalent across all masses \citep{Anderson:14, Main:17, Pellegrini:12, McNamara:07, McNamara:12}. The departures from self-similarity of the X-ray scaling relations of clusters, groups, and galaxies can be largely attributed to AGN feedback, which we discuss briefly below. 

\citet{Puchwein:08} performed hydrodynamic simulations that included radiative cooling, star formation, supernova feedback, and heating by the ultraviolet background. Their simulations have been applied to the systems with and without black hole growth \citep{Sijacki:07}. Without AGN heating, the X-ray luminosities of poor clusters and galaxies with kT$\leq$2-3 keV are overestimated. This result agrees with earlier studies (e.g. \citealt{Borgani:04, Khos:04, Sijacki:06, Nagai:07, Schaye:10, Gaspari:12, Gaspari:14, Anderson:14, Anderson:15}). Their simulated $L_X - T$ scaling relation also agrees with our observations of massive clusters and groups.  The models imply that AGN feedback removes hot gas from the cores of poor clusters and groups, suppressing the X-ray luminosity and thus steepening the $L_X - T$ relation.  The $L_X - T$ scaling relation obtained by \citet{Puchwein:08} agrees with observations over a wide range of mass scales, from massive clusters to small groups. 

%This conclusion is in line with our results. Our sampled cD, BCGs, and groups exhibit shallower slopes than the less massive galaxies. However, these simulations have were performed under specific conditions. They excluded very cold gas particles (3$\times$10$^4$ K), which have high density. Furthermore, their simulations have only been used to check the $L_X-T$ scaling relation and avoid $L_X-M$/$M-T$ relations. 

The influence of AGN feedback on the $L_X - T$ scaling relation of ellipticals has been recently studied by \citet{Gaspari:12, Gaspari:14}.  They used two feedback prescriptions: a quasar thermal blast and self-regulated kinetic feedback. Their hydrodynamic simulations include stellar evolution and cooling over the life of the galaxy. They successfully reproduced observed properties, including buoyant cavities, subsonic turbulence, and nuclear cold gas. They argued that AGN feedback in isolated galaxies should be both less efficient and powerful compared to ellipticals because the latter are influenced by gas in the intergalactic medium. They also concluded that both AGN feedback models describe well the observational $L_X - T$ scaling relation. However, the decreasing  X-ray luminosity below $\sim$ 0.5 keV in the model is not evident in our scaling relations. In other words, we do not observe a break in the $L_X - T$ relation. We observe a monotonically decreasing X-ray luminosity in the $L_X - T$ and $L_X - M$ relations toward the lower temperatures ($\sim$ 0.2 keV). This feature is reproduced by a gentle, self-regulated kinetic mechanism (\citealt{Anderson:14, Anderson:15}). Recently, \citet{Negri:14} predicted higher temperatures in low luminosity systems than  we and \citet{Kim:15} observe.

The effects of AGN winds and radiation on the temperatures and X-ray luminosities of elliptical galaxies were modeled using high-resolution, 1D hydrodynamical simulations (\citealt{Pellegrini:12}). Their simulations produce the large variations in $L_X$  observed in our sample. In addition, \citet{Choi:15} performed a set of particle hydrodynamic simulations, adding a pressure-entropy formulation to improve fluid mixing and the treatment of contact gaps. These simulations have been applied to 20 haloes with AGN feedback models that included no feedback, thermal feedback, and radiation and mechanical feedback. The feedback models successfully reproduced the $M_{BH} - \sigma$ relation.  They found that X-ray luminosity is determined primarily by galaxy mass, consistent with our $L_X - M$ scaling relation. 

\citet{Dave:02} examined the scaling of temperature, X-ray luminosity, and galaxy velocity dispersion for a sample of galaxy groups. They used a $\Lambda$CDM simulation that included prescriptions for gas dynamics, star formation, radiative cooling, and gravity. In agreement with our results, their $L_X - \sigma$ and $L_X - T$ relations steepen below kT$\approx$0.7 keV and $\sigma \approx$ 180 km/s. They argued that the breaks result from the increasing efficiency of radiative cooling in low-mass systems. Cooling affects both the density and the hot gas fraction. 

It is clear that a deeper understanding of AGN feedback is required. Unfortunately, a complete theoretical picture of AGN feedback is still under debate. The current theoretical models need improvements with a new input physics and our scaling relations can be easily used for comparison with a new generation of simulations.

\section{Conclusions}\label{sec_summary}

We derived atmospheric temperature, density, gas masses, and total masses for 94 early type galaxies using archival $Chandra$ observations.  We derived X-ray scaling relations for the largest sample of ETGs to date.  The main results and conclusions can be summarized as follows.
\begin{itemize}
\item We derived X-ray scaling relations between luminosity, temperature, mass, and $Y_X$. We find $L_X \propto T^{4.42\pm0.19}$, $M \propto T^{2.43\pm0.19}$, $L_X \propto M^{2.78\pm0.33}$, $M \propto Y_X^{0.45\pm0.04}$. 

\item Our results are significantly steeper than self-similar expectations. The steepening of the relations is likely due to AGN feedback. The tight $L_X - T$ correlation for low-luminosities systems (i.e., below 10$^{40}$ erg/s) are at variance with hydrodynamical simulations which generally predict higher temperatures for low luminosity galaxies. 

\item We investigated the structural and dynamical properties of ETGs over a wide range of temperature (0.2 -- 2.0 keV), X-ray luminosity (10$^{38}$ -- 10$^{43}$ erg/s), and total mass (10$^{12}$ -- 10$^{13} M_{\odot}$). We found no correlation between the gas-to-total mass fraction with temperature or total mass.

%\item We also found that X-ray cores are characterized by higher temperatures, luminosities, metallicities, and masses than coreless ETGs. This is consistent with optically derived cores.

\end{itemize}

\acknowledgments

BRM acknowledges funding from the Natural Sciences and Engineering Research Council of Canada, the University of Waterloo's Faculty of Science, and the Perimeter Institute. ACE acknowledges support from STFC grant ST/P00541/1.  We thank A. Alabi for useful discussions related to the effective radius measurements, and K. Arnaud for comments related to spectral modeling. This research has made use of data obtained from the Chandra Data Archive and the Chandra Source Catalog, and software provided by the Chandra X-ray Center (CXC) in the application packages CIAO, ChIPS, and Sherpa. We thank all the staff members involved in the Chandra project. This research has made use of the NASA/IPAC Extragalactic (NED), SIMBAD, and LEDA databases. We also thank all members involved into the $DSS$ and $Spitzer$ collaborations. Additionally we have used ADS facilities.

\bibliographystyle{apj}
\bibliography{hanna}

\begin{thebibliography}{}
\expandafter\ifx\csname natexlab\endcsname\relax\def\natexlab#1{#1}\fi

\bibitem[{{Akritas} \& {Bershady}(1996)}]{Akritas:96}
{Akritas}, M.~G., \& {Bershady}, M.~A. 1996, \apj, 470, 706

\bibitem[{{Alabi} {et~al.}(2016){Alabi}, {Forbes}, {Romanowsky}, {Brodie},
  {Strader}, {Janz}, {Pota}, {Pastorello}, {Usher}, {Spitler}, {Foster},
  {Jennings}, {Villaume}, \& {Kartha}}]{Alabi:16}
{Alabi}, A.~B., {Forbes}, D.~A., {Romanowsky}, A.~J., {et~al.} 2016, \mnras,
  460, 3838

\bibitem[{{Alabi} {et~al.}(2017){Alabi}, {Forbes}, {Romanowsky}, {Brodie},
  {Strader}, {Janz}, {Usher}, {Spitler}, {Bellstedt}, \&
  {Ferr{\'e}-Mateu}}]{Alabi:17}
---. 2017, ArXiv e-prints, arXiv:1701.05904

\bibitem[{{Anderson} {et~al.}(2015{\natexlab{a}}){Anderson}, {Gaspari},
  {White}, {Wang}, \& {Dai}}]{Anderson:14}
{Anderson}, M.~E., {Gaspari}, M., {White}, S.~D.~M., {Wang}, W., \& {Dai}, X.
  2015{\natexlab{a}}, \mnras, 449, 3806

\bibitem[{{Anderson} {et~al.}(2015{\natexlab{b}}){Anderson}, {Gaspari},
  {White}, {Wang}, \& {Dai}}]{Anderson:15}
---. 2015{\natexlab{b}}, \mnras, 449, 3806

\bibitem[{{Arnaud}(1996)}]{Arnaud:96}
{Arnaud}, K.~A. 1996, in Astronomical Society of the Pacific Conference Series,
  Vol. 101, Astronomical Data Analysis Software and Systems V, ed. G.~H.
  {Jacoby} \& J.~{Barnes}, 17

\bibitem[{{Arnaud} \& {Evrard}(1999)}]{Arnaud:99}
{Arnaud}, M., \& {Evrard}, A.~E. 1999, \mnras, 305, 631

\bibitem[{{Babyk}(2016)}]{Babyk:16}
{Babyk}, I. 2016, Astronomy Reports, 60, 542

\bibitem[{{Babyk} \& {Vavilova}(2014)}]{Babyk:14a}
{Babyk}, I., \& {Vavilova}, I. 2014, \apss, 349, 415

\bibitem[{{Babyk} {et~al.}(2014){Babyk}, {Del Popolo}, \&
  {Vavilova}}]{Babyk:14}
{Babyk}, I.~V., {Del Popolo}, A., \& {Vavilova}, I.~B. 2014, Astronomy Reports,
  58, 587

\bibitem[{{Babyk} \& {Vavilova}(2012)}]{Babyk:13}
{Babyk}, I.~V., \& {Vavilova}, I.~B. 2012, Odessa Astronomical Publications,
  25, 119

\bibitem[{{Bender} {et~al.}(1989){Bender}, {Surma}, {Doebereiner},
  {Moellenhoff}, \& {Madejsky}}]{Bender:89}
{Bender}, R., {Surma}, P., {Doebereiner}, S., {Moellenhoff}, C., \& {Madejsky},
  R. 1989, \aap, 217, 35

\bibitem[{{Beuing} {et~al.}(1999){Beuing}, {Dobereiner}, {Bohringer}, \&
  {Bender}}]{Beuing_sample:99}
{Beuing}, J., {Dobereiner}, S., {Bohringer}, H., \& {Bender}, R. 1999, \mnras,
  302, 209

\bibitem[{{B{\^i}rzan} {et~al.}(2004){B{\^i}rzan}, {Rafferty}, {McNamara},
  {Wise}, \& {Nulsen}}]{Birzan:04}
{B{\^i}rzan}, L., {Rafferty}, D.~A., {McNamara}, B.~R., {Wise}, M.~W., \&
  {Nulsen}, P.~E.~J. 2004, \apj, 607, 800

\bibitem[{{Booth} \& {Schaye}(2010)}]{Booth:10}
{Booth}, C.~M., \& {Schaye}, J. 2010, \mnras, 405, L1

\bibitem[{{Borgani} {et~al.}(2005){Borgani}, {Finoguenov}, {Kay}, {Ponman},
  {Springel}, {Tozzi}, \& {Voit}}]{Borgani:05}
{Borgani}, S., {Finoguenov}, A., {Kay}, S.~T., {et~al.} 2005, \mnras, 361, 233

\bibitem[{{Borgani} \& {Viel}(2009)}]{Borgani:09}
{Borgani}, S., \& {Viel}, M. 2009, \mnras, 392, L26

\bibitem[{{Borgani} {et~al.}(2004){Borgani}, {Murante}, {Springel}, {Diaferio},
  {Dolag}, {Moscardini}, {Tormen}, {Tornatore}, \& {Tozzi}}]{Borgani:04}
{Borgani}, S., {Murante}, G., {Springel}, V., {et~al.} 2004, \mnras, 348, 1078

\bibitem[{{Borgani} {et~al.}(2006){Borgani}, {Dolag}, {Murante}, {Cheng},
  {Springel}, {Diaferio}, {Moscardini}, {Tormen}, {Tornatore}, \&
  {Tozzi}}]{Borgani:06}
{Borgani}, S., {Dolag}, K., {Murante}, G., {et~al.} 2006, \mnras, 367, 1641

\bibitem[{{Boroson} {et~al.}(2011){Boroson}, {Kim}, \& {Fabbiano}}]{Boroson:10}
{Boroson}, B., {Kim}, D.-W., \& {Fabbiano}, G. 2011, \apj, 729, 12

\bibitem[{{Bryan} \& {Norman}(1998)}]{Bryan:98}
{Bryan}, G.~L., \& {Norman}, M.~L. 1998, \apj, 495, 80

\bibitem[{{Cappellari} {et~al.}(2011){Cappellari}, {Emsellem}, {Krajnovic},
  {McDermid}, {Scott}, {Verdoes Kleijn}, {Young}, {Alatalo}, {Bacon}, {Blitz},
  {Bois}, {Bournaud}, {Bureau}, {Davies}, {Davis}, {de Zeeuw}, {Duc}, \&
  {Khochfar}}]{Cappellari:11}
{Cappellari}, M., {Emsellem}, E., {Krajnovic}, D., {et~al.} 2011, \mnras, 413,
  813

\bibitem[{{Cash}(1979)}]{Cash:79}
{Cash}, W. 1979, ApJ, 228, 939

\bibitem[{{Cavaliere} \& {Fusco-Femiano}(1978)}]{Cavaliere:78}
{Cavaliere}, A., \& {Fusco-Femiano}, R. 1978, \aap, 70, 677

\bibitem[{{Choi} {et~al.}(2015){Choi}, {Ostriker}, {Naab}, {Oser}, \&
  {Moster}}]{Choi:15}
{Choi}, E., {Ostriker}, J.~P., {Naab}, T., {Oser}, L., \& {Moster}, B.~P. 2015,
  \mnras, 449, 4105

\bibitem[{{Churazov} {et~al.}(2008){Churazov}, {Forman}, {Vikhlinin},
  {Tremaine}, {Gerhard}, \& {Jones}}]{Churazov:08}
{Churazov}, E., {Forman}, W., {Vikhlinin}, A., {et~al.} 2008, \mnras, 388, 1062

\bibitem[{{Churazov} {et~al.}(2010){Churazov}, {Tremaine}, {Forman}, {Gerhard},
  {Das}, {Vikhlinin}, {Jones}, {B{\"o}hringer}, \& {Gebhardt}}]{Churazov:10}
{Churazov}, E., {Tremaine}, S., {Forman}, W., {et~al.} 2010, \mnras, 404, 1165

\bibitem[{{Dav{\'e}} {et~al.}(2002){Dav{\'e}}, {Katz}, \& {Weinberg}}]{Dave:02}
{Dav{\'e}}, R., {Katz}, N., \& {Weinberg}, D.~H. 2002, \apj, 579, 23

\bibitem[{{David} {et~al.}(1991){David}, {Forman}, \& {Jones}}]{David:91}
{David}, L.~P., {Forman}, W., \& {Jones}, C. 1991, \apj, 369, 121

\bibitem[{{David} {et~al.}(2006){David}, {Jones}, {Forman}, {Vargas}, \&
  {Nulsen}}]{David:06}
{David}, L.~P., {Jones}, C., {Forman}, W., {Vargas}, I.~M., \& {Nulsen}, P.
  2006, \apj, 653, 207

\bibitem[{{Deason} {et~al.}(2012){Deason}, {Belokurov}, {Evans}, \&
  {McCarthy}}]{Deason:12}
{Deason}, A.~J., {Belokurov}, V., {Evans}, N.~W., \& {McCarthy}, I.~G. 2012,
  \apj, 748, 2

\bibitem[{{Dickey} \& {Lockman}(1990)}]{Dickey:90}
{Dickey}, J.~M., \& {Lockman}, F.~J. 1990, \araa, 28, 215

\bibitem[{{Diehl} \& {Statler}(2007)}]{Diehl:07}
{Diehl}, S., \& {Statler}, T.~S. 2007, \apj, 668, 150

\bibitem[{{Ettori}(2000)}]{Ettori:00}
{Ettori}, S. 2000, \mnras, 318, 1041

\bibitem[{{Evrard} {et~al.}(1996){Evrard}, {Metzler}, \& {Navarro}}]{Evrard:96}
{Evrard}, A.~E., {Metzler}, C.~A., \& {Navarro}, J.~F. 1996, \apj, 469, 494

\bibitem[{{Fabbiano}(1989)}]{Fabbiano:89}
{Fabbiano}, G. 1989, \araa, 27, 87

\bibitem[{{Faber}(1973)}]{Faber:73}
{Faber}, S.~M. 1973, \apj, 179, 423

\bibitem[{{Fabian} {et~al.}(2017){Fabian}, {Walker}, {Russell}, {Pinto},
  {Sanders}, \& {Reynolds}}]{Fabian:17}
{Fabian}, A.~C., {Walker}, S.~A., {Russell}, H.~R., {et~al.} 2017, \mnras, 464,
  L1

\bibitem[{{Forbes} {et~al.}(2016){Forbes}, {Alabi}, {Romanowsky}, {Brodie},
  {Strader}, {Usher}, \& {Pota}}]{Forbes:16}
{Forbes}, D.~A., {Alabi}, A., {Romanowsky}, A.~J., {et~al.} 2016, \mnras, 458,
  L44

\bibitem[{{Forman} {et~al.}(2017){Forman}, {Churazov}, {Jones}, {Heinz},
  {Kraft}, \& {Vikhlinin}}]{Forman:17}
{Forman}, W., {Churazov}, E., {Jones}, C., {et~al.} 2017, \apj, 844, 122

\bibitem[{{Fukazawa} {et~al.}(2006){Fukazawa}, {Botoya-Nonesa}, {Pu}, {Ohto},
  \& {Kawano}}]{Fukazawa:06}
{Fukazawa}, Y., {Botoya-Nonesa}, J.~G., {Pu}, J., {Ohto}, A., \& {Kawano}, N.
  2006, \apj, 636, 698

\bibitem[{{Gaspari} {et~al.}(2012){Gaspari}, {Brighenti}, \&
  {Temi}}]{Gaspari:12}
{Gaspari}, M., {Brighenti}, F., \& {Temi}, P. 2012, \mnras, 424, 190

\bibitem[{{Gaspari} {et~al.}(2014){Gaspari}, {Brighenti}, {Temi}, \&
  {Ettori}}]{Gaspari:14}
{Gaspari}, M., {Brighenti}, F., {Temi}, P., \& {Ettori}, S. 2014, \apjl, 783,
  L10

\bibitem[{{Giodini} {et~al.}(2013){Giodini}, {Lovisari}, {Pointecouteau},
  {Ettori}, {Reiprich}, \& {Hoekstra}}]{Giodini:13}
{Giodini}, S., {Lovisari}, L., {Pointecouteau}, E., {et~al.} 2013, \ssr, 177,
  247

\bibitem[{{Goulding} {et~al.}(2016){Goulding}, {Greene}, {Ma}, {Veale},
  {Bogdan}, {Nyland}, {Blakeslee}, {McConnell}, \& {Thomas}}]{Goulding:16}
{Goulding}, A.~D., {Greene}, J.~E., {Ma}, C.-P., {et~al.} 2016, \apj, 826, 167

\bibitem[{{Hlavacek-Larrondo} {et~al.}(2015){Hlavacek-Larrondo}, {McDonald},
  {Benson}, {Forman}, \& {Allen}}]{Hlavacek:15}
{Hlavacek-Larrondo}, J., {McDonald}, M., {Benson}, B.~A., {Forman}, W.~R., \&
  {Allen}, S.~W. 2015, \apj, 805, 35

\bibitem[{{Irwin} \& {Sarazin}(1996)}]{Irwin:96}
{Irwin}, J.~A., \& {Sarazin}, C.~L. 1996, \apj, 471, 683

\bibitem[{{Irwin} \& {Sarazin}(1998)}]{Irwin:98}
---. 1998, \apjl, 494, L33

\bibitem[{{Kaiser}(1986)}]{Kaiser:86}
{Kaiser}, N. 1986, \mnras, 222, 323

\bibitem[{{Kalberla} {et~al.}(2005){Kalberla}, {Burton}, {Hartmann}, {Arnal},
  {Bajaja}, {Morras}, \& {P{\"o}ppel}}]{Kalberla:05}
{Kalberla}, P.~M.~W., {Burton}, W.~B., {Hartmann}, D., {et~al.} 2005, \aap,
  440, 775

\bibitem[{{Kelly}(2007)}]{Kelly:07}
{Kelly}, B.~C. 2007, \apj, 665, 1489

\bibitem[{{Khosroshahi} {et~al.}(2004){Khosroshahi}, {Raychaudhury}, {Ponman},
  {Miles}, \& {Forbes}}]{Khos:04}
{Khosroshahi}, H.~G., {Raychaudhury}, S., {Ponman}, T.~J., {Miles}, T.~A., \&
  {Forbes}, D.~A. 2004, \mnras, 349, 527

\bibitem[{{Kim} \& {Fabbiano}(2013)}]{Kim:13}
{Kim}, D.-W., \& {Fabbiano}, G. 2013, \apj, 776, 116

\bibitem[{{Kim} \& {Fabbiano}(2015)}]{Kim:15}
---. 2015, \apj, 812, 127

\bibitem[{{Kravtsov} {et~al.}(2006){Kravtsov}, {Vikhlinin}, \&
  {Nagai}}]{Kravtsov:06}
{Kravtsov}, A.~V., {Vikhlinin}, A., \& {Nagai}, D. 2006, \apj, 650, 128

\bibitem[{{Ma} {et~al.}(2014){Ma}, {Greene}, {McConnell}, {Janish},
  {Blakeslee}, {Thomas}, \& {Murphy}}]{Ma_sample}
{Ma}, C.-P., {Greene}, J.~E., {McConnell}, N., {et~al.} 2014, \apj, 795, 158

\bibitem[{{Main} {et~al.}(2017){Main}, {McNamara}, {Nulsen}, {Russell}, \&
  {Vantyghem}}]{Main:17}
{Main}, R.~A., {McNamara}, B.~R., {Nulsen}, P.~E.~J., {Russell}, H.~R., \&
  {Vantyghem}, A.~N. 2017, \mnras, 464, 4360

\bibitem[{{Markevitch} {et~al.}(1998){Markevitch}, {Forman}, {Sarazin}, \&
  {Vikhlinin}}]{Markevitch:98}
{Markevitch}, M., {Forman}, W.~R., {Sarazin}, C.~L., \& {Vikhlinin}, A. 1998,
  \apj, 503, 77

\bibitem[{{Mathews}(1990)}]{Mathews:90}
{Mathews}, W.~G. 1990, \apj, 354, 468

\bibitem[{{Mathews} \& {Brighenti}(2003)}]{Mathews:03}
{Mathews}, W.~G., \& {Brighenti}, F. 2003, \apj, 599, 992

\bibitem[{{Maughan} {et~al.}(2012){Maughan}, {Giles}, {Randall}, {Jones}, \&
  {Forman}}]{Maughan:12}
{Maughan}, B.~J., {Giles}, P.~A., {Randall}, S.~W., {Jones}, C., \& {Forman},
  W.~R. 2012, \mnras, 421, 1583

\bibitem[{{McNamara} \& {Nulsen}(2007)}]{McNamara:07}
{McNamara}, B.~R., \& {Nulsen}, P.~E.~J. 2007, \araa, 45, 117

\bibitem[{{McNamara} \& {Nulsen}(2012)}]{McNamara:12}
---. 2012, New Journal of Physics, 14, 055023

\bibitem[{{Mei} {et~al.}(2007){Mei}, {Blakeslee}, {C{\^o}t{\'e}}, {Tonry},
  {West}, {Ferrarese}, {Jord{\'a}n}, {Peng}, {Anthony}, \& {Merritt}}]{Mei:07}
{Mei}, S., {Blakeslee}, J.~P., {C{\^o}t{\'e}}, P., {et~al.} 2007, \apj, 655,
  144

\bibitem[{{Moore}(1994)}]{Moore:94}
{Moore}, B. 1994, \nat, 370, 629

\bibitem[{{Nagai} {et~al.}(2007){Nagai}, {Vikhlinin}, \& {Kravtsov}}]{Nagai:07}
{Nagai}, D., {Vikhlinin}, A., \& {Kravtsov}, A.~V. 2007, \apj, 655, 98

\bibitem[{{Navarro} {et~al.}(2010){Navarro}, {Ludlow}, {Springel}, {Wang},
  {Vogelsberger}, {White}, {Jenkins}, {Frenk}, \& {Helmi}}]{Navarro:10}
{Navarro}, J.~F., {Ludlow}, A., {Springel}, V., {et~al.} 2010, \mnras, 402, 21

\bibitem[{{Negri} {et~al.}(2014){Negri}, {Posacki}, {Pellegrini}, \&
  {Ciotti}}]{Negri:14}
{Negri}, A., {Posacki}, S., {Pellegrini}, S., \& {Ciotti}, L. 2014, \mnras,
  445, 1351

\bibitem[{{Nulsen} {et~al.}(2009){Nulsen}, {Jones}, {Forman}, {Churazov},
  {McNamara}, {David}, \& {Murray}}]{Nulsen:09}
{Nulsen}, P., {Jones}, C., {Forman}, W., {et~al.} 2009, in American Institute
  of Physics Conference Series, Vol. 1201, American Institute of Physics
  Conference Series, ed. S.~{Heinz} \& E.~{Wilcots}, 198--201

\bibitem[{{Nulsen} {et~al.}(2005){Nulsen}, {McNamara}, {Wise}, \&
  {David}}]{Nulsen:05}
{Nulsen}, P.~E.~J., {McNamara}, B.~R., {Wise}, M.~W., \& {David}, L.~P. 2005,
  \apj, 628, 629

\bibitem[{{O'Sullivan} {et~al.}(2001){O'Sullivan}, {Forbes}, \&
  {Ponman}}]{OSullivan:01}
{O'Sullivan}, E., {Forbes}, D.~A., \& {Ponman}, T.~J. 2001, \mnras, 328, 461

\bibitem[{{O'Sullivan} {et~al.}(2003){O'Sullivan}, {Ponman}, \&
  {Collins}}]{OSullivan_sample:03}
{O'Sullivan}, E., {Ponman}, T.~J., \& {Collins}, R.~S. 2003, \mnras, 340, 1375

\bibitem[{{Paturel} {et~al.}(1997){Paturel}, {Andernach}, {Bottinelli}, {di
  Nella}, {Durand}, {Garnier}, {Gouguenheim}, {Lanoix}, {Marthinet}, {Petit},
  {Rousseau}, {Theureau}, \& {Vauglin}}]{Paturel:97}
{Paturel}, G., {Andernach}, H., {Bottinelli}, L., {et~al.} 1997, \aaps, 124,
  astro-ph/9806140

\bibitem[{{Pellegrini} {et~al.}(2012){Pellegrini}, {Ciotti}, \&
  {Ostriker}}]{Pellegrini:12}
{Pellegrini}, S., {Ciotti}, L., \& {Ostriker}, J.~P. 2012, \apj, 744, 21

\bibitem[{{Pellegrini} \& {Fabbiano}(1994)}]{Pellegrini:94}
{Pellegrini}, S., \& {Fabbiano}, G. 1994, \apj, 429, 105

\bibitem[{{Pratt} {et~al.}(2009){Pratt}, {Croston}, {Arnaud}, \&
  {B{\"o}hringer}}]{Pratt:09}
{Pratt}, G.~W., {Croston}, J.~H., {Arnaud}, M., \& {B{\"o}hringer}, H. 2009,
  \aap, 498, 361

\bibitem[{{Puchwein} {et~al.}(2008){Puchwein}, {Sijacki}, \&
  {Springel}}]{Puchwein:08}
{Puchwein}, E., {Sijacki}, D., \& {Springel}, V. 2008, \apjl, 687, L53

\bibitem[{{Rafferty} {et~al.}(2006){Rafferty}, {McNamara}, {Nulsen}, \&
  {Wise}}]{Rafferty:06}
{Rafferty}, D.~A., {McNamara}, B.~R., {Nulsen}, P.~E.~J., \& {Wise}, M.~W.
  2006, \apj, 652, 216

\bibitem[{{Randall} {et~al.}(2015){Randall}, {Nulsen}, {Jones}, {Forman},
  {Bulbul}, {Clarke}, {Kraft}, {Blanton}, {David}, {Werner}, {Sun}, {Donahue},
  {Giacintucci}, \& {Simionescu}}]{Randall:15}
{Randall}, S.~W., {Nulsen}, P.~E.~J., {Jones}, C., {et~al.} 2015, \apj, 805,
  112

\bibitem[{{Revnivtsev} {et~al.}(2007{\natexlab{a}}){Revnivtsev}, {Churazov},
  {Sazonov}, {Forman}, \& {Jones}}]{Revnivtsev:07a}
{Revnivtsev}, M., {Churazov}, E., {Sazonov}, S., {Forman}, W., \& {Jones}, C.
  2007{\natexlab{a}}, \aap, astro-ph/0702578

\bibitem[{{Revnivtsev} {et~al.}(2008{\natexlab{a}}){Revnivtsev}, {Churazov},
  {Sazonov}, {Forman}, \& {Jones}}]{Revnivtsev:08a}
---. 2008{\natexlab{a}}, \aap, arXiv:0804.0319

\bibitem[{{Revnivtsev} {et~al.}(2008{\natexlab{b}}){Revnivtsev}, {Lutovinov},
  {Churazov}, {Sazonov}, {Gilfanov}, {Grebenev}, \& {Sunyaev}}]{Revnivtsev:08b}
{Revnivtsev}, M., {Lutovinov}, A., {Churazov}, E., {et~al.} 2008{\natexlab{b}},
  \aap, arXiv:0805.0259

\bibitem[{{Revnivtsev} {et~al.}(2007{\natexlab{b}}){Revnivtsev}, {Vikhlinin},
  \& {Sazonov}}]{Revnivtsev:07b}
{Revnivtsev}, M., {Vikhlinin}, A., \& {Sazonov}, S. 2007{\natexlab{b}}, \aap,
  astro-ph/0611952

\bibitem[{{Schaye} {et~al.}(2010){Schaye}, {Dalla Vecchia}, {Booth}, {Wiersma},
  {Theuns}, {Haas}, {Bertone}, {Duffy}, {McCarthy}, \& {van de
  Voort}}]{Schaye:10}
{Schaye}, J., {Dalla Vecchia}, C., {Booth}, C.~M., {et~al.} 2010, \mnras, 402,
  1536

\bibitem[{{Shin} {et~al.}(2016){Shin}, {Woo}, \& {Mulchaey}}]{Shin:16}
{Shin}, J., {Woo}, J.-H., \& {Mulchaey}, J.~S. 2016, \apjs, 227, 31

\bibitem[{{Sijacki} \& {Springel}(2006)}]{Sijacki:06}
{Sijacki}, D., \& {Springel}, V. 2006, \mnras, 371, 1025

\bibitem[{{Sijacki} {et~al.}(2007){Sijacki}, {Springel}, {Di Matteo}, \&
  {Hernquist}}]{Sijacki:07}
{Sijacki}, D., {Springel}, V., {Di Matteo}, T., \& {Hernquist}, L. 2007,
  \mnras, 380, 877

\bibitem[{{Stanek} {et~al.}(2006){Stanek}, {Evrard}, {B{\"o}hringer},
  {Schuecker}, \& {Nord}}]{Stanek:06}
{Stanek}, R., {Evrard}, A.~E., {B{\"o}hringer}, H., {Schuecker}, P., \& {Nord},
  B. 2006, \apj, 648, 956

\bibitem[{{Su} \& {Irwin}(2013)}]{Su:13}
{Su}, Y., \& {Irwin}, J.~A. 2013, \apj, 766, 61

\bibitem[{{Su} {et~al.}(2015){Su}, {Irwin}, {White}, \& {Cooper}}]{Su:15}
{Su}, Y., {Irwin}, J.~A., {White}, III, R.~E., \& {Cooper}, M.~C. 2015, \apj,
  806, 156

\bibitem[{{Thomas} {et~al.}(2001){Thomas}, {Muanwong}, {Pearce}, {Couchman},
  {Edge}, {Jenkins}, \& {Onuora}}]{Thomas:01}
{Thomas}, P.~A., {Muanwong}, O., {Pearce}, F.~R., {et~al.} 2001, \mnras, 324,
  450

\bibitem[{{Vavilova} {et~al.}(2015){Vavilova}, {Bolotin}, {Boyarsky},
  {Danevich}, {Kobychev}, {Tretyak}, {Babyk}, {Iakubovskyi}, {Hnatyk}, \&
  {Sergeev}}]{Babyk_book:15}
{Vavilova}, I.~B., {Bolotin}, Y.~L., {Boyarsky}, A.~M., {et~al.} 2015, {Dark
  matter: Observational manifestation and experimental searches}

\bibitem[{{Vikhlinin} {et~al.}(2006){Vikhlinin}, {Kravtsov}, {Forman}, {Jones},
  {Markevitch}, {Murray}, \& {Van Speybroeck}}]{Vikhlinin:06}
{Vikhlinin}, A., {Kravtsov}, A., {Forman}, W., {et~al.} 2006, \apj, 640, 691

\bibitem[{{Vikhlinin} {et~al.}(2009){Vikhlinin}, {Burenin}, {Ebeling},
  {Forman}, {Hornstrup}, {Jones}, {Kravtsov}, {Murray}, {Nagai}, {Quintana}, \&
  {Voevodkin}}]{Vikhlinin:09}
{Vikhlinin}, A., {Burenin}, R.~A., {Ebeling}, H., {et~al.} 2009, \apj, 692,
  1033

\bibitem[{{Voit}(2005)}]{Voitr:05}
{Voit}, G.~M. 2005, Reviews of Modern Physics, 77, 207

\bibitem[{{Voit} {et~al.}(2002){Voit}, {Bryan}, {Balogh}, \& {Bower}}]{Voit:02}
{Voit}, G.~M., {Bryan}, G.~L., {Balogh}, M.~L., \& {Bower}, R.~G. 2002, \apj,
  576, 601

\bibitem[{{Werner} {et~al.}(2012){Werner}, {Allen}, \&
  {Simionescu}}]{Werner:12}
{Werner}, N., {Allen}, S.~W., \& {Simionescu}, A. 2012, \mnras, 425, 2731

\bibitem[{{White} \& {Sarazin}(1991)}]{White:91}
{White}, III, R.~E., \& {Sarazin}, C.~L. 1991, \apj, 367, 476

\bibitem[{{Willingale} {et~al.}(2013){Willingale}, {Starling}, {Beardmore},
  {Tanvir}, \& {O'Brien}}]{Willingale:13}
{Willingale}, R., {Starling}, R.~L.~C., {Beardmore}, A.~P., {Tanvir}, N.~R., \&
  {O'Brien}, P.~T. 2013, \mnras, 431, 394

\bibitem[{{Wong} {et~al.}(2014){Wong}, {Irwin}, {Shcherbakov}, {Yukita},
  {Million}, \& {Bregman}}]{Wong:14}
{Wong}, K.-W., {Irwin}, J.~A., {Shcherbakov}, R.~V., {et~al.} 2014, \apj, 780,
  9

\bibitem[{{Wu} {et~al.}(1999){Wu}, {Xue}, \& {Fang}}]{Wu:99}
{Wu}, X.-P., {Xue}, Y.-J., \& {Fang}, L.-Z. 1999, \apj, 524, 22

\end{thebibliography}

\end{document}